\documentclass[12pt,a4paper]{article}

\usepackage{amssymb,amsmath,amsfonts,mathrsfs}
\usepackage{verbatim,amsthm,curves,graphics}
\usepackage{mathrsfs}
\usepackage{fullpage}
\usepackage{graphicx}
\usepackage{caption}
\usepackage[utf8]{inputenc}
\usepackage{tikz}
\usetikzlibrary{shapes.geometric}
\usetikzlibrary{decorations.markings}
\usetikzlibrary{decorations.pathmorphing}

\def\ben{\begin{equation}}
\def\een{\end{equation}}
\def\bena{\begin{eqnarray}}
\def\eena{\end{eqnarray}}
\def\non{\nonumber}

\renewcommand{\ell}{l}

\newcommand{\half}{\frac{1}{2}}
\newcommand{\X}{\mathscr{X}}
\renewcommand{\S}{{\mathscr S}}
\newcommand{\fz}{(\tfrac{\partial}{\partial z})}
\newcommand{\dl}{\tfrac{d}{d\lambda}}
\newcommand{\plo}{\tfrac{\partial}{\partial \lambda_1}}
\newcommand{\plt}{\tfrac{\partial}{\partial \lambda_2}}
\newcommand{\plot}{\tfrac{\partial^2}{\partial \lambda_1 \partial \lambda_2}}
\newcommand{\Dl}{\frac{d}{d\lambda}}

\theoremstyle{definition}
\newtheorem{thm}{Theorem}

\newtheorem{lemma}{Lemma}
\newtheorem{prop}{Proposition}

\newtheorem{defn}{Definition}[section]

\theoremstyle{definition}

\renewcommand{\H}{\mathscr{H}}
\newcommand{\M}{\mathscr{M}}
\renewcommand{\pounds}{{\mathscr L}}

\newcommand{\D}{\mbox{d}}

\newcommand{\bE}{\overline{{\mathcal E}}}

\newcommand{\mr}{\mathbb{R}}

\newcommand{\I}{\mathscr{I}}

\newcommand{\V}{\mathcal{V}}
\newcommand{\W}{\mathcal{W}}

\newcommand{\e}{{\rm e}}

\newcommand{\E}{{\mathcal{E}}}
\newcommand{\U}{{\mathcal{U}}}

\renewcommand{\L}{{\mathcal L}}

\begin{document}

\title{Stability of Black Holes and Black Branes}

\author{Stefan Hollands$^{1}$\thanks{\tt HollandsS@cardiff.ac.uk}
and Robert M. Wald$^{2}$\thanks{\tt rmwa@uchicago.edu}
\\ \\
{\it ${}^{1}$School of Mathematics,
     Cardiff University,} \\
{\it Cardiff, United Kingdom} \\
{\it ${}^{2}$Enrico Fermi Institute and Department of Physics, } \\
{\it The University of Chicago, Chicago, IL 60637, USA}}

\maketitle

\begin{abstract}

  We establish a new criterion for the dynamical stability of black
  holes in $D \geq 4$ spacetime dimensions in general relativity with
  respect to axisymmetric perturbations: Dynamical stability is
  equivalent to the positivity of the canonical energy, $\E$, on a
  subspace, $\mathcal T$, of linearized solutions that have vanishing
  linearized ADM mass, momentum, and angular momentum at infinity and
  satisfy certain gauge conditions at the horizon. This is shown by
  proving that---apart from pure gauge perturbations and perturbations
  towards other stationary black holes---$\E$ is nondegenerate on
  $\mathcal T$ and that, for axisymmetric perturbations, $\E$ has
  positive flux properties at both infinity and the horizon. We
  further show that $\E$ is related to the second order variations of
  mass, angular momentum, and horizon area by $\E = \delta^2 M -
  \sum_A \Omega_A \delta^2 J_A - \frac{\kappa}{8\pi} \delta^2 A$,
  thereby establishing a close connection between dynamical stability
  and thermodynamic stability. Thermodynamic instability of a family
  of black holes need not imply dynamical instability because the
  perturbations towards other members of the family will not, in
  general, have vanishing linearized ADM mass and/or angular
  momentum. However, we prove that for any black brane corresponding
  to a thermodynamically unstable black hole, sufficiently long
  wavelength perturbations can be found with $\E < 0$ and vanishing
  linearized ADM quantities. Thus, all black branes corresponding to
  thermodynmically unstable black holes are dynamically unstable, as
  conjectured by Gubser and Mitra. We also prove that positivity of
  $\E$ on $\mathcal T$ is equivalent to the satisfaction of a ``local
  Penrose inequality,'' thus showing that satisfaction of this local
  Penrose inequality is necessary and sufficient for dynamical
  stability. Although we restrict our considerations in this paper to
  vacuum general relativity, most of the results of this paper are
  derived using general Lagrangian and Hamiltonian methods and
  therefore can be straightforwardly generalized to allow for the
  presence of matter fields and/or to the case of an arbitrary
  diffeomorphism covariant gravitational action.

\end{abstract}

\tableofcontents

\section{Introduction}

It is of considerable interest to determine the linear stability of
black holes in $D$-dimensional general relativity. It is also of interest
to determine the linear stability of the corresponding black branes
in $(D+p)$-dimensions, i.e., spacetimes $\M \times {\mathbb T}^p$ with metric of the form
\ben\label{ds}
\D \tilde s^2_{D+p} = \D s^2_D + \sum_{i=1}^p \D z_i^2 \ .
\een
where $\D s^2_D$ is a black hole metric. One can analyze this issue
by writing out the
linearized Einstein equation off of the black hole or black brane
background spacetime. One way of establishing linear stability
is to find a suitable positive definite conserved norm
for perturbations. Linear
instability can be established by finding a solution for which some
gauge independent quantity has unbounded growth in time. However, even in the
very simplest cases---such as the Schwarzschild black hole~\cite{reggewheeler,zerilli}
and the Schwarzschild
black string~\cite{gregory}---it is highly nontrivial to carry out
the decoupling of equations and the fixing of gauge needed
to determine stability or instability directly from the equations of motion.
It is particularly difficult to analyze
stability when the background is stationary but not static.
In the case of a Kerr black hole
in $4$-dimensions, the Teukolsky formalism~\cite{teukolsky} reduces
the coupled system of linear perturbation equations to a
single equation for a gauge invariant, complex variable, but
there is no known formalism
of comparable power to decouple the equations
in other cases, including the higher dimensional Myers-Perry analogs of Kerr;
see~\cite{reall,reall2,dias} and references therein. Thus, it would be extremely useful to have
a stability criterion for black holes and black branes that does
not require one to perform a complete analysis of the linearized
perturbation equations.

For the case of black branes, a simple criterion for stability was
proposed in~\cite{gubser1,gubser2}, based on the analogy between laws
of black hole mechanics and the laws of thermodynamics.
We now describe their criterion. Let us assume for simplicity that we are given a
stationary, asymptotically flat black hole solution with compact event horizon cross section, in vacuum
general relativity (no matter fields) in $D$ spacetime dimensions.
By the rigidity theorem~\cite{hiw,isenberg},
we may assume that our solution is not only
stationary and asymptotically flat, but has additional commuting rotational
Killing fields $\psi^a_1, \dots, \psi^a_N$,
with associated angular momenta $J_1, \dots, J_N$, where $N \le \lfloor \frac{D-1}{2} \rfloor$. Suppose now that we do not just have
one such solution, but---as usually happens---a family of solutions parameterized
by the mass $M$, and the angular momenta\footnote{In the presence of matter fields,
there may be additional conserved charges.} $J_1, \dots, J_N$, such as the Myers-Perry family~\cite{myers}.
As usual, we identify the entropy as $S = \frac{1}{4}A$, where $A = A(M,J_1,\dots,J_N)$ is the area of the horizon.
Consider the Hessian matrix of $S$,
\ben
{\rm Hess}_S = \left(
\begin{matrix}
\frac{\partial^2 S}{\partial M^2} & \frac{\partial^2 S}{\partial J_A \partial M}\\
\frac{\partial^2 S}{\partial M \partial J_B} & \frac{\partial^2 S}{\partial J_A \partial J_B}\\
\end{matrix}
\right) \ .
\label{Hess}
\een
Then by analogy with the thermodynamical properties of ordinary laboratory type systems, the signs of the
eigenvalues of ${\rm Hess}_S$ might be expected to be related to the
stability of the system: For normal laboratory-type systems we expect the system to be unstable if the Hessian has a
positive eigenvalue, which would correspond to a negative heat capacity.
Therefore, one might expect that negativity of the Hessian,  eq.~(\ref{Hess}),
would be a necessary condition for black hole stability. However, this is clearly false:
For example, the $4$-dimensional Schwarzschild black hole has
negative heat capacity, since its area is
$A = 16\pi M^2$, so $\partial^2 S/\partial M^2 >0$. However,
despite its negative heat capacity, the Schwarzschild black hole
is well known to be linearly
stable. Thus, for black holes, thermodynamic stability
with respect to perturbations to other stationary black holes cannot be
necessary for dynamical stability.
Nevertheless, there is no obvious counterexample to the possibility that
thermodynamic stability could be necessary for black brane stability, since,
in particular,
the Schwarzschild black string is known to be unstable~\cite{gregory}.
Therefore, suppose we exclude consideration of black holes and restrict attention to
black brane spacetimes of the form~(\ref{ds}), where the $z_i$
coordinates are periodically identified with length $l$, so that horizon cross-sections
are compact and have a well defined area.
The Gubser-Mitra conjecture states that any such black brane spacetime is unstable for sufficiently
large $l$ if the Hessian matrix (of the original $D$-dimensional black hole family) has a
positive eigenvalue. In this paper, we will show that this conjecture follows as a consequence of
a more fundamental dynamical stability criterion that we shall establish.

Another simple possible stability criterion that is applicable to black holes
is the ``local Penrose inequality,''
discussed recently in~\cite{reallfig}. We reformulate this criterion as follows:
Suppose, as above, we have a family
of stationary, axisymmetric black hole
spacetimes with mass $M$ and angular momenta
$J_1, \dots, J_N$. Since surface gravity cannot be negative,
it follows immediately from the first law of black hole mechanics that
the horizon area, $\bar A$,
must be a non-decreasing function of $M$ at fixed $J_1, \dots, J_N$.
Let $g_{ab}(\lambda)$ be an arbitrary smooth one-parameter family
of axisymmetric metrics with $g_{ab}(0)$ being equal to a member of
this family with non-zero surface gravity. Let $B$ denote the bifurcation
surface of the Killing horizon of $g_{ab}(0)$.
Let $\mathcal{A}(\lambda)$
denote the area of the apparent horizon of $g_{ab}(\lambda)$ on an
initial slice passing through $B$.
As will be shown in subsection 2.1 below, to first order in
$\lambda$ the event horizon coincides with the apparent horizon. To
second order in $\lambda$, the event horizon need not coincide with the
apparent horizon---assuming cosmic censorship, it cannot lie inside the
apparent horizon but it may lie outside of it. Nevertheless,
to second order, the only contribution to
the difference between the area, $A$, of the
event horizon and $\mathcal{A}$ arises from the (second order)
displacement of the event horizon relative to the apparent horizon in
the background metric $g_{ab}(0)$. However, in the background
metric, $B$ is an extremal
surface\footnote{This follows from the fact
that both its outgoing and ingoing expansion vanish.}. It follows that
to second order, we have
$A(\lambda) = \mathcal{A}(\lambda)$.
Now, let $\bar{g}_{ab}(\lambda)$ be the one
parameter family of metrics in the black hole family such that
$\bar{M}(\lambda),\bar{J}_1(\lambda), \dots, \bar{J}_N(\lambda)$ agrees with the corresponding
ADM mass
and angular momenta $M(\lambda), J_1(\lambda), \dots, J_N(\lambda)$
of $g_{ab}(\lambda)$.
Then, by the first law of black hole mechanics together with the fact
that the bifurcation surface of $g_{ab}(0)$ has vanishing extrinsic curvature,
we automatically have
$(d\mathcal{A}/d \lambda)|_{\lambda =0} = (d \bar{\mathcal A}/d \lambda)|_{\lambda =0}$.
However,
suppose that $\delta^2 \mathcal{A} \equiv (d^2\mathcal{A}/d \lambda^2)|_{\lambda =0} > \delta^2 \bar{ \mathcal A} \equiv (d^2 \bar{\mathcal A}/d \lambda^2)|_{\lambda =0}$.
Then, since horizon area is nondecreasing with time (assuming the validity of cosmic censorship),
$M$ is non-increasing with time (by positivity of Bondi flux), and $J_A$ is conserved (by
axisymmetry), we obtain a contradiction with the possibility that $g_{ab}(\lambda)$ could
``settle down'' at late times to a black hole solution in the original family. This strongly suggests
that satisfaction of the ``local Penrose inequality''
\ben
\delta^2 \mathcal{A} \leq \delta^2 \bar{\mathcal A}
\label{lpe}
\een
should be a necessary condition for stability of any black hole member of a family. This argument
does not suggest that satisfaction of the local Penrose
inequality should be sufficient for
stability with respect to axisymmetric perturbations, but
evidence in favor of this conjecture
has been presented in~\cite{reallfig}.
In this paper, we will prove that our more fundamental
stability criterion implies that satisfaction of the local Penrose
inequality is necessary and sufficient for black hole
stability with respect to axisymmetric
perturbations.

We now describe our main results. We consider stationary,
asymptotically flat black holes in $D \geq 4$ spacetime dimensions
in vacuum general relativity. In section 5, we will also consider
the corresponding black branes in $D + p$ dimensions.
By the rigidity theorem~\cite{hiw,isenberg}, the event horizon of the black hole
must be a Killing horizon,
and, if rotating, the spacetime
must be axisymmetric (possibly with respect to
multiple rotational planes if $D \geq 5$).
We restrict consideration
to the case where the surface gravity,
$\kappa$, of the event horizon is non-vanishing.
In the case of a
nonrotating black hole, our results apply to arbitrary, smooth, asymptotically flat
solutions of the linearized field equations, but for a rotating stationary, axisymmetric
black hole, some of our results apply only to perturbations
that preserve axisymmetry in the rotational planes that have
nonvanishing angular velocity, $\Omega_A$,
of the horizon.
For perturbations of the black hole,
we define a quantity $\E$---called the {\it canonical energy}---which is quadratic in the perturbation
and given by an integral over a
Cauchy surface, $\Sigma$, for the exterior region. (Thus, $\Sigma$
extends from spatial infinity to the bifurcation surface, $B$, of the
black hole.)  We will establish the following
properties of $\E$: (1) Without loss of generality, certain gauge
conditions can be imposed near the horizon of the black hole
that, in particular, ensure that the location of the true event horizon
does not change to first order. When restricted to axisymmetric perturbations with vanishing linearized
linear momenta, $\delta P_i = 0$,
and vanishing linearized change of horizon area, $\delta A = 0$,
the quantity $\E$ is invariant under
axisymmetric gauge transformations that
satisfy these conditions at the horizon and approach arbitrary asymptotic
symmetries at infinity. (2) $\E$ is conserved in the sense that it takes the same value
if evaluated on another
Cauchy surface $\Sigma'$ extending from spatial infinity to $B$.
(3) For axisymmetric perturbations $\E$ is related to the second
order variations of mass, $M$, horizon area, $A$, and angular momentum,
$J_A$ by\footnote{Although $\delta^2 M$, $\delta^2 A$, and $\delta^2 J_A$
individually depend upon the second order metric perturbation,
it follows from the first law of black hole mechanics that the combination
appearing on the right side of \eqref{thermo} is independent of the
choice of second order metric perturbation.}
\ben \E = \delta^2 M - \frac{\kappa}{8
  \pi} \delta^2 A - \sum_A \Omega_A \ \delta^2 J_A \, .
\label{thermo}
\een
(4) A Hilbert space $\V$ can be defined whose elements are square integrable initial
data that (weakly) satisfy the linearized constraints, satisfy our horizon gauge conditions and, in
addition, at spatial infinity have vanishing linearized ADM mass, linear momentum, and
angular momentum with respect to the rotational Killing fields of the
background spacetime. There is a dense domain, $\mathcal T$, of smooth elements of $\V$ on which
$\E$ can be defined as a quadratic form.
$\E$ is symmetric on ${\mathcal T} \times \mathcal T$
and is degenerate precisely on the perturbations
that are ``perturbations toward stationary black holes.''  Thus, if we define a
new space $\mathcal T'$ by modding out by
perturbations toward stationary black holes (which, of course, includes all pure gauge
perturbations), $\E$ is non-degenerate.  Consequently, on this space,
either (a) $\E$ is positive definite or (b) there is a $\Psi \in \mathcal T'$
such that $\E(\Psi, \Psi) < 0$.  (5) Consider a
foliation by Cauchy surfaces, $\Sigma(t)$ for the exterior region such
that each $\Sigma(t)$ is composed of a hypersurface, $\mathscr{S}(t)$,
that extends from a cross-section $C(t)$ of future null
infinity\footnote{In order to make use of the machinery of null
  infinity, we must restrict consideration to even dimensional
  spacetimes, since $\I$ does not exist in odd dimensions~\cite{hw}. We do not believe that this is an essential
  restriction, i.e., we believe that our results on the positivity of
  flux at infinity will also hold in odd dimensions.} $\I^+$ to a
cross-section $B(t)$ of the black hole horizon $\H^+$, together with the
portions of future null infinity and the future horizon that lie to
the past of $C(t)$ and $B(t)$ respectively.
For perturbations
that preserve axisymmetry in the rotational planes that have
nonvanishing angular velocity, $\Omega_A$,
of the horizon, the
canonical energy
on \textcolor{red}{$\Sigma(t)$} (drawn as the
broken \textcolor{red}{red} line in the figure)
can be expressed as the sum of a
manifestly non-negative contribution
from $\I^+$ (equal to the Bondi energy flux), a
manifestly non-negative contribution from $\H^+$ (given by an integral involving
the square of the linearized shear), and a term $\overline{\E}(t)$
which differs from the canonical energy evaluated on $\mathscr{S}(t)$
only by boundary contributions from $C(t)$ and $B(t)$. Furthermore,
the boundary contribution from $C(t)$ vanishes if the Bondi news
vanishes on $C(t)$ and the boundary contribution from $B(t)$ vanishes
if the perturbed shear of the horizon vanishes on $B(t)$.

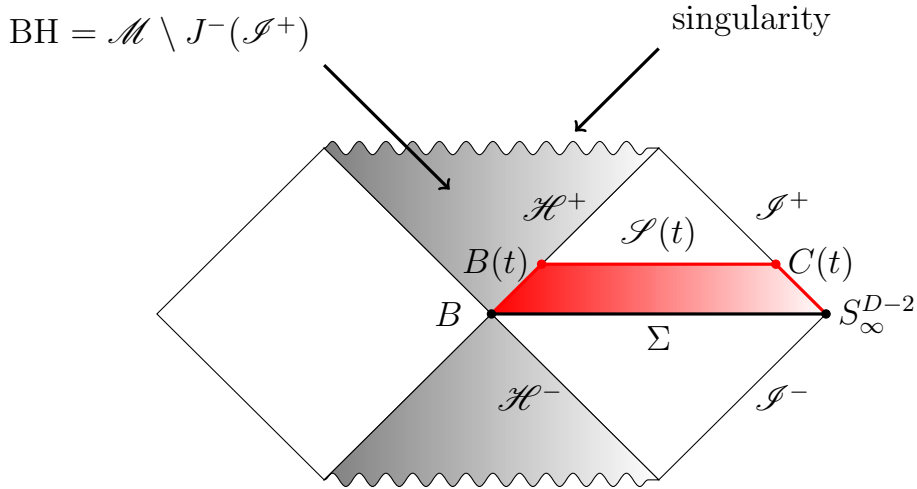
\begin{center}
\begin{tikzpicture}[scale=1.1, transform shape]
\shade[left color=red] (0,0) -- (0.6,0.6) -- (3.4,.6) -- (4,0);
\draw (0,0) -- (-2,2) -- (-4,0)  -- (-2,-2) -- (0,0);
\draw (4,0) node[right]{$S^{D-2}_{\infty}$};
\draw (0,0) -- (2,2) -- node[above right] (c) {$\I^+$}  (4,0) -- node[right] (a) {$\I^-$} (2,-2) -- (0,0);
\shade[left color=gray] (0,0) -- (-2,2) decorate[decoration=snake] {-- (2,2)} --  (0,0);
\draw (0,0) -- (-2,2) decorate[decoration=snake] {-- (2,2)} -- (0,0);
\shade[left color=gray] (0,0) -- (-2,-2) decorate[decoration=snake] {-- (2,-2)} -- (0,0);
\draw (0,0) -- (-2,-2) decorate[decoration=snake] {-- (2,-2)} -- (0,0);
\draw[->, very thick] (2,3.2) node[above right]{singularity} -- (1, 2.2);
\draw[very thick, red] (.6,.6) -- (3.4,.6);
\draw[very thick, red] (0,0) -- (.6,.6);
\draw[very thick, red] (3.4,.6) -- (4,0);
\draw[very thick, black] (0,0) -- (4,0);
\draw (2,0) node[below]{$\Sigma$};
\draw (.6,.6) node[above, left]{$B(t)$};
\draw (3.4,.6) node[above, right]{$C(t)$};
\filldraw[color=red] (.6,.6)  circle (.05cm);
\filldraw[color=red] (3.4,.6)  circle (.05cm);
\filldraw (0,0) circle (.05cm);
\filldraw (4,0) circle (.05cm);
\draw (2,.6) node[above]{${\mathscr S}(t)$};
\draw[->, very thick] (-2,3) node[above left] {BH $=\M \setminus J^{-}(\mathscr{I}^{+})$} -- (-.5,1.5);
\draw (1.3,1.3) node[black,above,left]{$\H^+$};
\draw (1,-1) node[black,below,left]{$\H^-$};
\draw(-0.2,0) node[black, left]{$B$};
\end{tikzpicture}
\end{center}

A stability criterion for black holes and black branes with respect to axisymmetric perturbations
can now be formulated as follows:
Suppose that $\E$ is positive definite on $\mathcal T'$ (case (a) of point (4) above).
Since $\E$ is conserved, we obtain a positive definite conserved norm on ${\mathcal T}'$. Since
${\mathcal T}'$
excludes from ${\mathcal T}$ only those perturbations that are
towards stationary black holes---and stationary
perturbations are automatically stable---this establishes
stability of perturbations in ${\mathcal T}$, at least in the sense of establishing the nonexistence of
``growing modes''. Furthermore,
if the black hole or black brane whose stability is being investigated
is a member of a family\footnote{If the black hole or black brane in question is not a member of a general family, one
would have to separately establish the
existence of a stable perturbation
for given variations of $M, J_1, \dots, J_N$.}
of stationary black holes/branes
containing arbitrary variations of
 $M, J_1, \dots, J_N$, then a general perturbation can be written as the sum of a perturbation in ${\mathcal T}$ and
 a perturbation to another member of the family,
thereby establishing stability for general perturbations.

On the other hand, suppose $\E$ fails to be positive definite on ${\mathcal T}'$
and hence, according to point (4) above, there exists a perturbation in
${\mathcal T}'$ for which $\E < 0$ (case (b)).  We obtain a
contradiction with the stability of the black hole or black brane by
the following type of argument used previously in~\cite{friedman1, friedman, schutzfried,  friedman2}:
By property (5), the quantity $\overline{\E}(t)$ is
non-increasing with time, and, since $\overline{\E} = \E$ on $\Sigma$, for
this perturbation we have $\overline{\E} (t) \leq \E <
0$. If the black hole/brane were stable, then the Bondi flux at $\I^+$
and the perturbed shear at $\H^+$ should go to zero at late times for this
perturbation. Hence, $\overline{\E} (t)$ should approach the canonical
energy evaluated on $\mathscr{S}(t)$. However,
we have just seen that the canonical energy is
bounded away from zero on $\mathscr{S}(t)$ at late times. Consequently, the
perturbation cannot be approaching a solution that is a perturbation towards
a stationary black hole since the linearized mass, linear momentum, and
angular momentum must remain zero\footnote{Mass and momentum can be radiated at
quadratic order; angular momentum cannot be radiated at all for axisymmetric spacetimes.} and
$\E$ vanishes for perturbations towards
a stationary black hole with vanishing linearized mass, linear momentum, and
angular momentum. On the
other hand, as already stated, any stable perturbation should become
``non-radiating'' at late times.  However, this should be impossible,
since non-stationary, non-radiating linearized perturbations should
not exist.

Our argument for stability in case (a) is a genuine proof that
perturbations remain bounded in a suitable norm, although it does not
establish pointwise boundedness and does not establish
decay\footnote{In the stable case, the arguments of the previous
  paragraph strongly suggest that perturbations in $\mathcal T$ must decay to
  solutions that are perturbations towards stationary black holes.}.  On the other
hand, our argument for instability in case (b) has a status closer to
that of a plausibility argument than a complete and rigorous proof. At the end of
subsection 4.2 below, we shall indicate some possible strategies for improving
upon these results.

At a conceptual level, our above criterion for black hole
and black brane stability is
remarkably simple: Evaluate the canonical energy $\E$ of initial data
for perturbations that (i) satisfy the constraints, (ii) have
vanishing linearized ADM mass, momentum, and angular momentum at spatial infinity, and
(iii) satisfy our gauge conditions at the horizon. If $\E \geq 0$ for
all such perturbations, then the black hole/brane is stable; if not,
then it is unstable. Thus, $\E$ provides a ``variational
principle'' for black hole and black brane stability, and one can test
for stability by plugging in ``trial initial data'' into the formula
for $\E$. Of course, the trial initial data must be chosen so as to
satisfy the linearized constraints, so this variational principle
cannot be used as readily as unconstrained variational
principles\footnote{For the case of spherically symmetric
  perturbations of static, spherically symmetric spacetimes with
  appropriate matter fields, the linearized constraints can be
  solved. In that case, our variational principle (when generalized to include matter)
  would reduce to the
  unconstrained variational principle given in~\cite{seifert}.}. Nevertheless, the use of our variational
principle---even if it requires solving the constraints---represents
an enormous simplification over performing a complete analysis of the
dynamical behavior of solutions to the full set of perturbation
equations.

If we have a family of black holes or black branes parameterized by $M, J_1, \dots, J_N$,
then we shall show at the beginning of section 5 that positivity of the right side of \eqref{thermo} for
all second order variations within this family is equivalent to the negative
definiteness of the Hessian \eqref{Hess}. Thus, thermodynamic stability
of the family as defined above by the Hessian criterion
is equivalent to positivity of $\E$ for perturbations within the family. However, our
criterion for dynamical stability is positivity of $\E$
{\it for the case of perturbations with vanishing
linearized ADM mass, momentum, and angular momentum at spatial infinity}, and this excludes all perturbations
within the family. Thus, for a family of black holes,
the thermodynamic stability of the family provides absolutely no information about dynamical
stability. In particular, this explains why the Schwarzschild
black hole can be dynamically
stable despite the fact that it is
thermodynamically unstable: The thermodynamically
unstable perturbations of Schwarzschild
have a nonvanishing linearized ADM mass and thus do not ``count''
for the analysis of dynamical stability.
However, the situation is very different for a family of black branes. If the corresponding family
of black holes is thermodynamically
unstable---i.e., if we can find a perturbation to another member of the black hole family that makes
$\E < 0$---then we shall show in section 5
that one can find a sufficiently long wavelength perturbation of the black brane
for which $\E < 0$ but
the linearized ADM mass, momentum, and angular momentum vanish. Consequently, the thermodynamic
instability of the corresponding family of black holes implies dynamical instability of the black branes,
as conjectured by Gubser and Mitra\footnote{Another argument in favor of the Gubser Mitra conjecture based on a ``Wick rotation'', applicable to certain {\em static} black branes, has previously been given by~\cite{reall3}.}.

In this paper, we will restrict attention to black holes (and black branes)
in vacuum general relativity.
However, one of the strengths of our approach is that it is based upon the Lagrangian formulation
of general relativity and, hence, can be applied to non-vacuum general relativity as well
as other diffeomorphism covariant theories of gravity. In particular, the notion of canonical
energy can be readily generalized to these cases, and it
appears likely that analogs of properties (1)-(4) above will hold. Furthermore, property
(5) should hold in non-vacuum general relativity with matter satisfying appropriate energy
conditions, whereas the satisfaction of property (5) in other theories of gravity should be
intimately related to the positivity of energy flux at infinity and the satisfaction of the
second law of black hole mechanics in these theories. Thus, it is likely
that most of our stability analysis can be
applied to much more general cases. However, we shall defer
to a subsequent publication the analysis of
black hole and black brane stability in non-vacuum
general relativity and other theories of gravity.

Finally, we should emphasize that, as stated above, for rotating black holes our stability analysis
is restricted to perturbations that are axisymmetric in the planes in which $\Omega_A$
is nonvanishing. This restriction appears to be essential since,
for rotating black holes, one should easily be able to find
non-axisymmetric perturbations for which $\E < 0$
and for which the flux of canonical energy through $\H^+$ (or $\I^+$ if one instead
uses the horizon Killing field
to define $\E$) is negative. These properties should preclude the possibility of making either a stability
or an instability argument of the type described above.

In the next section, we impose our horizon gauge conditions, define canonical energy, and establish
many of its properties. In section 3 we prove our results on the flux of canonical energy at infinity and
through the horizon. Our stability/instability arguments---including the construction of the
Hilbert space $\V$ and the domain, $\mathcal T$, of $\E$---are given in section 4.
In section 5, we present a proof that the thermodynamic instability of a family of black hole implies
the existence of a long wavelength perturbation of the corresponding black branes
with negative $\E$ that
has vanishing
linearized ADM mass, momentum, and angular momentum. This implies instability of the
black brane, as conjectured by Gubser and Mitra.
In section 6, we shall show that for a family of black holes, satisfaction of the local Penrose inequality~(\ref{lpe}) is equivalent to positivity of $\E$
on perturbations with vanishing
linearized ADM mass, momentum, and angular momentum,
and, consequently, for black holes,
the satisfaction of the
local Penrose inequality is equivalent to linear stability.

\section{Canonical Energy}

\subsection{Gauge choice near the horizon}\label{gauge}

In this section, we consider one-parameter families of metrics $g_{ab}(\lambda)$ (not
necessarily stationary) in $D \geq 4$ dimensions, that solve the field equations $G_{ab}(\lambda)
= 0$, are jointly smooth in the parameter $\lambda \in \mr$ and spacetime point, and which, for
$\lambda=0$, reduce to a stationary, asymptotically flat black hole spacetime
$g_{ab}=g_{ab}(0)$, with bifurcate Killing
horizon $\H$, bifurcation surface $B \subset \H$, and surface
gravity $\kappa>0$.
We also will consider corresponding families of black
brane spacetimes, but the extension of our considerations to black brane spacetimes
is straightforward, so we will restrict attention to black holes until section~5.

We shall require $g_{ab}(\lambda)$ to be asymptotically flat in the sense that
there exist asymptotically Minkowskian coordinates $x^\mu$ such that
$g_{\mu \nu} (\lambda)$ differs from $\eta_{\mu \nu}$ by $O(\rho^{-(D-3)})$
as $\rho \rightarrow \infty$---where $\rho=(x_1^2+\dots+x_{D-1}^2)^\half$---and $N$th derivatives
of $g_{\mu \nu} (\lambda)$ are $O(\rho^{-(D-3+N)})$.
In $D=4$, we also require the Regge-Teitelboim
parity conditions (see appendix~A of~\cite{reggeteit}) to hold.
In section 3, we will, in addition, impose asymptotic flatness
conditions on the metric at $\I^+$.

We shall now impose certain gauge conditions on the metric
near the horizon that will be used in the remainder of this paper.
These conditions involve no loss of generality.
Let $\H$ denote the black hole event horizon for the metric
$g_{ab}(\lambda = 0)$, and $\H^\pm = \H \cap I^\pm(\I^\mp)$ the
future and past component. When $\lambda > 0$, the surface lying at the
same coordinate location as $\H^+$ need not coincide with the actual event horizon
of $g_{ab}(\lambda)$---indeed,
there may not even be an event horizon when $\lambda>0$. It will be
very useful to require that our coordinates be chosen so that, when there is a horizon
for $\lambda > 0$,
the surface lying at the same coordinate location as $\H^+$ (we could equally
choose $\H^-$) continues
to coincide with the actual event horizon, at least to first order in
$\lambda$. Since the event horizon is
a global concept and cannot easily be located
exactly in any spacetime with $\lambda > 0$, this might appear to be a hopelessly
difficult task. However, as we shall see, the event horizon can be easily located to first
order in $\lambda$, and the desired gauge condition can be imposed as follows:

First, by using Gaussian null coordinates, we can ensure that
for all $\lambda$, the surface located at the same coordinates
as $\H^+$ is a null surface.
Specifically, we may assume without loss of generality, that,
near $\H^+$, each metric in our family can be written in the form
\ben
\label{gaussian}
g_{ab}(\lambda) = 2 \ \nabla_{(a} u \ [ \nabla_{b)}
r - r^2 \ \alpha(\lambda) \ \nabla_{b)} u - r \ \beta_{b)}(\lambda)] + \mu_{ab}(\lambda)
\een
where $r,u$ are functions defined in a
neighborhood of $\H^+$ that are {\em independent} of $\lambda$.
The location of $\H^+$ for any value of $\lambda$ is given by
$r=0$, and on $\H^+$, $u$ is an affine parameter for any metric in our
family\footnote{Note that in~\cite{hiw}, $u$ was chosen to be a
  Killing parameter rather than an affine parameter, resulting in some
  differences in the metric form in the case of a non-degenerate
  horizon.}. The tensor fields $\beta_a, \mu_{ab}$
are orthogonal, for all $\lambda$, to the normal bundle of the $(D-2)$-dimensional
surfaces $B(u,r)$, which are the joint level sets of
$u,r$. That is, if
\ben\label{nala}
n^a = \left(\frac{\partial}{\partial u} \right)^a \ ,
\quad l^a = \left(\frac{\partial}{\partial r}\right)^a \ ,
\een
then we have $\mu_{ab} n^a = \mu_{ab} l^a = \beta_a n^a = \beta_a l^a = 0$.
For all $\lambda$, the integral curves of $l^a$ are null geodesics with affine parameter $r$.
For more details on Gaussian null coordinates, see e.g.~\cite{hiw}.
We assume that $u$ is chosen such that $B\equiv B(0,0)$ is the
bifurcation surface of the original metric $g_{ab}(0)$. The location of
the past horizon at $\lambda = 0$ is then given by $u=0$.
The above form
of the metric near $\H^+$ is merely a choice of gauge, which we will impose for
the rest of this paper.  It does not use any information coming
from Einstein's equation, nor any symmetry assumptions.

Although the above gauge choice requires $\H^+$ to be a null surface for all $\lambda$,
it does not impose the desired condition that $\H^+$ coincide with the actual (future) event horizon
to first order in $\lambda$. To impose this condition, we note that the future event horizon
is uniquely characterized as being the outermost outgoing null surface
with compact cross-sections
whose expansion $\vartheta$ vanishes asymptotically at late times. However, the linearized
Raychaudhuri equation implies that, to first order in $\lambda$, the expansion must
be constant along the horizon. Thus, the perturbed expansion
$\delta \vartheta$ must vanish on the horizon at all times, and, thus, must vanish on any
cross-section. Since
the requirement that $\H^+$ is null has already be imposed, we can ensure that
$\H^+$ coincides with the event horizon to first order in $\lambda$ by simply imposing the
requirement that its perturbed expansion vanishes on any one cross-section, such as
$B$. Thus, our desired additional gauge condition is~\cite{sorkin}
\ben
\delta \vartheta |_B = 0 \  .
\label{varB}
\een
We now wish to argue that this condition can always be imposed.
Let $\phi_s$ be the 1-parameter family of diffeomorphisms generated by the vector field
$X^a = (1/\sqrt{2}) \ fl^a$, where $f$ is some smooth function. Let
$B_s = \phi_s[B]$ be the deformed surface, and let $\vartheta(s)$ be the
expansion [of the unperturbed metric $g_{ab}=g_{ab}(0)$] of $B_s$ in the
outward going future null directions $k^a$, normalized by $k^a l_a=1$. The vacuum Einstein
equations allow one to show that the
following equation holds on $B$~\cite{andersson,galoway}:
\bena\label{ch}
\frac{d}{ds} \vartheta(0) &=& -D^a D_a f + \beta^a D_a f + \half \left[R(\mu) - \frac{1}{2} \  \beta^a \beta_a + D^a \beta_a - \frac{1}{2} \ (\pounds_n \mu_{ab}) \pounds_n \mu^{ab} \right]f \non\\
&=:& C(f) \, ,
\eena
where $R(\mu)$ is the curvature scalar of $\mu_{ab}$, and $D_a$ is the derivative operator of $\mu_{ab}$. Since the background spacetime is stationary, we have $\pounds_n \mu_{ab} = 0$ on $\H^+$. As discussed in~\cite{andersson}, although $C$ is clearly not self-adjoint,
it admits a strictly positive principal eigenfunction $\psi$. As argued in~\cite{andersson,galoway}, we must have $\lambda_1 \ge 0$---otherwise there would be trapped surfaces in the exterior, which is not possible. We now show that $\lambda_1$ is strictly positive, $\lambda_1>0$. (For another argument leading to the same conclusion in the more general context of ``non-evolving horizons'', see Prop.~3 of~\cite{mars2}.)
To do so, we use the fact that in the background metric, $B$ is the bifurcation surface
of a bifurcate Killing horizon
with horizon Killing field $K^a$. We claim first that in our Gaussian null
coordinates, $K^a$ takes the remarkably simple form
\ben\label{kaaff}
K^a = \kappa \left[
u \left( \frac{\partial}{\partial u} \right)^a
-
r \left( \frac{\partial}{\partial r} \right)^a
\right] \ ,
\een
where $\kappa$ is the surface gravity of the black hole. To prove this,
we note that the standard relation between the
affine parameter $u$ on $\H^+$ and the flow parameter
of the Killing field $K^a$ on $\H^+$ implies that on $\H^+$ (i.e.,
when $r=0$) we have
$K^a = \kappa u \partial/\partial u$. It follows that on $\H^+$, flow under
$K^a$ by parameter $\alpha$ maps $u$ to $u \exp(\kappa \alpha)$
and $n^a$ to $n^a \exp(\kappa \alpha)$. Since $l^a$ on $\H^+$ is uniquely
determined up to scale by orthogonality to $(du)_a$, it follows that
flow under $K^a$ maps $l^a$ to $l^a \exp(-\kappa \alpha)$ on $\H^+$. Since
flow under $K^a$ maps null geodesics to null geodesics, it follows
that flow under
$K^a$ by parameter $\alpha$ acts the Gaussian null coordinates
as follows throughout the domain of validity of the coordinates:
$u \rightarrow u \exp(\kappa \alpha)$, $r \rightarrow r \exp(-\kappa \alpha)$,
and $x^A \rightarrow x^A$, where $x^A$ denotes the remaining Gaussian
null coordinates. The infinitesimal form of this transformation is just
\eqref{kaaff}, as we desired to show.

Now consider the above 1-parameter
family of surfaces $B_s$ with $f = \psi$. For $s_1, s_2 > 0$, the surface $B_{s_2}$
is obtained from the surface $B_{s_1}$ by Killing transport with Killing parameter
$\frac{1}{\kappa} \log(s_2/s_1)$. Furthermore, $\tilde k^{a} = (1/r) \ k^a$ is
Lie-transported by $K^a$. It follows immediately that the expansion $\tilde \vartheta(s)$
with respect to $\tilde k^{a}$ is independent of $s$. From this, it follows that
the expansion $\vartheta(s)$ with respect to $k^a$ is given by $\vartheta(s) = s\vartheta(1)$, where $\vartheta(1)$ denotes the expansion of the surface $B_1$. But we have $\frac{d}{ds} \vartheta(0) = \lambda_1 \psi$. Consequently, if $\lambda_1=0$, we would have $\vartheta(1)=0$, i.e. $B_1$ would be a MOTS lying outside the black hole, which is impossible. This establishes that $\lambda_1 > 0$.

Thus, we have shown that the stability operator $C$ on $B$, defined by~\eqref{ch},
has a strictly positive principal eigenvalue. But the principal eigenvalue of
the adjoint $C^*$ is equal to that of $C$ (see lemma~4.1 of~\cite{andersson}), so
$C^*$ cannot have a kernel. This implies that we can uniquely
solve the equation $C(f) = j$ for
any smooth $j$. Solving this equation with $j=-\delta \vartheta|_B$,
we find that the 1-parameter family of
metrics $\phi_\lambda^* g_{ab}(\lambda)$ will have
$\vartheta|_B = 0$ to (zeroth and-) first order in $\lambda$. Thus,
we have proven that for any metric perturbation
$\gamma_{ab} = \frac{d}{d\lambda} g_{ab}(0)$, we can apply a first order
gauge transformation $\gamma_{ab} \to \gamma_{ab} + \pounds_X g_{ab}$ of the form  $X^a = (1/\sqrt{2}) \ fl^a$ that imposes
eq.~(\ref{varB}).

We are still free, without destroying our previous gauge choices, to apply a 1-parameter family of diffeomorphisms $\phi_s$
leaving $r,u$ invariant. Such a diffeomorphism is locally (near $\H^+$) uniquely
determined by its action on the bifurcation surface $B$, and at the linearized level
corresponds to a gauge vector field $\xi^a$ tangent to $B$ satisfying $\pounds_\xi r = 0 =
\pounds_\xi u$.
It is possible to use this freedom to impose the additional gauge
condition
\ben\label{addgauge}
\mu^{ab} \delta \mu_{ab} |_B = {\rm const}.  \ ,
\een
or, equivalently,
\ben
\delta \epsilon|_B = \frac{\delta A}{A} \ \epsilon|_B
\label{deleps}
\een
where $\epsilon$ denotes the background volume form on $B$ and $\delta
A$ denotes the perturbed area of $B$. To prove this, we note that
$\delta \mu_{ab}$ changes under a linear gauge transformation of the
kind we just described by $2D_{(a} \xi_{a)}$, where $D_a$ denotes the
derivative operator on $B$.  Thus, to impose the desired gauge, we
must solve $2D^a \xi_a = - \mu^{ab} \delta \mu_{ab} + {\rm const}$.
The necessary and sufficient condition for obtaining a solution to
this equation is that the right
side integrates to zero over $B$, which can be achieved by an appropriate
choice of the constant. It is easily seen that \eqref{addgauge} is equivalent
to \eqref{deleps}.

We summarize our gauge choices in the following lemma:
\begin{lemma}\label{lem1}
Let $(\M, g_{ab})$ a stationary black hole spacetime satisfying
the vacuum Einstein equations, and let $g_{ab}(\lambda)$
a 1-parameter family of solutions perturbing $g_{ab} = g_{ab}(0)$.
Then near $\H^+ \subset \M$,
the metric can be brought into the form~\eqref{gaussian}
with $\delta \vartheta |_B = 0$ (and, hence,
the perturbed expansion vanishes on all of $\H^+$). Furthermore, the gauge
condition \eqref{deleps} can be imposed at $B$.
\end{lemma}

\noindent
{\bf Remark:} The gauge freedom that remains on a perturbation
$\gamma_{ab}$ after the imposition of the gauge conditions of lemma~\ref{lem1} is
$\gamma_{ab} \rightarrow \gamma_{ab} + \pounds_X g_{ab}$ where $X^a$
is a smooth vector field that is an asymptotic symmetry near infinity, is tangent
to $\H^+$, satisfies $\pounds_n X^a |_{\H^+} = f n^a$
with $n^a \nabla_a f = 0$, and satisfies $(\mu^{ab} \nabla_a X_b)|_B = 0$.

\bigskip

Finally, the black hole background metric $g_{ab}(0)$ is stationary,
with Killing field $t^a$. By the rigidity theorem, we know that $t^a$
can be decomposed as
\ben
t^a = K^a - \sum_{A=1}^N \Omega_A \psi_A{}^a \ ,
\label{rig}
\een
where, for $\lambda=0$, $K^a$ is a Killing field normal to $\H$,
$\psi^a_A$ are Killing fields with closed orbits, and $\Omega_A$ are
constants called the ``angular velocities of the
horizon''. On the future horizon, we have from~\eqref{kaaff}
\ben
K^a|_{\H^+} = \kappa \ u \left( \frac{\partial}{\partial u} \right)^a \, .
\label{Kgau}
\een
In our analysis, we will later
restrict attention to the axisymmetric case, although we do not make any such
restriction now.  More
precisely, we will assume later that for each $A$ appearing in the sum in
(\ref{rig}), there is a corresponding Killing field $\psi^a_A
(\lambda)$ of $g_{ab}(\lambda)$ for $\lambda > 0$. Without loss of
generality, we may assume that $\psi^a_A (\lambda) = \psi^a_A$, i.e.,
it is compatible with the gauge choices we have made above to
additionally require that our gauge is chosen so that the
$\psi^a_A$ do not vary with $\lambda$. For $\lambda > 0$,
$g_{ab}(\lambda)$ is not, in general, stationary, so there are no
corresponding Killing fields $t^a(\lambda)$ or
$K^a(\lambda)$. Nevertheless, it will be useful to consider the
$\lambda$-independent vector fields $t^a$ and $K^a$ in the spacetime
$(\M, g_{ab}(\lambda))$, and we will do so in the next
subsection. However, we caution the reader that, in the event that
$g_{ab}(\lambda)$ is a stationary black hole, with Killing fields
$t^a(\lambda)$ and $K^a(\lambda)$, we may {\em not} assume that
$t^a(\lambda)=t^a$ or $K^a(\lambda)=K^a$. Indeed, if
$g_{ab}(\lambda)$ is a stationary black hole with surface gravity or
horizon angular velocities differing from that of $g_{ab}(0)$, it
would be manifestly inconsistent with eqs.(\ref{rig}) and (\ref{Kgau}) for
$t^a(\lambda)=t^a$ and $K^a(\lambda)=K^a$. Thus, we will have to
exercise care at the end of section~2.3 when identifying perturbations
toward stationary black holes, since we cannot assume that these
perturbations are manifestly stationary in our gauge.

\subsection{First variation formulas}

In this subsection, we will develop the machinery needed to define the canonical energy and
establish its properties.

The Lagrangian for vacuum general relativity is
\begin{equation}
  L_{a_1 \dots a_D} = \frac{1}{16 \pi} R \ \epsilon_{a_1 \dots a_D},
\label{vacgr}
\end{equation}
where $R$ is the Ricci scalar and $\epsilon_{a_1 \dots a_D}$ is the positively oriented\footnote{We choose orientations in this paper is as follows: The orientation of $\M$
is defined by declaring the Gaussian null coordinates $(r,u,x^1,\dots,x^{D-1})$ to be
right-handed. The orientation of $\Sigma$ is defined by declaring $(r,x^1,\dots,x^{D-1})$
to be left-handed, and the orientation of $B$ is defined by declaring $(x^1,\dots,x^{D-1})$ to be right-handed.} volume form of a $D$-dimensional
spacetime with metric $g_{ab}$. We will consider the variations of $L$ and other
quantities under smooth, one-parameter variations of the metric $g_{ab}(\lambda)$, and,
occasionally, under two-parameter variations $g_{ab}(\lambda_1, \lambda_2)$. We will write
\ben
\delta g_{ab} = \dl g_{ab} |_{\lambda = 0} \ , \quad \delta L = \dl L |_{\lambda=0} \ , \quad
\delta^2 L = \tfrac{d^2}{d\lambda^2} L |_{\lambda = 0} \ \quad \text{etc.}
\een
for the derivative(s) at $\lambda = 0$.
In cases where our variational formulas hold for all $\lambda$, we will use the notation
``$d/d\lambda$" rather than ``$\delta$."

The variation of $L$ can be written as
\begin{equation}
  \label{eq:L}
  \dl L(g) = E(g) \cdot \dl g + \D \theta(g; \dl g),
\end{equation}
where $E=0$ are the field equations\footnote{More precisely, ${E^{ab}}_{c_1 \dots c_D} = - \frac{1}{16 \pi} \epsilon_{c_1 \dots c_D} \left(R^{ab} - \frac{1}{2} R g^{ab} \right)$} and $\theta$ corresponds to the boundary term that would arise if the variation were performed under an integral sign, namely
\begin{equation}
  \label{thetadef}
  \theta_{a_1 \dots a_{d-1}} = \frac{1}{16 \pi } v^c \epsilon_{c a_1 \dots a_{d-1}},
\end{equation}
where
\begin{equation}\label{vdef}
  v^a = g^{ac} g^{bd} (\nabla_d \dl g_{bc} - \nabla_c \dl g_{bd}).
\end{equation}
The symplectic current $(D-1)$-form $\omega$ is defined as
\begin{equation}
  \label{eq:omega}
  \omega(g; \plo g, \plt g) = \plo \theta(g; \plt g) - \plt \theta(g; \plo g).
\end{equation}
Thus, $\omega$ depends on the unperturbed metric and
a pair of perturbations $(\plo g, \plt g)$, and it
is antisymmetric in $(\plo g, \plt g)$.
If both perturbations satisfy the linearized equations of motion, then it follows by
taking a second, antisymmetrized variation of \eqref{eq:L} that $\omega$
is closed,
\ben
\D \omega = 0 \, .
\een
Concretely, $\omega(g; \gamma_1, \gamma_2)$ is given by
\begin{equation}
  \label{omegadef}
  \omega_{a_1 \dots a_{D-1}} =  \frac{1}{16 \pi } w^c \epsilon_{c a_1 \dots a_{D-1}},
\end{equation}
where
\begin{equation}\label{wadef}
  w^a = P^{abcdef} (\gamma_{2 \ bc} \nabla_d \gamma_{1 \ ef} - \gamma_{1 \ bc} \nabla_d \gamma_{2 \ ef})
\end{equation}
and
\begin{equation}
  P^{abcdef}
  = g^{ae} g^{fb} g^{cd} - \frac{1}{2} g^{ad} g^{be} g^{fc} - \frac{1}{2} g^{ab} g^{cd} g^{ef}
  - \frac{1}{2} g^{bc} g^{ae} g^{fd} + \frac{1}{2} g^{bc} g^{ad} g^{ef}.
\end{equation}

The symplectic form $W_\Sigma(g; \gamma_1, \gamma_2)$ is obtained
by integrating $\omega$ over a Cauchy surface $\Sigma$,
\ben
W_\Sigma(g; \gamma_1, \gamma_2) \equiv \int_\Sigma \omega(g; \gamma_1, \gamma_2) \, .
\label{symform}
\een
In the case mostly considered in this paper---where $g_{ab} = g_{ab} (0)$ is a stationary black hole with bifurcate
Killing horizon (see the previous subsection)---we take $\Sigma$ to be a Cauchy surface for the exterior
region, i.e., a spacelike
slice that goes from the black hole bifurcation surface $B$
to spatial infinity. For $D>4$ this integral manifestly converges for
our asymptotic conditions on the metric near spatial infinity
(see the second paragraph of the previous subsection).
For $D=4$, the Regge-Teitelboim parity conditions assure convergence.
If $(\gamma_1, \gamma_2)$ both satisfy the linearized field equations around a solution $g_{ab}$ to
Einstein's equations, then
then it follows from $\D \omega = 0$ and our asymptotic conditions near spatial infinity
that
\ben
W_{\Sigma'}(g; \gamma_1, \gamma_2) = W_\Sigma(g; \gamma_1, \gamma_2)
\een
where $\Sigma'$ is another
Cauchy surface extending from $B$ to spatial infinity.

We can bring the symplectic form into a more
recognizable form by performing a space-time split of the spacetime metric
into the canonically
conjugate variables $(h_{ab}, p^{ab})$, where
$h_{ab}$ is the induced metric on $\Sigma$ and
\ben
p^{ab} = h^\half (K^{ab} - h^{ab} K) \, ,
\een
with $K_{ab}$
the extrinsic curvature of $\Sigma$.
Our asymptotic conditions near spatial infinity imply that\footnote{Here it is assumed
that $\Sigma$ approaches a surface of
constant $x^0$ in our asymptotically Minkowskian coordinates.}
\ben\label{decay}
h_{ab} = \delta_{ab} + O(\rho^{-(D-3)}) \ , \qquad
p^{ab} = O(\rho^{-(D-2)}) \ ,
\een
with the $N$-th spatial derivatives of $h_{ab}$ and $p^{ab}$
falling off faster by a factor of $\rho^{-N}$.
Note that we have followed the standard practice of incorporating the volume element
of $\Sigma$ into the definition of $p^{ab}$, as is necessary in order that
$(h_{ab}, p^{ab})$ be canonically conjugate. Thus, $p^{ab}$ is a density of weight $\half$.
Our integrals below involving $p^{ab}$ will be taken
with respect to a fixed, nondynamical (e.g., coordinate) volume element
$e_0$ on $\Sigma$, such that the volume element
$\nu^a \epsilon_{ab_1 \dots b_{D-1}}$ (where $\nu^a$ is the future directed unit normal to $\Sigma$) associated with $h_{ab}$, is equal to
$h^\half \ (e_0)_{b_1 \dots b_{D-1}}$.
A perturbation $\delta g_{ab}$ satisfying the linearized
Einstein equations corresponds to a pair $(\delta h_{ab}, \delta p^{ab})$ satisfying the
linearized constraints and the
linearized Hamiltonian equations of motion. In terms of $(\delta h_{ab}, \delta p^{ab})$
the symplectic form as defined by eq.~(\ref{symform}) is simply\footnote{The pullback of \eqref{omegadef}
to a spacelike hypersurface was computed in eq.~(4.14) of~\cite{burnett}.
There is an additional ``boundary term'' present
in their expression for $W_\Sigma$, which is easily seen to vanish on account of our boundary conditions
at $B$ and our asymptotic conditions at infinity.}
\ben\label{Wdef}
W_\Sigma(g; \delta_1 g, \delta_2 g) = -\frac{1}{16\pi} \int_\Sigma (\delta_1 h_{ab} \delta_2 p^{ab} - \delta_2 h_{ab} \delta_1 p^{ab}) \ .
\een
Here, and in the following, we omit the reference volume form $e_0$.

For an arbitrary vector field $X^a$, the associated Noether current
$\mathcal J_X$ is defined by
\ben
\mathcal{J}_X(g) = \theta(g; \pounds_X g) - i_X L(g)
\een
where $i_X$ denotes contraction of $X^a$ into the first index of a differential form such as
$L_{a_1 \dots a_D}$. A simple calculation~\cite{iyerwald2} shows that
the first variation of $\mathcal{J}_X$ satisfies
\ben
\dl \mathcal{J}_X = - i_X (E \cdot \dl g) + \omega(g;\dl g, \pounds_X g)
+ \D [i_X \theta(g,\dl g)] \, ,
\label{dJ}
\een
where, in this formula, it has not been assumed that $g_{ab}$ satisfies
the field equations nor that $d g_{ab}/d \lambda$ satisfies the linearized
field equations. Furthermore, it can be shown~\cite{iyerwald} that
$\mathcal J_X$ can be written in the form
\ben
\mathcal{J}_X = C_X + \D Q_X \, ,
\label{noecon}
\een
where $C_X\equiv C_a X^a$ are the constraints of the theory~\cite{seifert} and $Q_X$
is the Noether charge. In the case of general relativity
in vacuum, eq.~(\ref{vacgr}), we have
\ben
(C_X)_{a_1 \dots a_{D-1}} = \frac{1}{8\pi} \ X^a {G_a}^b \epsilon_{b a_1 \dots a_{D-1}}
\label{constraints}
\een
and
\ben\label{Qdef}
(Q_X)_{a_1\dots a_{D-2}} = -\frac{1}{16\pi} \ \nabla_b X_c \epsilon^{bc}{}_{a_1\dots a_{D-2}} \ .
\een
Combining eqs.~(\ref{dJ}) and~(\ref{noecon}), we obtain
\ben
\omega(g;\dl g, \pounds_X g) = i_X (E(g) \cdot \dl g) + \dl C_X(g)
+ \D \left[ \dl Q_X(g) - i_X \theta(g;\dl g) \right]
\label{fundid}
\een
It should be emphasized that eq.~(\ref{fundid}) is an identity that
holds for arbitrary $X^a$ and $g_{ab}(\lambda)$.

We now impose the condition that our one-parameter family $g_{ab}(\lambda)$
is composed of asymptotically flat solutions to Einstein's equation, so $E = C_a = 0$.
We also require that $X^a$ be an asymptotic symmetry near infinity, i.e.,
we require $\pounds_X g_{ab}$ to satisfy the same asymptotic conditions
at infinity as $\delta g_{ab}$, so that
$X^a$ suitably approaches a Killing field of the Minkowskian background
as $\rho \rightarrow \infty$. Integrating
eq.~(\ref{fundid}) over a compact region, $K$, of $\Sigma$,
extending from $B$ to a surface, $S$, in the asymptotic region,
we obtain, for all $\lambda$
\ben
\int_K \omega(g;\dl g, \pounds_X g) =
\int_S [\dl Q_X(g) - i_X \theta(g; \dl g)]
-\int_{B} [\dl Q_X(g) - i_X \theta(g; \dl g)] \, .
\een
Taking the limit as $S \rightarrow \infty$, we obtain
\ben
W_\Sigma(g;\dl g, \pounds_X g) =
\int_{\infty} [\dl Q_X(g) - i_X \theta(g; \dl g)]
-\int_{B} [\dl Q_X(g) - i_X \theta(g; \dl g)] \, ,
\label{fundid2}
\een
where $\int_{\infty} [\dl Q_X(g) - i_X \theta(g; \dl g)]
\equiv \lim_{S \rightarrow \infty}
\int_S [\dl Q_X(g) - i_X \theta(g; \dl g)]$. For $D>4$ or for the case
where $X$ is an asymptotic translation in $D=4$, this
limit exists independently of how $S$
approaches infinity because the integral
defining $W_\Sigma(g;\dl g, \pounds_X g)$ converges absolutely by virtue of our
assumed asymptotic conditions. For asymptotic boosts and rotations in $D=4$, the limit
must be taken where $S$ approaches a two-sphere asymptotically.

The formula
\ben
\dl H_X = \int_{\infty} [\dl Q_X(g) -
i_X \theta(g; \dl g)]  \ ,
\label{HX}
\een
can be shown to define~\cite{zoupaswald} a conserved quantity $H_X$. If
$X^a \to (\partial/\partial t)^a$, then $H_X$ is equal to the ADM-mass $M$;
if $X^a \to (\partial/\partial x^i)^a$ (asymptotic translation),
then $-H_X$ is equal\footnote{The map
$X \rightarrow H_X$ is a linear functional and thus is a ``covector''. To get the $D$-dimensional energy-momentum vector, we have to ``raise the index'' with the Minkowskian metric. This accounts for the relative minus sign between $M$ as compared with
$P_i$ and $J_{[ij]}$  in their definitions in terms of $H_X$.}
to the linear ADM-momentum $P_i$; if
$X^a \to x^i (\partial/\partial x^{j})^a - x^j (\partial/\partial x^{i})^a$
(asymptotic rotation), then $-H_X$
is equal to the ADM-angular momentum $J_{[ij]}$ in the $ij$-plane\footnote{The
notation $J_A$ used above
refers to $N$ distinguished mutually orthogonal 2-planes
defined by the $N$ rotational Killing fields
$\psi_A^a$.}; if
$X^a \to t(\partial/\partial x^i)^a + x^i(\partial/\partial t)^a$
(asymptotic boost), then $H_X$ is equal
to the ADM-center of mass, $C_i$. Together, the $\half D(D+1)$
quantities $(M, {\bf P}, {\bf J}, {\bf C})$ will
be referred to as ``ADM conserved quantities'' associated with
the spacetime $(\M,g_{ab})$.

Now consider the case $X^a = K^a$, where $K^a$ is the horizon Killing
field of $g_{ab}(0)$. Then $K^a$ takes the coordinate form (\ref{Kgau})
at the horizon for all $\lambda$. In particular, $K^a|_B = 0$ and,
thus, $i_K \theta = 0$ on $B$. On the other hand, using
\eqref{gaussian}, we obtain
\ben
\int_B Q_K(g(\lambda)) = \frac{\kappa}{8 \pi} A(\lambda)
\een
where $A(\lambda)$ is the area of $B$ in the metric $g_{ab}(\lambda)$.
Thus, we obtain
\ben
\int_{B} [\dl Q_K(g) - i_K \theta(g; \dl g)] =
\frac{\kappa}{8 \pi} \frac{dA}{d\lambda} \, ,
\label{KintB}
\een
which holds for the metric $g_{ab} \equiv g_{ab}(\lambda)$.
By eqs.~(\ref{rig}) and (\ref{HX}), we have
\ben
\int_{\infty} [\dl Q_K(g) - i_K \theta(g; \dl g)]
= \Dl M - \sum_A \Omega_A \Dl J_A \, .
\label{Kintinfty}
\een
Combining eqs.(\ref{fundid2}), (\ref{KintB}), and (\ref{Kintinfty}),
we obtain
\ben
W_\Sigma\bigg(g;\Dl g, \pounds_K g\bigg) = \Dl M - \sum_A \Omega_A \Dl J_A
- \frac{\kappa}{8 \pi} \Dl A \, .
\label{fundid3}
\een
This equation holds for all $\lambda$, with $\kappa$ and $\Omega_A$
fixed constants, equal, respectively, to the surface gravity
and angular velocities of the stationary black hole metric $g_{ab}(0)$.
Since $K^a$ is a Killing
field of $g_{ab}(0)$, when $\lambda=0$ we have $\pounds_K g = 0$ and
eq.~(\ref{fundid3}) reduces to the first law of black hole mechanics
\ben
0 = \delta M - \sum_A \Omega_A \delta J_A
- \frac{\kappa}{8 \pi} \delta A \, ,
\label{fundid4}
\een
where we remind the reader that our use of the notation "$\delta$" here
(rather than "$d/d\lambda$") indicates that this equation holds only at $\lambda = 0$.

Finally, we consider the gauge dependence of $W_\Sigma$.  It is clear
that $W_\Sigma(g; \delta_1 g, \delta_2 g)$ will be gauge invariant if
and only if $W_\Sigma(g; \delta g, \pounds_\xi g) = 0$ for all allowed
$\delta g_{ab}$ and $\xi^a$. It follows immediately from integrating
\eqref{fundid} over $\Sigma$ (with $E=0$) that if $\delta g_{ab}$ satisfies the linearized
constraints, then $W_\Sigma(g; \delta g,
\pounds_\xi g) = 0$ for all $\xi^a$ of compact support\footnote{Conversely,
if $W_\Sigma(g; \delta g,
\pounds_\xi g) = 0$ for all $\xi^a$ of compact support away from $B$,
then $\delta g_{ab}$ satisfies the linearized
constraints. These facts, which are manifestations of the familiar statement
  that in general relativity ``the constraints generate gauge
  transformations,'' will be exploited heavily in subsection 4.1 below.}
  away from $B$. Thus, for solutions, $W_\Sigma(g; \delta_1 g,
\delta_2 g)$ is gauge invariant for gauge transformations of compact
support.  However, if $\xi^a$ is a smooth vector field such that
$\pounds_\xi g_{ab}$ satisfies our asymptotic conditions at infinity
(i.e., $\xi^a$ is an asymptotic symmetry at infinity) and our boundary
conditions at $B$, then $W_\Sigma(g; \delta g, \pounds_\xi g)$
need not vanish for all $\delta g$ that satisfy our asymptotic
conditions at infinity and our boundary conditions at $B$. The
following lemma characterizes the extent to which $W_\Sigma$ is gauge
invariant. This lemma will play a critical role in our analysis.

\begin{lemma}\label{lemma2}
Let $\delta g_{ab}$ be a solution to the linearized Einstein equations
around our stationary black hole background $g_{ab}$
satisfying our asymptotic flatness conditions and our gauge conditions
\eqref{varB} and \eqref{deleps} at $B$. Suppose in addition that
$\delta A = 0$ (so that, by \eqref{deleps}, we have $\delta \epsilon|_B = 0$)
and that $\delta H_X = 0$ for some asymptotic symmetry $X^a$. Then
$W_\Sigma(g; \delta g, \pounds_\xi g) = 0$
for all smooth $\xi^a$ such that (i) $\xi^a|_B$ is tangent to the generators
of $\H$ and
(ii) $\xi^a$ approaches a multiple of $X^a$ as $\rho \rightarrow \infty$.
Conversely, if $\delta g_{ab}$ is smooth and asymptotically flat
and if $W_\Sigma(g; \delta g, \pounds_\xi g) = 0$ for all such $\xi^a$, then
$\delta g_{ab}$ is a solution to the linearized Einstein equation
with  $\delta \vartheta|_B = \delta \epsilon|_B = 0$ at $B$ and with
$\delta H_X = 0$.
\end{lemma}

A proof of this lemma is given in appendix~A.

\subsection{Second variations and canonical energy}

Returning to eq.~(\ref{fundid3}), we write out the $\lambda$-dependence explicitly as
\ben
W_\Sigma \left(g(\lambda); \frac{d}{d\lambda} g(\lambda), \pounds_K g(\lambda) \right) =
\frac{d}{d\lambda} M(\lambda) - \sum_A \Omega_A \ \frac{d}{d\lambda} J_A(\lambda)
- \frac{\kappa}{8 \pi} \ \frac{d}{d\lambda} A(\lambda) \, ,
\een
where it should be emphasized that $\kappa$ and $\Omega_A$ do {\em not} depend on $\lambda$.
We now take a $\lambda$-derivative
of this equation, and then
set $\lambda=0$. Using the fact that $\pounds_K g_{ab}(0) = 0$, we get
\ben
W_\Sigma \left(g; \gamma, \pounds_K \gamma \right) =
\frac{d^2}{d\lambda^2} M(\lambda) \bigg|_{\lambda=0}
- \sum_A \Omega_A \frac{d^2}{d \lambda^2} J_A(\lambda)\bigg|_{\lambda=0}
- \frac{\kappa}{8 \pi} \frac{d^2}{d \lambda^2} A(\lambda)\bigg|_{\lambda=0}  \, ,
\label{canid}
\een
where $\gamma_{ab} = d g_{ab}/d\lambda |_{\lambda=0} = \delta g_{ab} (0)$ is the first
order perturbation of the
black hole spacetime. Writing
$K^a = t^a + \sum_A \Omega_A \ \psi_A{}^a$, the left side may be written
as
\ben
W_\Sigma \left(g; \gamma, \pounds_K \gamma \right) = \E +
\sum_A \Omega_A \ W_\Sigma \left(g; \gamma, \pounds_{\psi_A} \gamma \right) \, ,
\een
where we have defined the {\em canonical energy} by
\ben
\E \equiv W_\Sigma \left(g; \gamma, \pounds_t \gamma \right) \, .
\label{canonen}
\een
Since $t^a$ is a Killing field of the background, $\pounds_t$ commutes with the linearized
Einstein operator, so
$\pounds_t \gamma_{ab}$ satisfies the linearized field equations whenever $\gamma_{ab}$ does.
$\pounds_t \gamma_{ab}$ also satisfies our asymptotic conditions
at infinity (with faster fall-off) and our boundary conditions at $B$.
Hence, it follows from the conservation of symplectic product that {\it $\E$ is conserved
for all solutions $\gamma_{ab}$}
in the sense that it takes the same value if evaluated on another Cauchy surface
$\Sigma'$ extending from $B$ to spatial infinity.
We may rewrite \eqref{canid} as
\ben\label{variation}
\E = - \sum_A \Omega_A \ W_\Sigma \left(g; \gamma, \pounds_{\psi_A} \gamma \right)
+ \delta^2 M - \sum_A
\Omega_A \, \delta^2 J_A - \frac{\kappa}{8 \pi} \delta^2 A \, ,
\een
where we have written $\delta^2 M = d^2M/d\lambda^2 |_{\lambda=0}$, etc. Note that $\E$ and
$W_\Sigma \left(g; \gamma, \pounds_{\psi_A} \gamma \right)$
depend only upon $\gamma_{ab} = \delta g_{ab}$, whereas $\delta^2 M$, $\delta^2 J_A$,
and $\delta^2 A$ depend upon the second order perturbation
$\delta^2 g_{ab} = d^2 g_{ab}/d \lambda^2 |_{\lambda=0}$. However, the difference between
two second order perturbations will be a homogeneous solution
of the first order perturbation equations, so its net contribution to
(\ref{variation}) will vanish by the first law of black hole
mechanics \eqref{fundid4}.

We summarize our main result thus far as follows:

\begin{prop}
(Second variation formula):
Let $(\M,g_{ab}(\lambda))$ be a one-parameter family of smooth, asymptotically
flat vacuum solutions such that $g_{ab}(0)$ is a
stationary black hole
spacetime with
bifurcate Killing horizon. Suppose that the gauge
condition~\eqref{gaussian} has been imposed near the horizon $\H^+$.
Then the canonical energy \eqref{canonen} of $\gamma_{ab}$ satisfies
\eqref{variation}, where $\Omega_A$ and $\kappa$ are the angular
velocities and surface gravity of the horizon of
the background solution $g_{ab}(0)$.
\end{prop}

It is useful to view the canonical energy as a bilinear form on
perturbations, defined by
\ben
\E(\gamma_1, \gamma_2) \equiv
W_\Sigma \left(g; \gamma_1, \pounds_t \gamma_2 \right) \, ,
\label{canonen2}
\een
so that the canonical energy
originally defined by \eqref{canonen} is given by
$\E = \E(\gamma,\gamma)$. Although it is not manifest from its
definition, it is easily seen that $\E(\gamma_1, \gamma_2)$ is symmetric in $(\gamma_1, \gamma_2)$:

\begin{prop} Let $\gamma_1$ and $\gamma_2$ be smooth,
asymptotically flat solutions to
the linearized field equations on the stationary black hole background $g_{ab}(0)$. Then, we have
\ben
\E(\gamma_1, \gamma_2) = \E(\gamma_2, \gamma_1) \, .
\label{canonen4}
\een
\end{prop}

\noindent
{\em Proof:}
Since the background metric $g$ is stationary,
it follows immediately from the
antisymmetry of the symplectic current $\omega$ that
\ben
\omega(g; \gamma_1, \pounds_t \gamma_2) -
\omega(g; \gamma_2, \pounds_t \gamma_1) = \pounds_t \omega(g; \gamma_1, \gamma_2) \ .
\een
Now we use the standard identity
$\pounds_X = i_X \ \D + \D  \ i_X$ for the Lie-derivative acting on forms,
together with the fact that
$\D \omega  =0$ when $\gamma_1$ and $\gamma_2$
are linearized solutions. Integrating over
$\Sigma$, we obtain
\ben
\E(\gamma_1, \gamma_2) - \E(\gamma_2, \gamma_1)
= \int_\infty i_t \omega(g; \gamma_1, \gamma_2)
- \int_B i_t \omega(g; \gamma_1, \gamma_2)  \ .
\een
However, the contribution from infinity vanishes on account of our
asymptotic conditions, whereas the contribution from $B$ vanishes
because $t^a$ is tangent to $B$. Thus, $\E$ is symmetric. \qed

\bigskip

In our analysis of the next sections, we will be concerned with the
positivity properties of $\E$ as well as fluxes of $\E$ through the
horizon and at null infinity. Unfortunately, for a rotating black hole,
the term $\sum_A \Omega_A W_\Sigma
\left(g; \gamma, \pounds_{\psi_A} \gamma \right)$ will spoil any
positivity properties of $\E$. Furthermore, the fact that, for the
case of a rotating black hole, $t^a$ is spacelike at the horizon will
preclude the possibility of having a positive flux of canonical energy
through the horizon\footnote{We could have defined a canonical energy
$\E_K$ with respect
to the horizon Killing field $K^a$ instead of $t^a$, in which case the term
$\sum_A \Omega_A W_\Sigma \left(g; \gamma, \pounds_{\psi_A} \gamma \right)$ would not have
appeared in the expression for $\E_K$ and the flux through the horizon would
be positive. However, in the absence of axisymmetry, the spacelike character
of $K^a$ near infinity for a rotating black hole
will similarly spoil the positivity properties of $\E_K$ and preclude
the possibility of its having a positive flux  through null infinity}.
Fortunately, both difficulties can be resolved by
restricting consideration to axisymmetric perturbations
\ben
\pounds_{\psi_A} \gamma = 0 \, ,
\label{axisym}
\een
and we shall impose this
restriction later in our analysis.  For axisymmetric perturbations,
\eqref{variation} reduces to
\ben
\label{variation2}
\E = \delta^2 M - \sum_A \Omega_A \, \delta^2 J_A
- \frac{\kappa}{8 \pi} \ \delta^2 A \, .
\een

An important property of axisymmetric perturbations is given by the
following proposition:

\begin{prop}
The canonical energy is gauge invariant with respect to gauge
transformations that preserve the gauge conditions of subsection 2.1
when restricted to the space of axisymmetric perturbations that satisfy
the linearized field equations and have vanishing linearized ADM linear momenta,
$\delta P_i = 0$, and vanishing change of area, $\delta A = 0$.
\end{prop}

\noindent
{\em Proof:} The gauge transformations that preserve the gauge conditions
of subsection 2.1 and  axisymmetry are $\gamma_{ab} \rightarrow
\gamma_{ab} + \pounds_X g_{ab}$ where $X^a$ satisfies the conditions
in the remark below lemma~\ref{lem1} of subsection 2.1 and, in addition,
satisfies $\pounds_{\psi_A} X^a = 0$ everywhere. Since, by Proposition 2,
$\E(\gamma_1,\gamma_2)$ is symmetric
in $(\gamma_1,\gamma_2)$, gauge invariance will be proven if we can
show that
\ben
\E(\gamma, \pounds_X g) = W_\Sigma (\gamma, \pounds_t \pounds_X g) = 0
\een
for all solutions $\gamma_{ab}$ for which $\delta P_i = 0$ and $\delta A = 0$. However,
we have
\ben
\pounds_t \pounds_X g_{ab}
= (\pounds_t \pounds_X - \pounds_X \pounds_t)g_{ab}
= \pounds_{[t,X]} g_{ab} \ ,
\label{sasg}
\een
since $\pounds_t g_{ab} = 0$. Since $X^a$ is an asymptotic symmetry, $[t,X]^a$ is an
asymptotic spatial translation. Furthermore, since
$t^a = K^a - \sum \Omega_A {\psi^a}_A$ and
$[\psi_A, X]^a = \pounds_{\psi_A} X^a = 0$, it follows
that $[t,X]^a|_B$ is normal to $\H^+$ at $B$. The proposition now follows immediately
from lemma~\ref{lemma2} applied to $\xi^a = [t,X]^a$. \qed

\bigskip

In our later arguments below we will need to consider perturbations
towards other stationary black holes.
We now explain
the subtleties involved in this notion and give a precise definition of it.
It might appear to be obvious from its definition \eqref{canonen}
that $\E$ must vanish for stationary perturbations. However, this is not the case
because, as already
indicated at the end of subsection 2.1, we cannot assume that the
stationary Killing field $t^a (\lambda)$ of the varied spacetime coincides
with $t^a$. Indeed, the
example of the one-parameter family of Schwarzschild metrics of mass
$M = M_0 + \alpha \lambda$ in $4$-dimensions
illustrates that the canonical energy need not vanish for perturbations
to other stationary black holes. For this
family, we have $\delta^2 M = 0$, but
$A(\lambda) = 16 \pi (M_0 + \alpha \lambda)^2$,
so $\delta^2 A = 32 \pi \alpha^2$. Thus, by \eqref{variation2}, we have
(using $\kappa = 1/4 M_0$)
\ben
\E = - \alpha^2/M_0 < 0
\label{sce}
\een

To see why this result is compatible with \eqref{canonen} and to find
the appropriate conditions to express the notion that
a perturbation $\gamma_{ab}$ represents a perturbation towards another
stationary black hole,
let $g_{ab}(\lambda)$ be a one-parameter family
of stationary black holes, with stationary Killing field
$t^a(\lambda)$, horizon Killing field $K^a(\lambda)$, and
axial Killing fields $\psi_A{}^a (\lambda)$ associated with a nonvanishing
angular velocity of the horizon. As already stated at
the end of subsection 2.1, it is compatible with our other gauge
conditions to assume that the axial
Killing fields $\psi_A{}^a (\lambda)$ are independent of $\lambda$, so that
$\psi_A{}^a (\lambda) = \psi_A{}^a$. We may also assume that near infinity,
we have $t^a(\lambda) = t^a$. However, since $K^a$ takes the form
\eqref{Kgau} on $\H^+$, we must have $K^a(\lambda) \neq K^a$ near $\H^+$
if the surface gravity, $\kappa(\lambda)$, of $g_{ab}(\lambda)$
differs from $\kappa$. Nevertheless, we may assume that
$\kappa K^a(\lambda) = \kappa(\lambda)K^a$ near $\H^+$. If we make this
choice, then, near $\H^+$ we have
\ben
t^a(\lambda) = \frac{\kappa(\lambda)}{\kappa} K^a - \sum_A \Omega_A (\lambda) \psi_A{}^a \, ,
\label{t}
\een
so $t^a (\lambda)$ must differ from $t^a$ near $\H^+$ if there is any change in the
surface gravity or angular velocities of the horizon.

Thus, if $g_{ab}(\lambda)$ is
a one parameter family of stationary black holes, then we can choose a gauge
compatible with our previous choices such that
$\psi_A{}^a (\lambda) = \psi_A{}^a$,
$t^a(\lambda) = t^a$ near infinity, and $t^a(\lambda)$ is given by \eqref{t} near $\H^+$.
The perturbation $\gamma_{ab} = d g_{ab}/d \lambda|_{\lambda=0}$ must therefore
satisfy $\pounds_{\psi_A} \gamma_{ab} = 0$ for all $\psi_A{}^a$ appearing in \eqref{t} and
\ben
\pounds_t \gamma_{ab} + \pounds_{\delta t} g_{ab} = 0
\een
where,
near infinity, we have $\delta t^a = 0$
whereas near $\H^+$, we have
\ben
\delta t^a = c \ t^a_{} + \sum_A b_A^{} \ \psi_A{}^a
\label{bkvf}
\een
where $c$ and $b_A$ are constants. Thus, near $\H^+$,
$\delta t^a$ is of the form of a Killing field of the background
metric\footnote{The sum on the right side of \eqref{bkvf}
is allowed to include rotational Killing fields $\psi_A{}^a$ of the background that may not
have appeared in \eqref{rig} because
$\Omega_A (\lambda = 0) = 0$.}.

Thus, rather than having to vanish, we see that $\pounds_t  \gamma_{ab}$
must take the form $\pounds_Y g_{ab}$ where $Y^a$ vanishes near infinity
and is a linear combination
of Killing fields \eqref{bkvf} near $\H^+$. Equivalently, writing
$Z^a = Y^a - c t^a - \sum_A b_A^{} \psi_A{}^a$
and noting that $\pounds_Y g_{ab} = \pounds_Z g_{ab}$, we see that
a perturbation towards a stationary black hole can be put in a gauge
compatible with our gauge conditions such that
\ben
\pounds_t \gamma_{ab} = \pounds_Z g_{ab} \ ,
\een
where $Z^a$ vanishes near $\H^+$ and near infinity takes the form
\ben
Z^a = c \  t^a + \sum_A b_A^{} \ \psi_A{}^a \, ,
\label{bkvf2}
\een
and where, furthermore, $\pounds_{\psi_A} \gamma_{ab} = 0$
for all $\psi_A{}^a$ appearing
in \eqref{rig} and/or \eqref{bkvf2}. In a general gauge satisfying
our gauge conditions, we have
\ben
\pounds_t \gamma_{ab} = \pounds_\xi g_{ab} + \pounds_t \pounds_X g_{ab} \ ,
\een
where $X^a$ satisfies the conditions of the remark below lemma~\ref{lem1}.
Thus, taking account of \eqref{sasg} and the
remarks below that equation, we have
have motivated
the following definition:
\begin{defn}
A smooth, asymptotically flat, axisymmetric
solution $\gamma_{ab}$ of the linearized field equations
is said to be a {\it perturbation towards a stationary black hole} if
\ben
\pounds_t \gamma_{ab} = \pounds_\xi g_{ab} \ .
\een
where  $\xi^a|_B$ is normal to $\H^+$ and near infinity takes the form
\ben
\xi^a = c \ t^a + \sum_A b_A^{} \ \psi_A{}^a + \sum_i a_i
\left(\frac{\partial}{\partial x^i} \right)^a \, .
\label{bkvf3}
\een
\end{defn}

\bigskip

Note that although the linearization of any one-parameter family of stationary
black holes must satisfy definition 2.1,
our definition does not require that there actually exist a
one-parameter family $g_{ab}(\lambda)$ of stationary black holes
corresponding to $\gamma_{ab}$. In any case, the key point about perturbations
towards a stationary black hole is that they are stationary and thus
are benign with regard to linear stability of the background black
hole. An important property of perturbations towards a stationary black holes is
the following:

\begin{prop}
Let $\gamma_1$ be a perturbation for which $\delta M = \delta J_A = \delta P_i = 0$ (and hence,
by the first law of black hole mechanics $\delta A = 0$), and let $\gamma_2$
be a perturbation towards a stationary black hole. Then
$\E (\gamma_1, \gamma_2) = 0$.
\end{prop}

\noindent
{\it Proof:} We have
\ben
\E (\gamma_1, \gamma_2)  = W_\Sigma(g; \gamma_1, \pounds_t \gamma_2)
= W_\Sigma(g; \gamma_1, \pounds_\xi g) = 0 \, ,
\label{canen0}
\een
where the second equality merely substitutes the definition of a
perturbations towards a stationary black hole and the last equality
is an immediate consequence of lemma~\ref{lemma2}. \qed

\bigskip

In subsection~4.1 below, we will considerably strengthen this result
by showing that, when considered as
a quadratic form on a suitable space of
smooth, axisymmetric solutions to the linearized field
equations for which $\delta M = \delta J_A = \delta P_i =
0$, the canonical energy $\E$ will be degenerate precisely on the
perturbations towards stationary black holes.
This will play a key role in our stability arguments.

\subsection{Evaluation of canonical energy}\label{sec:can}

In the previous section, we obtained a simple formula for $\E$ in terms
of second order variations of mass, angular momentum, and area. However,
in order to use this formula to evaluate $\E$, we must calculate second
order perturbation $d^2g_{ab}/d \lambda^2|_{\lambda=0}$,
even though $\E$ really only depends
on the first order perturbation $\gamma_{ab} = d g_{ab}/d\lambda|_{\lambda=0}$.
It is useful to have a formula for $\E$ that
expresses it directly in terms of $\gamma_{ab}$,
and/or the initial data, $(\delta h_{ab}, \delta p^{ab})$, for $\gamma_{ab}$.
Such a formula can be obtained from the original definition~(\ref{canonen2}),
\ben
\E(\gamma_1, \gamma_2) = W_\Sigma(g; \gamma_1, \pounds_t \gamma_2) = -\frac{1}{16\pi} \int_\Sigma (\delta_1 h_{ab} \ \pounds_t \delta_2 p^{ab} -
\delta_1 p^{ab} \ \pounds_t \delta_2 h_{ab} ) \ ,
\een
by substituting the explicit expressions~\eqref{omegadef} and~\eqref{wadef}.
However, to put the right side in a more useful form,
we will follow a different strategy and return to the fundamental
identity~\eqref{fundid}, using the fact that the variation is being taken
about a solution, so $E=0$. We obtain at $\lambda = 0$
\ben
\omega(g;\delta g, \pounds_X g) = \delta C_X(g)
+ \D \left[ \delta Q_X - i_X \theta(g;\delta g) \right] \, .
\label{newfundid}
\een
The constraints $C_X = X^a C_a$
are given by eq.~(\ref{constraints}) above. In terms of the
variables $(h_{ab}, p^{ab})$, the constraints take the form\footnote{Here and below,
we omit writing the factor of the non-dynamical
coordinate $(D-1)$-form $e_0$ on $\Sigma$ in the expression for $C_a$
[see the discussion below \eqref{decay}].}
\ben
 C_a = \left(
 \begin{matrix}
 C_a \nu^a \\
 C_b {h_a}^b
 \end{matrix}
 \right)
 := \frac{1}{16\pi} \ h^\half
  \left(
  \begin{matrix}
   -R(h) + h^{-1} p_{ab} p^{ab} - \frac{1}{D-2} h^{-1} p^2 \\
    -2D_b(h^{-\half} {p_a}^b)
    \end{matrix}
\right) = 0  \ ,
\label{constraintformula}
\een
where $\nu^a$ denotes the future-directed unit normal to $\Sigma$.
The first line of this equation corresponds to the Hamiltonian constraint and the second line
corresponds to the momentum constraint.
The linearized constraints  $\delta C_a$ may be viewed as the
result of acting on $(\delta h_{ab}, \delta p^{ab})$ by a linear operator, $\L$,
\ben
\delta C_a =  \L\left(
\begin{matrix}
\delta h_{ab}\\
\delta p^{ab}
\end{matrix}
\right)
\label{L}
\een
where $\L$ is explicitly given by
\ben\label{delc}
16 \pi \L\left(
\begin{matrix}
\delta h_{ab}\\
\delta p^{ab}
\end{matrix}
\right) = \
\left(
\begin{matrix}
h^\half(D^a D_a \delta h_c{}^c - D^a D^b \delta h_{ab} + R^{ab}(h) \delta h_{ab})+\\
h^{-\half}(-\delta h_c{}^c p^{ab} p_{ab}
+ 2p_{ab} \delta p^{ab} + 2p^{ac} p^b{}_a \delta h_{bc}+\\
\frac{1}{D-2} p^c{}_c p^d{}_d \delta h^a{}_a
-\frac{2}{D-2} p^a{}_a \delta p^b{}_b - \frac{2}{D-2}\delta h_{ab} p^{ab} p_c{}^c) \\
\\
-2h^\half D^b(h^{-\half} \delta p_{ab}) +  D_a \delta h_{cb} p^{cb} -
2D_c \delta h_{ab} p^{bc}
\end{matrix}
\right) \ .
\een
Since $\L$ is a differential operator that
maps the pair $(\delta h_{ab}, \delta p^{ab})$ consisting of a symmetric tensor, $\delta h_{ab}$,
and a symmetric tensor density, $\delta p^{ab}$, on $\Sigma$ into a pair $(\tilde N, \tilde N_a)$ consisting of a scalar density
and dual vector density on $\Sigma$, its adjoint
differential operator, $\L^*$, maps a pair $(N, N^a)$ consisting of
a scalar and vector field on $\Sigma$ into a pair $(\delta \tilde h^{ab}, \delta \tilde p_{ab})$ consisting of a symmetric tensor density and symmetric tensor on $\Sigma$. In other words, $\L, \L^*$ are maps
\ben
\begin{split}
&\L : C^\infty(\Sigma, (T^* \Sigma)^{\vee 2}) \oplus C^\infty(\Sigma, (T\Sigma)^{\vee 2} \otimes \Lambda^\half) \to
C^\infty(\Sigma, \Lambda^\half) \oplus C^\infty(\Sigma, T^* \Sigma \otimes \Lambda^\half) \\
&\L^* : C^\infty(\Sigma, \mr) \oplus C^\infty(\Sigma, T\Sigma) \to
C^\infty(\Sigma, (T \Sigma)^{\vee 2} \otimes \Lambda^\half) \oplus C^\infty(\Sigma, (T^*\Sigma)^{\vee 2})
\end{split}
\een
where $\Lambda^\half$ is the line bundle of densities of weight $\half$, and $\vee$ is the symmetric tensor product.
$\L^*$ is uniquely determined as a differential
operator by the requirement
that for all smooth $(\delta h_{ab}, \delta p^{ab})$ of compact support on $\Sigma \setminus B$
and all smooth $(N, N^a)$, we have
\ben\label{adjoint}
\left\langle \L^*\left(
\begin{matrix}
N\\
N^a
\end{matrix}
\right)
 \bigg| \left(
\begin{matrix}
\delta h_{ab}\\
\delta p^{ab}
\end{matrix}
\right) \right\rangle  =
\left\langle  \left(
\begin{matrix}
N\\
N^a
\end{matrix}
\right) \bigg| \ \L \left(
\begin{matrix}
\delta h_{ab}\\
\delta p^{ab}
\end{matrix}
\right) \right\rangle \ .
\een
Here, the
angle brackets denote the natural $L^2$-type inner product on the
appropriate tensor/tensor density pairs, e.g.
\ben\label{innprod}
\left\langle \left(
\begin{matrix}
N\\
N^a
\end{matrix}
\right)
 \bigg| \left(
\begin{matrix}
\tilde N\\
\tilde N_a
\end{matrix}
\right) \right\rangle = \int_\Sigma (N\tilde N + N^a \tilde N_a) \ .
\een
One can straightforwardly calculate that $\L^*$ is given by
\ben\label{delc*}
16 \pi \L^*  \left(
\begin{matrix}
N\\
N^a
\end{matrix}
\right) = \left(
\begin{matrix}
h^\half(-D^c D_c N h^{ab} + D^a D^b N  + R^{ab}(h) N)+\\
h^{-\half}( - h^{ab} p^{cd} p_{cd} N + 2 p^{(a}{}_c p^{b)c} N + \frac{1}{D-2} h^{ab} p^c{}_c p^d{}_d N\\
-\frac{2}{D-2}p^{ab}p^c{}_c N - p^{ab} D_c N^c + 2D_c N^{(a} p^{b)c})\\
\\
h^{-\half}(2p_{ab}N
- \frac{2}{D-2} h_{ab} p^c{}_c N)
+2 D_{(a} N_{b)}
\end{matrix}
\right)
\een
Comparing \eqref{adjoint} with the integral of \eqref{newfundid} over $\Sigma$
with $(\delta h, \delta p)$ of compact support on $\Sigma \setminus B$, we see that
\ben
\label{ham1}
W_\Sigma (g;\delta g, \pounds_X g) =
\left\langle \L^* (X) \bigg| \  \left(
\begin{matrix}
\delta h\\
\delta p
\end{matrix}
\right) \right\rangle \ .
\een
Here by $\L^*(X)$, we mean $\L^*(N,N^a)$, where
\ben
X^a = N \nu^a + N^a
\een
is the usual decomposition of a vector into its lapse and shift.
Since neither side of \eqref{ham1} contains derivatives of
$(\delta h, \delta p)$, we can solve
for $\pounds_X g$, obtaining
\ben\label{timeevolution0}
\pounds_X \left(
\begin{matrix}
 h \\
 p
\end{matrix}
\right)
= 16\pi \ \sigma \L^* (X)
\ , \quad
\sigma \equiv \left(
\begin{matrix}
0 & h_{ac} h_{bd}\\
-h^{ac} h^{bd} & 0
\end{matrix}
\right)
\een
which corresponds to the usual ADM evolution equations. Note
that---in view of \eqref{timeevolution0}---\eqref{ham1} continues to hold
even when $(\delta h, \delta p)$
is not of compact support on $\Sigma \setminus B$,
provided only that the integrals defining the left and right sides converge.
Note also that when $(\delta h, \delta p)$
is not of compact support on $\Sigma \setminus B$,
integration of \eqref{newfundid} over $\Sigma$ yields at $\lambda = 0$
\bena\label{adjoint1}
\left\langle \L^*(X) \bigg| \left(
\begin{matrix}
\delta h\\
\delta p
\end{matrix}
\right) \right\rangle &=&
\left\langle X \bigg| \ \L \left(
\begin{matrix}
\delta h\\
\delta p
\end{matrix}
\right) \right\rangle + \int_\infty
[ \delta Q_X(g)
- i_X \theta(g;\delta g)] \nonumber \\
&& - \int_B
[ \delta Q_X(g)
- i_X \theta(g;\delta g)]
\eena

To derive the desired formula for $\E(\gamma_1, \gamma_2)$, we now
consider a two-parameter family of metrics $g_{ab}(\lambda_1, \lambda_2) =
g_{ab} + \lambda_1 \gamma_{1 \ ab} +\lambda_2 \gamma_{2 \ ab}$, where $g_{ab}$ is the
background metric and $\gamma_{1 \ ab}$ and $\gamma_{2 \ ab}$ solve the linearized
equations. Although $g_{ab}(\lambda_1, \lambda_2)$ is {\em not}
a solution, it does satisfy the field equations to first order in
both $\lambda_1$
and $\lambda_2$. We find that the corrections to \eqref{ham1} away from $\lambda_1=\lambda_2 = 0$
are of the form
\ben
W_\Sigma \bigg( g(\lambda_1,\lambda_2);\plo g(0, \lambda_2), \pounds_X g(\lambda_1,\lambda_2) \bigg) =
\left\langle \L^* (X) \bigg| \  \left(
\begin{matrix}
\plo h\\
\plo p
\end{matrix}
\right) \right\rangle
+ O(\lambda_1) +O(\lambda^2_2)\, .
\label{corrW}
\een
Taking the derivative of this equation with respect to $\lambda_2$,
setting $\lambda_1=\lambda_2=0$, and then also setting $X^a=t^a$, we obtain
\ben\label{eq1}
\E(\gamma_1, \gamma_2) =
W_\Sigma \left(g; \gamma_1, \pounds_t \gamma_2 \right) =
\left\langle \plt  \L^* (t) \bigg| \  \left(
\begin{matrix}
\plo h\\
\plo p
\end{matrix}
\right) \right\rangle \bigg|_{\lambda_1 = \lambda_2 = 0} \ .
\een
The right side of this equation can be computed
by taking the variation of the right side
of \eqref{delc*}, although we shall
not explicitly write out the resulting formula here.
Note that although $\delta t^a=0$, it is not compatible with the gauge conditions
we have already imposed near the horizon to require $\delta \nu^a = 0$,
so variations of the lapse and shift must be taken into account in the calculation of $\plt \L^*(t)|_{\lambda_1 = \lambda_2 = 0}$.

An alternative formula for $\E$ can be derived by noting that, to the desired accuracy,
one may substitute \eqref{adjoint1} on the right side of \eqref{corrW}.
Taking the derivative of the resulting equation with respect to $\lambda_2$,
and again setting $\lambda_1=\lambda_2=0$ as well as $X=t$, we obtain
\ben
\begin{split}
\E(\gamma_1, \gamma_2) &=
\left\langle t \bigg| \ \plt \L \left(
\begin{matrix}
\plo h\\
\plo p
\end{matrix}
\right) \right\rangle \Bigg|_{\lambda_1 = \lambda_2 = 0} \\
&+ \bigg( \int_{\infty} - \int_B \bigg) \bigg(
\plot Q_t - i_t \plt \theta(g; \plo g ) \bigg)
 \, .
\end{split}
\een
It is easily seen that the boundary contribution from
infinity vanishes.
Furthermore, the pullback to $B$ of $i_t \plt \theta(g;\plo g)$
vanishes for $\lambda_1 = \lambda_2 = 0$,
since $t^a$ is tangent to $B$. Thus, we obtain
\ben
\E(\gamma_1, \gamma_2) =
\left\langle t \bigg| \  \plt \L \left(
\begin{matrix}
\plo h\\
\plo p
\end{matrix}
\right) \right\rangle
- \int_{B} \plot Q_t  \ ,
\een
where evaluation at $\lambda_1 = \lambda_2 = 0$ is understood again.
On the other hand, we have
\ben
\left\langle t \bigg| \ \plt \L \left(
\begin{matrix}
\plo h\\
\plo p
\end{matrix}
\right) \right\rangle \Bigg|_{\lambda_1 = \lambda_2 = 0} = \frac{\partial^2}{\partial \lambda_1 \partial \lambda_2}
\int_\Sigma t^a C_a(g+\lambda_1 \delta_1 g+\lambda_2 \delta_2 g) \ \Bigg|_{\lambda_1=\lambda_2=0} \, .
\een
Since
$t^a C_a = -\frac{1}{8\pi} G_{ab} \nu^a t^b \ h^\half$
the right side may be computed by taking second order variations
of the Einstein tensor. This gives
\ben
\E = \E(\gamma, \gamma) = -\frac{1}{8\pi}
\int_\Sigma h^\half \ \frac{d^2}{d\lambda^2} G_{ab}
(g+\lambda \gamma)  \ \nu^a t^b \bigg|_{\lambda=0}
- \int_B \frac{d^2}{d\lambda^2} Q_t(g+\lambda\gamma)\bigg|_{\lambda=0} \ .
\een
Explicit evaluation of the two terms on the right side
(see~\cite{habison} for the first term, and use~\eqref{kaaff}, \eqref{rig}, \eqref{Qdef}, \eqref{gaussian} for the
second term) gives:
\bena
\E &=&\frac{1}{8 \pi} \int_\Sigma h^\half \bigg(
\gamma^{cd} \nabla_a \nabla_b \gamma_{cd} - 2\gamma^{cd} \nabla_c \nabla_{(a} \gamma_{b)d} +
\half (\nabla_a \gamma_{cd}) \nabla_b \gamma^{cd} + 2 (\nabla^d \gamma^c{}_b) \nabla_{[d} \gamma_{c]a} \non\\
&& + \nabla_d (\gamma^{dc} \nabla_c \gamma_{ab}) - \half (\nabla^c \gamma) \nabla_c \gamma_{ab}
-2(\nabla_d \gamma^{cd} - \half \nabla^c \gamma) \nabla_{(a} \gamma_{b)c} - \half g_{ab}
({\rm trace})
\bigg) \nu^a t^b \  \ , \non \\
&& - \frac{\kappa}{16\pi} \int_B \mu^\half \ \delta \mu_{ab} \delta \mu^{ab} \ ,
\eena
where $\delta \mu_{ab}$ is, as before, equal to the pullback of
$\gamma_{ab}$ to $B$, where ``trace'' denotes the trace of the
preceding terms, and where use has been made
of the gauge conditions at $B$ in the computation of the boundary
term. For later use, we also quote the
lengthy expression for $\E$ in terms of the variables $(\delta h_{ab}, \delta p^{ab})$
and the lapse and shift $(N,N^a)$ of $t^a$, which can be obtained by taking
a second variation of \eqref{constraintformula}:
\bena
\E &=& \frac{1}{16\pi} \int_\Sigma N \bigg(h^{\half}\bigg\{ \half \ R_{ab}(h) \delta h_c{}^c \delta h^{ab}
- 2 \ R_{ac}(h) \delta h^{ab} \delta h_b{}^c -\half \ \delta h^{ac} D_a D_c \delta h_d{}^d - \non\\
&&\half \ \delta h^{ac} D^b D_b \delta h_{ac}
+ \delta h^{ac} D^b D_a \delta h_{cb} -  \frac{3}{2} \ D_a(\delta h^{bc} D^a \delta h_{bc}) - \frac{3}{2} \ D_a(\delta h^{ab} D_b \delta h_c{}^c) + \non\\
&& \half \ D_a(\delta h_d{}^d D^a \delta h_c{}^c) + 2 \ D_a(\delta h^a{}_c D_b \delta h^{cb}) + D_a(\delta h^b{}_c D_b \delta h^{ac}) - \half \ D^a (\delta h_c{}^c D^b \delta h_{ab}) \bigg\}+\non\\
&& h^{-\half}\bigg\{ 2 \ \delta p_{ab} \delta p^{ab} +
\half \ p_{ab} p^{ab} (\delta h_a{}^a)^2 -  p_{ab} \delta p^{ab} \delta h_c{}^c -
3 \ p^a{}_b p^{bc} \delta h_d{}^d \delta h_{ac} -
\non\\
&&\frac{2}{D-2} \ (\delta p_a{}^a)^2 +  \frac{3}{D-2}  \ p_c{}^c \delta p_b{}^b \delta h_a{}^a + \frac{3}{D-2} \  p_d{}^d p^{ab} \delta h_c{}^c \delta h_{ab} +
8 \ p^c{}_b \delta h_{ac} \delta p^{ab} + \non\\
&&p_{cd} p^{cd} \delta h_{ab} \delta h^{ab} +
2 \ p^{ab} p^{dc} \delta h_{ac} \delta h_{bd} -\frac{1}{D-2} \  (p_c{}^c)^2 \delta h_{ab} \delta h^{ab}
- \frac{1}{2(D-2)} \ (p_b{}^b)^2 (\delta h_a{}^a)^2 -\non\\
&& \frac{4}{D-2}  \ p_c{}^c \delta p^{ab} \delta h_{ab} -
\frac{2}{D-2} \ (p^{ab} \delta h_{ab})^2 - \frac{4}{D-2} \ p_{ab} \delta p_c{}^c \delta h^{ab}
\bigg\} \bigg) - \non\\
&& \frac{1}{16\pi} \int_\Sigma N^a \bigg( -2 \ \delta p^{bc} D_a \delta h_{bc} + 4 \ \delta p^{cb} D_b \delta h_{ac} +2 \ \delta h_{ac} D_b \delta p^{cb} - \non\\
&&2 \ p^{cb} \delta h_{ad}D_b \delta h_c{}^d + p^{cb} \delta h_{ad} D^d \delta h_{cb}
\bigg) +\frac{\kappa}{16\pi} \int_B \mu^\half \bigg( \delta \mu_{ab} \delta \mu^{ab} - \half \ \delta \mu_a{}^a \delta \mu_b{}^b \bigg) \ \ .
\label{Eexpr}
\eena

\section{Flux of canonical energy at infinity and the horizon}

As shown in the previous section, canonical energy $\E(\gamma)$
is conserved for all solutions $\gamma_{ab}$ of the linearized Einstein equation
in the sense that it is independent
of the choice of Cauchy surface $\Sigma$ extending from spatial infinity to $B$. However,
if we evaluate $\E$ on a slice ${\S}(t)$ that extends from a cross-section, $C(t)$, of
future null infinity $\I^+$,
(rather than from spatial infinity) to a cross section, $B(t)$, of the future horizon, $\H^+$,
(rather than to $B$)
\ben
\E(\gamma, {\S}(t)) \equiv W_{{\S}(t)}(\gamma, \pounds_t \gamma)
= \int_{{\S}(t)} \omega(g;\gamma, \pounds_t \gamma)  \, ,
\een
then, of course, in general we will find that $\E(\gamma, {\S}(t)) \neq \E(\gamma)$.
In this section we study the time evolution of $\E(\gamma, {\S}(t))$
as we march the slice ${\S}(t)$ forward in $t$. We will show that, up to boundary terms,
$\E(\gamma, {\S}(t))$ decreases with $t$.

In order to be able to make use of the machinery of null infinity, we will restrict consideration
in this section to even dimensional spacetimes---$\I$ does not exist for odd dimensional
spacetimes~\cite{hw}. However, we do not believe that this is an essential
restriction, i.e., we believe that our results hold in odd dimensions, with a suitable notion of
asymptotic flatness for that case, see e.g.~\cite{tanabe}.

We work in the ``Bondi gauge'' for the background metric $g_{ab}$, so the unphysical
background metric $\tilde g_{ab}=\Omega^2 g_{ab}$ near $\I^+$ takes the form
\ben\label{bondigauge}
\tilde g_{ab} = 2 \tilde \nabla_{(a} \Omega \tilde \nabla_{b)} \tilde u + \tilde \mu_{ab} + O(\Omega)
\een
with $\tilde u$ a future directed affine parameter on the null geodesic generators of $\I^+$. On $\I^+$
we have $(\partial/\partial \tilde u)^a = \tilde g^{ab} \tilde \nabla_b \Omega \equiv \tilde n^a$, and
$\tilde \mu_{ab}$ is the unit round metric on the $S^{D-2}$  cross-sections of $\I^+$, with $\tilde \mu_{ab} \tilde{n}^a = 0 = \tilde \mu_{ab} (\partial/\partial \Omega)^a$.
The asymptotically timelike Killing field $t^a$ can be extended continuously to a vector field
$\tilde t^a$ on $\I^+$ which is proportional to $\tilde n^a$, where we consequently have
\ben\label{tana}
\tilde t^a = (\tilde t^c \tilde \nabla_c \tilde u) \ \tilde n^a \ ,
\een
with $\quad \tilde t^c \tilde \nabla _c \tilde u >0$ and constant on $\I$.
If the initial data for $\gamma_{ab}$ on $\Sigma$ are of compact support\footnote{\label{footref} Below,
we will consider perturbations whose initial data are in a space $\mathcal T$. Elements in this space
may not have compact support, but it is shown in lemma~\ref{lemma5}
of Appendix C that the initial
data of compact support are dense in this space.}, then it has been shown in~\cite{hi}
that there exists a gauge\footnote{In $D=4$, this is the Geroch-Xanthopoulos gauge,
whereas in even $D>4$, it is the transverse-traceless gauge, see~\cite{hi} for further
details and other subtle differences between $D=4$ and the rest.} near future null infinity in which $\gamma_{ab}$
is asymptotically flat at  future null-infinity $\I^+$ in the sense that
\ben
\tilde \gamma_{ab} := \Omega^{-(D-6)/2} \gamma_{ab}
\een
is smooth on $\I^+$, and that $\tilde \gamma_{ab} \tilde g^{ab} = O(\Omega), \tilde \gamma_{ab} \tilde n^a = O(\Omega)$.

It is also assumed throughout this section that $\gamma_{ab}$ is axisymmetric
with respect to all of the axial Killing fields appearing in~\eqref{rig}
\ben
\pounds_{\psi_A} \gamma_{ab} = 0 \quad \text{for all $A=1, \dots,N$} \  \ .
\label{axi}
\een
This restriction is essential because
although we will obtain positivity of flux results at $\I^+$ for
canonical energy, we will obtain positivity of flux results at $\H^+$ for the
analogous quantity defined by replacing $t^a$ by the horizon Killing field $K^a$. It is only
in the presence of axisymmetry \eqref{axi} that these quantities are equal.

Let $\I_{12}$ denote
the portion of future null-infinity bounded by cross-sections $C(t_1)$ and $C(t_2)$, and let
$\H_{12}$ the portion of the horizon bounded by the cross-sections $B(t_1)$ and $B(t_2)$, see
the following figure.

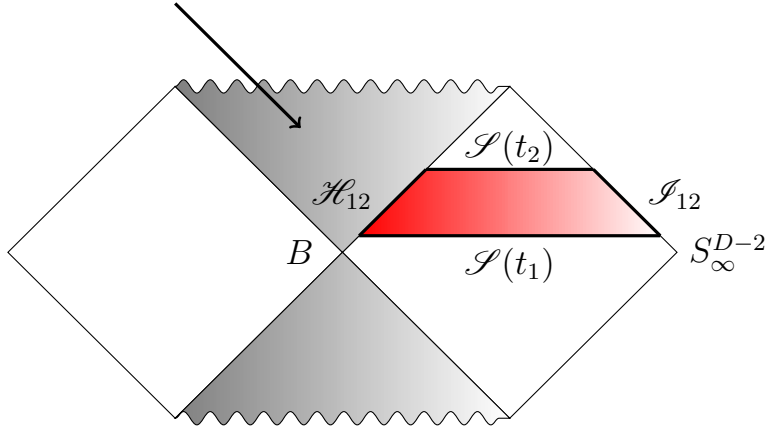
\begin{center}
\begin{tikzpicture}[scale=1.1, transform shape]
\shade[left color=red] (0.2,0.2) -- (3.8,.2) -- (3,1) -- (1,1);
\draw (0,0) -- (-2,2) -- (-4,0)  -- (-2,-2) -- (0,0);
\draw (4,0) node[right]{$S^{D-2}_{\infty}$};
\draw (0,0) -- (2,2) --   (4,0) --  (2,-2) -- (0,0);
\shade[left color=gray] (0,0) -- (-2,2) decorate[decoration=snake] {-- (2,2)} --  (0,0);
\draw (0,0) -- (-2,2) decorate[decoration=snake] {-- (2,2)} -- (0,0);
\shade[left color=gray] (0,0) -- (-2,-2) decorate[decoration=snake] {-- (2,-2)} -- (0,0);
\draw (0,0) -- (-2,-2) decorate[decoration=snake] {-- (2,-2)} -- (0,0);
\draw[very thick, black] (.2,.2) -- (3.8,.2);
\draw[very thick, black] (.2,.2) -- (1,1);
\draw[very thick, black] (3.8,.2) -- (3,1);
\draw[very thick, black] (1,1) -- (3,1);
\draw (2,0.2) node[below]{${\small \mathscr S}(t_1)$};
\draw (2,.9) node[above]{${\small \mathscr S}(t_2)$};
\draw (3.5,.7)  node[above,right]{$\I_{12}$};
\draw (.5,.7) node[above,left]{$\H_{12}$};
\draw[->, very thick] (-2,3) node[above left] {BH $=\M \setminus J^{-}(\mathscr{I}^{+})$} -- (-.5,1.5);
\draw(-0.2,0) node[black, left]{$B$};
\end{tikzpicture}
\end{center}

Clearly
\ben\label{decomp}
{\S}(t_{1}) \cup \H_{12} \cup \I_{12} \cup {\S}(t_2)
\een
bounds a compact region of the conformally compactified spacetime.
By conservation of canonical energy, we have
\ben
\E(\gamma, {\S}(t_2)) = \E(\gamma, {\S}(t_1))
-W_{\I_{12}}(\gamma, \pounds_t \gamma) - W_{\H_{12}}(\gamma, \pounds_t \gamma) \ .
\een
Thus, $\E(\gamma, {\S}(t_2))$ differs from $\E(\gamma, {\S}(t_1))$ by the ``flux
terms'' $W_{\I_{12}}(\gamma, \pounds_t \gamma)$ and $W_{\H_{12}}(\gamma, \pounds_t \gamma)$,
corresponding, respectively, to the canonical energy
radiated to null infinity and into the black hole.

We now evaluate $W_{\I_{12}}(\gamma, \pounds_t \gamma)$.
Using a formula from~\cite{hi} based on the Einstein equations for the
background and perturbations, we have
\ben
W_{\I_{12}}(\gamma_1 , \gamma_2 ) = \frac{1}{32\pi} \int_{\I_{12}} (\tilde \gamma_{1 \ cd} \delta_2 \tilde N^{cd}
- \tilde \gamma_{2 \ cd} \delta_1 \tilde N^{cd}) \  \ ,
\een
where here and in the following, the natural integration measure on $\I^+$ coming from $\tilde g_{ab}$ is understood\footnote{In
local coordinates $\Omega,u,x_1,\dots,x_{D-2}$ adapted to~\eqref{bondigauge}, this would be
$\tilde \mu^\half \D u \D x_1 \cdots \D x_{D-2}$.}, and where indices on tilde tensor fields are
always raised and lowered with $\tilde g_{ab}$.
The Bondi news tensor is defined by\footnote{The form of the second term depends in general on
the the conformal factor chosen near $\I$; the trace-type form is consistent with our gauge in which the
cross sections of $\I$ are round spheres relative to the unphysical metric.}
\ben
\tilde N_{cd} = \tilde \mu^a{}_c \tilde \mu^b{}_d \ \Omega^{-\frac{D-4}{2}} \
\left[\frac{2}{D-2}\tilde R_{ab} - \frac{1}{(D-1)(D-2)} \tilde R \tilde g_{ab} \right]- \frac{1}{D-2} \tilde \mu_{cd} ({\rm trace})
\een
where ``trace'' denotes the contraction with $\tilde g^{cd}$ of the preceding terms.
As shown in~\cite{hi}, both $N_{ab}$ and $\delta \tilde N_{ab} = \frac{d}{d\lambda} \tilde N_{ab}(0)$ are
smooth at $\I$ in our gauge. In fact, eq.~(56) from~\cite{hi}, which uses the linearized
Einstein equations, gives that, at $\I^+$,
\ben\label{bondirelation}
\delta \tilde N_{ab} = - \pounds_{\tilde n} \tilde \gamma_{ab} + D \ \Omega^{-1} \tilde n_{(a} \tilde \gamma_{b)c}
\tilde n^c - \frac{D-2}{2} \ \Omega^{-1} \tilde n^c \tilde n_c \tilde \gamma_{ab} \
 \ .
\een
From this, eq.~\eqref{tana}, and using also $\tilde n^c \tilde n_c = O(\Omega^2)$ since $\tilde u$ is an
affine parameter on scri, we obtain
\ben
W_{\I_{12}}(\gamma, \pounds_t \gamma) =
\frac{1}{16 \pi} \int_{\I_{12}} (\tilde t^a \tilde \nabla_a \tilde u) \ \delta \tilde N_{cd} \delta \tilde N^{cd}
+{\mathcal C}(t_2) -  {\mathcal C}(t_1) \ ,
\een
where
\ben
{\mathcal C}(t) := \frac{1}{32\pi} \int_{C(t)} (\tilde t^a \tilde \nabla_a \tilde u) \ \tilde \gamma^{cd}  (\pounds_{\tilde n} \tilde \gamma_{cd}) \ ,
\een
and where the natural integration element on $C(t)$ induced by $\tilde \mu_{ab}$ is understood.

Next, we evaluate  $W_{\H_{12}}(\gamma, \pounds_t \gamma)$.
The Raychaudhuri equation on $\H^+$ yields
\ben
\frac{d}{du} \vartheta(\lambda) = -\frac{1}{D-2} \ \vartheta(\lambda)^2 - \sigma_{ab}(\lambda) \sigma^{ab}(\lambda) - R_{ab}(\lambda) n^a n^b \ ,
\een
where, as before,
\ben
\vartheta = \half \mu^{ab} \pounds_n \mu_{ab}
\een
denotes the expansion
of the generators of $\H^+$, whereas
$\sigma_{ab}$ denotes their shear
\ben\label{pnmab}
\half \pounds_n \mu_{ab} = \sigma_{ab} + \frac{1}{D-2} \mu_{ab} \vartheta \, .
\een
Here, the quantities $\mu_{ab}$ and $n^a= (\partial/\partial u)^a$ refer
to the horizon metric; see~\eqref{gaussian}.
We take a $\lambda$-derivative of this equation and evaluate at $\lambda=0$. Then,
since $\vartheta(0) = 0 = \sigma_{ab}(0)$ for the background metric $g_{ab}(0)$, and since $\delta R_{ab} = dR_{ab}/d\lambda(0)=0$ by the
linearized Einstein equations, we have $d \delta \vartheta/du = 0$. Since $\delta \vartheta(0)=0$
on $B$ by our choice of gauge (see subsection 2.1),
it follows that $\delta \vartheta = 0$ on the entire horizon, as we have previously claimed. This implies that
$\pounds_n (g^{ab} \gamma_{ab}) = 0$. Using this fact,
one can show that
\ben
W_{\H_{12}}(\gamma, \pounds_t \gamma) = \frac{1}{32\pi} \int_{\H_{12}}
(\pounds_K \gamma^{cd}  \pounds_n \gamma_{cd}
- \gamma^{cd} \pounds_n
\pounds_K \gamma_{cd} )  \  \ ,
\een
where we have used the axisymmetry~\eqref{axi} of $\gamma_{ab}$ to replace
$\pounds_t \gamma_{cd}$ by $\pounds_K \gamma_{cd}$, where $K^a$ is the Killing field
\eqref{rig} normal to the horizon. Using $K^a = \kappa u n^a$, we obtain
\ben
W_{\H_{12}}(\gamma, \pounds_t \gamma) = \frac{1}{4\pi} \int_{\H_{12}}
(K^a \nabla_a u) \ \delta \sigma_{cd} \delta \sigma^{cd}
+{\mathcal B}(t_2) -  {\mathcal B}(t_1)
\een
where
\ben
{\mathcal B}(t) \equiv
\frac{1}{32\pi} \int_{B(t)} (K^a \nabla_a u) \ \gamma^{cd} (\pounds_n \gamma_{cd}) \ ,
\een
and where the natural integration element on $B(t)$ induced by $\mu_{ab}$ is understood.
The above calculations motivate the definition of a {\em modified canonical energy}
$\bE(\gamma, {\S}(t))$ given by
\bena
\label{modcan}
&&\bE(\gamma, {\S}(t)) := \E(\gamma, {\S}(t)) \\
&&- \frac{1}{32\pi} \int_{C(t)} (\tilde t^a \tilde \nabla_a \tilde u) \ \tilde \gamma^{cd}  (\pounds_{\tilde n} \tilde \gamma_{cd}) \  - \frac{1}{32\pi} \int_{B(t)} (K^a \nabla_a u) \ \gamma^{cd} (\pounds_n \gamma_{cd}) \ \ . \non
\eena
Thus, the modified canonical energy $\bE(\gamma, {\S}(t))$ differs from
$\E(\gamma, {\S}(t))$ only
by the above boundary terms ${\mathcal C}(t)$ and ${\mathcal B}(t)$. Note that ${\mathcal C}(t)$
vanishes when the perturbed Bondi news, $\delta N_{ab}$, vanishes. This is seen using
eq.~\eqref{bondirelation}, the fact~\cite{hi} that $\tilde \gamma_{ab} \tilde n^a = O(\Omega)$, and that $\tilde n^a \tilde n_a = O(\Omega^2)$, which in turn follows since $\tilde u$ is by construction an affine parameter in $\I^+$.
Also, ${\mathcal B}(t)$ vanishes
when the perturbed shear, $\delta \sigma_{ab}$, vanishes, since $\delta \sigma_{ab} = \half \pounds_n \gamma_{ab}$
on $\H^+$, by~\eqref{pnmab}, $\delta \vartheta = 0 = \vartheta$ on $\H^+$, and by expression eq.~\eqref{lgaussian} for $\gamma_{ab}$. In addition, ${\mathcal B}$
vanishes at $B$ since $K^a|_B = 0$. Since
the perturbed Bondi news vanishes as one approaches spatial infinity\footnote{\label{footref1} This is
manifestly true because we consider perturbations having compact support on $\Sigma$; see footnote~\ref{footref}.}
it follows that $\bE(\gamma, {\S}(t)) \rightarrow \E(\gamma, \Sigma)$ as
${\S}(t) \rightarrow \Sigma$.

The above results establish the following theorem:

\begin{thm}\label{thm}
Let $\gamma_{ab}$ be an axisymmetric linearized perturbation that, in addition to the properties assumed in the
previous section, is asymptotically flat at future null infinity. Then the modified canonical energy
\eqref{modcan}
has the property that, for $t_1< t_2$
\bena\label{flux}
&&\bE(\gamma, {\S}(t_2))-\bE(\gamma, {\S}(t_1)) \\
&&=-\frac{1}{16 \pi} \int_{\I_{12}} (\tilde t^a \tilde \nabla_a \tilde u) \ \delta \tilde N_{cd} \delta \tilde N^{cd}
 - \frac{1}{4\pi} \int_{\H_{12}} (K^a \nabla_a u) \ \delta \sigma_{cd} \delta \sigma^{cd} \
  \le 0 \ . \non
\eena
\end{thm}

We remark that Habisohn~\cite{habison} has considered, in $D = 4$ dimensions, the second order Einstein tensor, and has derived a balance law for the flux through an infinitely extended timelike tube approaching null infinity that is similar to that given above in~\eqref{flux}, although his balance law does not refer to the flux between fixed times $t_1$ and $t_2$ as above, but to the total flux for all times.

\section{Stability and instability}

As seen is section~2, the canonical energy, $\E$, can be viewed as a
quadratic form defined on the vector space of smooth, linearized
solutions that satisfy our asymptotic conditions at spatial infinity
and our gauge conditions near the horizon (see subsection 2.1). If
$\E$ were positive definite on this vector space, it would provide a
conserved norm that could be used to argue for stability. On the other
hand, if $\E<0$ for some perturbation, then the flux results of the
previous section show that the modified canonical energy $\bar{\E}$ on
the slice ${\S}(t)$ can only become more negative with time, which
suggests instability. However, there are a number of obvious
difficulties with making such arguments. In particular:

\begin{itemize}

\item
As already
noted [see~\eqref{sce}], $\E$ is negative for the ``change of mass''
perturbation of the Schwarzschild black hole, yet Schwarzschild is
known to be linearly stable. Such ``trivial'' perturbations must be
eliminated from the analysis.

\item
$\E$ will, in general, be degenerate
on some perturbations, so, even if positive,
it cannot be expected to be positive definite. Thus, the degeneracies
of $\E$ must be carefully analyzed.

\end{itemize}

In order to
properly analyze the degeneracies of $\E$, it will be very useful to
view $\E$ as a densely defined quadratic form on a Hilbert space. In
the next subsection, we will introduce such a Hilbert space $\V$---whose construction
and properties may be of some interest in their own right---and
analyze the degeneracies of $\E$. It should be noted that the perturbations in $\V$
will have vanishing linearized ADM mass, linear momentum, and
angular momentum with respect to rotational Killing fields,
$\delta M = \delta P_i = \delta J_A  = 0$, so they will automatically
eliminate the ``trivial'' perturbations corresponding to variations of the mass
or angular momentum in a family of stationary black holes.
We will then make our
stability/instability arguments in subsection 4.2.

\subsection{The Hilbert space $\V$ and the degeneracies of $\E$}

We start\footnote{Although we will restrict consideration to axisymmetric
perturbations when we make our stability arguments, the constructions
of this subsection do not require axisymmetry, and we will not
impose this restriction until the end of this subsection.} with the
real Hilbert space
$${\mathcal K} = L^2(\Sigma, (T^*\Sigma)^{\vee 2}; h^\half e_0) \oplus L^2(\Sigma, (T\Sigma)^{\vee 2} \otimes \Lambda^{\half}; h^{-\half} e_0)$$
of all square integrable
linearized initial data, not necessarily satisfying the constraints. Here
$\Lambda^{\half}$ is the line bundle over $\Sigma$ of densities of weight one half. Thus, the first summand
 denotes the Hilbert space of square integrable
symmetric tensors on $\Sigma$, whereas the second denotes the Hilbert space of
square integrable symmetric tensor
densities of weight $\half$
over $\Sigma$. In other words, elements of ${\mathcal K}$ consist of pairs $(\delta h_{ab}, \delta p^{ab})$
with inner product
\ben
\left\langle \left(
\begin{matrix}
\delta_1 h\\
\delta_1 p
\end{matrix}
\right)
\bigg|  \left(
\begin{matrix}
\delta_2 h\\
\delta_2 p
\end{matrix}
\right) \right\rangle_{\mathcal K}
:= \int_\Sigma h^\half \ \delta_1 h_{ab} \delta_2 h^{ab} + h^{-\half} \delta_1 p_{ab} \delta_2 p^{ab} \ .
\een
Note that for $D=4,5$, the requirement of square integrability will impose faster fall-off
conditions at infinity on $\delta h_{ab}$ (but not $\delta p^{ab}$) than assumed
in section 2 and will
exclude the possibility of
having a nonvanishing $\delta M$, but for $D \geq 6$, ${\mathcal K}$ will include all
smooth initial data satisfying the asymptotic conditions at spatial infinity stated at the beginning
of subsection 2.1. We may view the symplectic form \eqref{Wdef} as a bounded linear map
$S: {\mathcal K} \rightarrow {\mathcal K}$ defined by\footnote{\label{footnote1}
Note that, in subsec.~\ref{sec:can},
the angles $\langle \ | \ \rangle$ as e.g. in~\eqref{adjoint}
denote the dual pairing between tensors and densities, whereas
$\langle \ | \ \rangle_{\mathcal K}$ is an inner product between objects of the same density.
As a consequence, the definition of $\L^*$, defined relative to $\langle \ | \ \rangle_{\mathcal K}$, used from now on,
differs from the earlier one in~\eqref{delc*} by a factor of $diag(h^{-\half},h^\half)$. This also
accounts for the difference between $S$ and $\sigma$ in~\eqref{timeevolution0}.
}
\ben\label{Wdef1}
16 \pi \ W_\Sigma(\delta_1 g, \delta_2 g) = \left\langle \left(
\begin{matrix}
\delta_1 h\\
\delta_1 p
\end{matrix}
\right)
\bigg| S \left(
\begin{matrix}
\delta_2 h\\
\delta_2 p
\end{matrix}
\right) \right\rangle_{\mathcal K}
\een
where
\ben
S \equiv \left(
\begin{matrix}
0 & - h^{-\half} h_{ac} h_{bd}\\
h^{\half} h^{ac} h^{bd} & 0
\end{matrix}
\right) \ .
\een
 Note that $S^*= -S$ and
$S^2 = - I$, so, in particular, $S$ is an orthogonal map, $S^* S = I$.

We would now like to pass to the subspace of ${\mathcal K}$ that
satisfies the constraints and satisfies our gauge conditions. At first
sight, it might appear that this would be a difficult task, given that
generic elements of ${\mathcal K}$ are not even
differentiable. However, the desired subspace $\V$ can be defined
straightforwardly as follows: Let
\bena
\W &:=& \{ \pounds_X g_{ab} \in {\mathcal K} \mid X^a \in C^\infty,
\text{near infinity $X^a$ coincides with a rotational Killing} \non\\
&& \text{field plus an asymptotic translation,
$X^a |_B$ tangent to generators of $\H^+$} \}
\label{W}
\eena
Here, by $\pounds_X g_{ab}$ we mean the corresponding initial data
$((\pounds_X g)_h, (\pounds_X g)_p)$.
The space of interest for us is the space $\V$ of initial data symplectically orthogonal to $\W$:
\ben
\V=\W^{\perp_S} =\Bigg\{ (\delta h, \delta p) \in {\mathcal K} \  : \
\left\langle \left(
\begin{matrix}
\delta h\\
\delta p
\end{matrix}
\right)
\bigg| S \left(
\begin{matrix}
\delta h'\\
\delta p'
\end{matrix}
\right) \right\rangle_{\mathcal K}
 = 0 \quad
\text{for all $(\delta h', \delta p') \in \W$}\Bigg\} \, .
\een
Thus, $\V$ is a closed subspace of ${\mathcal K}$ and, thus, is itself
a Hilbert space.
Note that since $S^* = -S$, we have $\V^{\perp_S} = \overline{\W}$, where
$\overline{\W}$ denotes the closure of $\W$.

In the following, we will need to work with tensor fields that
are in the weighted Sobolev spaces $W^k_\rho\equiv W^k_\rho(\Sigma)$.
These spaces denote the weakly differentiable
tensor fields $u$ on $\Sigma$ such that $\rho^n \ D_{(a_1} \dots D_{a_n)} u \in L^2$
for all $n \le k$, where $\rho>0$ is a
smooth function on $\Sigma$ which interpolates
between $1$ in a neighborhood of $B$ and
$(x_1^2+\dots+x_{D-1}^2)^\half$ in a neighborhood of infinity.
The weight factor $\rho^{n}$
for the $n$-th derivative is inserted so as to
force weak derivatives to fall off
faster by an appropriate power.
The weighted Sobolev norm for $u \in W^k_\rho$ is defined by
\ben
\| u \|_{W^k_{\rho}} := \left\{ \sum_{n=0}^k \int_\Sigma h^\half \ \rho^{2n} (D_{(a_1} \cdots D_{a_n)} u)
D^{(a_1} \cdots D^{a_n)} u  \right\}^\half  \ .
\een
We will use the notation $C^\infty_0(\Sigma)$ to
denote tensor fields
that are smooth and of compact support on $\Sigma \setminus B$. We will use the
notation $C^\infty(\overline \Sigma)$ to denote smooth tensor fields on
$\Sigma$ that can be smoothly extended across $B$.

We denote by $\mathcal U$ the space
\ben
\mathcal U = \bigcap_k \Big( W^k_\rho(\Sigma) \oplus W^k_\rho(\Sigma) \Big)
\label{U}
\een
Clearly, $\U$ is a subspace of $\mathcal K$, and
it follows from the Sobolev embedding theorems that $\U \subset
C^\infty(\overline \Sigma) \oplus C^\infty(\overline \Sigma)$.

The proof of property (3) of proposition~5 below
will require the following lemma:
\begin{lemma}\label{lemma3}
Let $Y=(N_0^{},N^a_0) \in C^\infty_0 \oplus C^\infty_0$. Then there exists a solution
$X$ in the space $\cap_k (\rho^2 W^{k+1}_\rho(\Sigma) \oplus \rho W^k_\rho(\Sigma)) \subset C^\infty(\overline \Sigma) \oplus C^\infty(\overline \Sigma)$
to the following boundary value problem:
\ben\label{pde1}
\L\L^*(X) = Y \qquad \text{in $\Sigma$,}
\een
and $X = (N,N^a)$ satisfies
\ben\label{bcon1}
N^a = N \eta^a \ , \quad \delta \vartheta = \delta \epsilon = 0 \quad \text{on $B=\partial \Sigma$.}
\een
Here $\L$ and $\L^*$ are given by \eqref{delc} and \eqref{delc*}
and $\delta \vartheta, \delta \epsilon$ are the
perturbed expansion/volume element
on $B$ associated with the perturbation
$(\delta h, \delta p)=\L^*(X)$, given by
\bena\label{bndy}
\delta \vartheta |_B &=&h^{-\half}(h^{ab}-\eta^a \eta^b)
\Big(\delta p_{ab} -\half \delta h_c{}^c p_{ab}
+2p_{c(a} \delta h_{b)}{}^c  + \half h^{\half} \ \pounds_\eta \delta h_{ab} \non\\
&& - \frac{1}{D-2} (-\half p h_{ab} \delta h_c{}^c
+p^{cd} \delta h_{cd} h_{ab} + p \delta h_{ab} + \delta p_c{}^c
h_{ab}
)
\Big) \Big|_B \\
\delta \epsilon |_B &=& \half (h^{ab}-\eta^a \eta^b) \delta h_{ab}
\Big|_B \ ,
\label{bndy2}
\eena
where $\eta^a$ is the unit inward normal to $B$ within $\Sigma$.
Furthermore, if $D \ge 5$,
the solution is
unique, whereas for $D=4$, the solution is also unique unless $t^a$ is
tangent to the generators of the horizon (i.e., the black hole is nonrotating),
in which case the solution is unique up to $X^a \to X^a + ct^a$.
\end{lemma}
\noindent
A proof of this lemma is given in appendix~B.

\medskip
\noindent
The key properties of $\V$ are now summarized in the following proposition:

\begin{prop}\label{prop2}
\begin{enumerate}
\item Let $(\delta h, \delta p) \in \V$. Then $(\delta h, \delta p)$ is
a distributional solution to the linearized constraints in the sense that
\ben
\left\langle \L^*(X) \bigg| \left(
\begin{matrix}
\delta h\\
\delta p
\end{matrix}
\right) \right\rangle_{\mathcal K} = 0 \ ,
\een
for all $X^a \in C^\infty_0$.

\item If $(\delta h, \delta p)$
is a smooth element of $\V$, then
$\delta M = \delta J_A =
\delta P_i = 0$ and
$\delta \epsilon |_B = \delta\vartheta |_B = 0$.
Conversely, if $(\delta h, \delta p) \in {\mathcal K}$ is a smooth
solution of the
linearized constraints satisfying $\delta M = \delta J_A =
\delta P_i = 0$ and
$\delta \epsilon |_B = \delta\vartheta |_B = 0$, then
$(\delta h, \delta p) \in \V$.

\item The space $\U \cap \V$ is dense in $\V$, where $\U$ was defined
by \eqref{U}.
\end{enumerate}
\end{prop}

\noindent
{\em Proof:} Property (1) is an immediate consequence of \eqref{ham1}
together with the fact that for any $X^a \in C^\infty_0$, we clearly
have $\pounds_X g_{ab} \in \W$. Property (2) is an
immediate consequence\footnote{Our
asymptotic conditions at infinity are now that
$(\delta h, \delta p) \in {\mathcal K}$ rather than
the conditions that follow from those
stated in the second paragraph of subsection 2.1. It
is easily verified that the proof of lemma~\ref{lemma2}
continues to hold with our present
asymptotic conditions.}
of lemma~\ref{lemma2}.

Our strategy for proving property (3) is to define a
bounded projection operator $\Pi:
{\mathcal K} \to \V$ which takes $C^\infty_0 \oplus C^\infty_0 \subset L^2 \oplus L^2 \equiv \mathcal K$ to elements in
$\U \cap \V$.
Since $C^\infty_0 \oplus C^\infty_0$ is dense in $\mathcal K$,
it will then follow that $\U \cap \V$
is dense in $\V$.
We will produce the desired operator $\Pi$ in two steps. First
we define an orthogonal
projector $\Pi_0: {\mathcal K} \to \V_0$ onto a closed
subspace $\V_0 \subset {\mathcal K} \equiv L^2 \oplus L^2$ containing $\V$. The
space $\V_0$
``imposes the constraints'' and the ``correct boundary conditions''
$\delta \epsilon |_B = \delta\vartheta |_B = 0$ at $B$,
but does not impose $\delta M = \delta J_A =
\delta P_i = 0$. We will then compose $\Pi_0$ with a finite co-rank
projection operator $\Pi_1$ that commutes with $\Pi_0$ to obtain the
desired projection operator $\Pi$. Since orthogonal projectors and finite
co-rank projectors are bounded, $\Pi$ is bounded.

Our prescription for the operator $\Pi_0$ is as follows.
Let $(\delta h_0, \delta p_0) \in C^\infty_0 \oplus C^\infty_0$.
Then, obviously, $Y:= \L(\delta h_0, \delta p_0)$ is
smooth and of compact support. Let $X$ be the solution to
\ben\label{EL}
\L \L^*(X) = Y
\een
given by lemma~\ref{lemma3} above. We define\footnote{See footnote \ref{footnote1}.}
\ben\label{pidef}
\Pi_0 \left(
\begin{matrix}
\delta h_0\\
\delta p_0
\end{matrix}
\right) = \left(
\begin{matrix}
\delta h_0\\
\delta p_0
\end{matrix}
\right) - \L^*(X) \ .
\een
It follows immediately that $\Pi_0(\delta h_0, \delta p_0) \in \U$,
that it satisfies the linearized constraints, and
that it satisfies
$\delta \vartheta |_B = 0 = \delta \epsilon |_B$ [cf. eqs.~\eqref{bndy}
and \eqref{bndy2}].
Since we have
\bena
&&\|\L^*(X)\|^2_{\mathcal K} = \langle X | Y \rangle
= \left\langle X \ \bigg| \  \L \left(
\begin{matrix}
\delta h_0\\
\delta p_0
\end{matrix}
\right) \right\rangle = \non\\
&&\left\langle \L^* (X) \bigg| \left(
\begin{matrix}
\delta h_0\\
\delta p_0
\end{matrix}
\right) \right\rangle_{\mathcal K} \leq
\| \L^*(X) \|_{\mathcal K} \left\| \left(
\begin{matrix}
\delta h_0\\
\delta p_0
\end{matrix}
\right) \right\|_{\mathcal K}
\eena
(where eq.~\eqref{LaxMilgr} of Appendix B was used in the first equality)
it follows that $\Pi_0$ is bounded and, hence, its action
can be extended to $\mathcal K$. We now show that
$\Pi_0^{} = \Pi^*_0$ and $\Pi^2_0 = \Pi_0^{}$,
so $\Pi_0$ is an orthogonal projection.
To show that $\Pi^*_0 = \Pi_0^{}$, we let $\Psi_0:=(\delta h_0, \delta p_0)$
and $\Psi_0':=(\delta h_0', \delta p_0')$ be smooth and of compact support. Then
\bena\label{pidef1}
&&\left\langle \Pi_0^{} \Psi_0^{} |
\Psi_0'
\right\rangle_{\mathcal K}
= \left\langle \Psi_0^{} |
\Psi_0'
\right\rangle_{\mathcal K} -
\left\langle \L^*(X) |
\Psi_0 \right\rangle_{\mathcal K} \non\\
&=&\left\langle \Psi_0^{} |
\Psi_0'
\right\rangle_{\mathcal K}
- \langle X | Y' \rangle_{\mathcal K} = \left\langle \Psi_0^{} |
\Psi_0'
\right\rangle_{\mathcal K}
- \langle \L^*(X) | \L^*(X') \rangle_{\mathcal K}
\eena
where $X,X',Y,Y'$ are defined as above and,
in the last step, we again used~\eqref{LaxMilgr}.
The expression on the right is manifestly symmetric, thus
proving the claim. Similarly, we have
\bena
\left\langle \Pi_0^{} \Psi_0^{}|
\Pi_0^{} \Psi_0'
\right\rangle_{\mathcal K}
&=& \left\langle \Psi_0^{}|
\Pi_0^{} \Psi_0'
\right\rangle_{\mathcal K} -
\left\langle \L^*(X) \ |
\Pi_0^{} \Psi_0' \right\rangle_{\mathcal K} \nonumber \\
&=& \left\langle \Psi_0^{} |
\Pi_0^{} \Psi_0'
\right\rangle_{\mathcal K}
\eena
thus showing that $\Pi^2_0 = \Pi_0^{}$.

Let $\V_0$ denote the closure in $\mathcal K$
of the image of $C^\infty_0 \oplus C^\infty_0$
under $\Pi_0$. We now show that $\V \subset \V_0$.
Suppose that
$(\delta h, \delta p) \in \V$ is orthogonal to $\Pi_0(C^\infty_0
\oplus C^\infty_0)$. Then inserting~\eqref{pidef}, we see that
\ben
0=\left\langle \left(
\begin{matrix}
\delta h\\
\delta p
\end{matrix}
\right) \bigg| \left(
\begin{matrix}
\delta h_0\\
\delta p_0
\end{matrix}
\right) \right\rangle_{\mathcal K} -
\left\langle
\left(
\begin{matrix}
\delta h\\
\delta p
\end{matrix}
\right) \Bigg| \ \L^*(X) \right\rangle_{\mathcal K} \, .
\een
The second term
vanishes, because $X \in \rho^2 W^2_\rho \oplus \rho W^1_\rho$, in
which $C^\infty_0(\overline \Sigma) \oplus C^\infty_0(\overline
\Sigma)$ are dense. Then, from the boundary conditions of $X$ (see
lemma~\ref{lemma3}), $X$ can be approximated by $X_n \in \W$ such
that $\L^* X_n \to \L^* X$ in $L^2 \oplus L^2$. Hence, since $(\delta
h, \delta p)$ in $\W^{\perp_S}$, it follows that $\langle (\delta h,
\delta p) | \ \L^*(X) \rangle_{\mathcal K} =0$.  Thus, we see that $0=\langle
(\delta h, \delta p) | (\delta h_0, \delta p_0) \rangle_{\mathcal K} $ for any
compactly supported $(\delta h_0, \delta p_0)$, which clearly means
that $(\delta h, \delta p) = 0$. Thus, the closure in $\mathcal K$ of
$\Pi_0(C^\infty_0 \oplus C^\infty_0)$ contains $\V$.

On the other hand, since all elements $(\delta h, \delta p) \in
\Pi_0(C^\infty_0 \oplus C^\infty_0)$ are in $\U$ (and, hence, are
smooth), satisfy the linearized constraints, and satisfy $\delta
\vartheta |_B = 0 = \delta \epsilon |_B$, it follows from property (2)
of this proposition that $(\delta h, \delta p) \in \V$ if and only if
$\delta M = \delta J_A = \delta P_i = 0$. Therefore, to obtain
the desired projection map we need only compose $\Pi_0$ with a finite
co-rank projector $\Pi_1$ defined as follows: Let $\xi_I^a$, $I=1,\dots,k$
denote a finite collection of smooth vector fields that vanish in
a neighborhood of $B$ and coincide, respectively, with
$t^a$, $\psi_A{}^a$, and $(\partial/\partial x^i)^a$ in a neighborhood
of infinity. Let $\Psi_I:=(\delta_I h, \delta_I p)$, $i=1,\dots,k$ be an
(arbitrarily chosen) collection of perturbations in
$\Pi_0(C^\infty_0 \oplus C^\infty_0)$ for which
$\delta H_{\xi_I} (\Psi_J) = \delta_{IJ}$. The desired
projector $\Pi_1$ is defined by
\ben
\Pi_1\left(
\begin{matrix}
\delta h\\
\delta p
\end{matrix}
\right) =
\left(
\begin{matrix}
\delta h\\
\delta p
\end{matrix}
\right) - \sum_{I=1}^k
|\Psi_I \rangle \
\left\langle
\left(
\begin{matrix}
\delta h\\
\delta p
\end{matrix}
\right)
 \Bigg| \L^*(\xi_I) \right\rangle_{\mathcal K} \
\een
and the desired projector $\Pi$ onto $\V$ is then defined by
$\Pi = \Pi_1\Pi_0$. By our arguments above, the image of
$\Pi (C^\infty_0 \oplus C^\infty_0)$
under $\Pi$ is contained in $\U$ and is dense in $\V$
thus proving that $\U \cap \V$ is dense in $\V$.
\qed

\bigskip

\noindent
{\bf Remark 1:} It is possible to strengthen
property (3) using gluing techniques
to show that even $(C^\infty_0(\overline \Sigma) \oplus C^\infty_0(\overline \Sigma)) \cap \V$ is dense in $\V$, where by definition $C^\infty_0(\overline \Sigma) \oplus C^\infty_0(\overline \Sigma)\subset \U$
consists of all smooth initial data that vanishes in a
neighborhood of spatial infinity. A statement and proof of this strengthened
result is given in Appendix C.

\bigskip

\noindent
{\bf Remark 2:} All elements of $\W$ are smooth solutions to the
linearized constraints
and satisfy $\delta M = \delta J_A =
\delta P_i = 0$ and
$\delta \epsilon |_B = \delta\vartheta |_B = 0$. Thus, by property (2)
of the above Proposition, we have $\W \subset \V$ and, hence,
$\overline{\W} \subset \V$. On the other hand, since $\overline{\W}$
and $\V$ are symplectic complements of each other in $\mathcal K$, it follows
immediately that an element $\gamma = (\delta h, \delta p) \in \V$
is such that $W_\Sigma (g; \gamma', \gamma) = 0$ for all $\gamma' \in \V$
if and only if $\gamma \in \overline{\W}$. Using the
inequality~\eqref{poincare}, it can be shown that smooth elements
of $\overline{\W}$ must lie in $\W$. Thus, for any $\gamma \in
\U \cap \V$, we have $W_\Sigma (g; \gamma', \gamma) = 0$
for all $\gamma' \in \V$ if and only if $\gamma \in \W$.

\bigskip

We now restrict consideration to axisymmetric perturbations,
$\pounds_{\psi_A} \gamma = 0$ (see \eqref{axisym}). We can redefine
the spaces $\mathcal K$, $\W$, and $\V$ with the word
``axisymmetric'' suitably inserted, and all of the results of this
subsection and their proofs continue to hold without modification.
In order not to make our notation more cumbersome than necessary, we shall
continue to use $\mathcal K$, $\W$, and $\V$ to denote the axisymmetric
versions of these spaces. Thus, even when not stated explicitly,
axisymmetry should be understood in all statements below.

We now view the canonical energy $\E$, defined in section 2, as a quadratic
form on the (axisymmetric)
Hilbert space $\V$, with dense domain ${\mathcal T} \equiv \U \cap \V$. We are
interested in finding
all of the elements $\gamma \in {\mathcal T}$ on which
$\E$ is degenerate, i.e., for which
\ben
\E(\gamma', \gamma) = 0 \quad \text{for all $\gamma' \in {\mathcal T}$}
\een
Since $\E(\gamma', \gamma) = W_\Sigma (g; \gamma', \pounds_t \gamma)$,
it follows
immediately from Remark 2 above that $\E$ is degenerate on
$\gamma \in {\mathcal T}$ if and only if $\pounds_t \gamma \in \W$.
But, in the presence of axisymmetry, this is precisely the
condition that $\gamma$ is a perturbation
towards a stationary black hole (see Definition 2.1 at the end of subsection
2.3). Let $\mathcal T'$ denote the space of equivalence
classes of elements of $\mathcal T$, where two elements are equivalent
if and only if they differ by a perturbation towards a stationary black hole.
We have proven the following: {\em The canonical energy $\E$ is well defined,
non-degenerate quadratic form on $\mathcal T' \times \mathcal T'$.}

\subsection{Stability and instability arguments}

For axisymmetric perturbations of a stationary black hole,
we have just shown that $\E$ is well
defined as a quadratic form on the space $\mathcal T'$ of
smooth solutions satisfying our gauge conditions
with $\delta M = \delta J_A = \delta P_i = 0$ modulo perturbations towards a stationary black hole. Since
$\E$ is
non-degenerate on $\mathcal T'$, only the
following two cases can occur:

\begin{itemize}

\item {\bf Case (a):} $\E$ is positive semi-definite on $\mathcal T$
and hence is positive definite on $\mathcal T'$.

\item {\bf Case (b):} There exists $\gamma \in {\mathcal T}$
such that $\E(\gamma, \gamma) < 0$.

\end{itemize}
We now analyze the behavior of perturbations in these two cases.

\medskip

\noindent {\bf Case (a):} $\E$ provides a positive definite conserved
norm on $\mathcal T'$. This precludes existence of ``growing modes''
in $\mathcal T'$. But elements of $\mathcal T$ can be expressed as the
sum of a representative element of $\mathcal T'$ and a perturbation
towards a stationary black hole. Since the latter perturbations are
manifestly stable, this shows stability for perturbations in $\mathcal
T$. However, by Proposition 5, $\mathcal T$ is dense in $\mathcal V$.
Now, a general axisymmetric solution to the
constraints in $\mathcal K$
can be written as a sum of an element of ${\mathcal V}$ and a finite linear
combination of representative
``change of mass,''
``change of angular momentum,'' and ``change of linear momentum''
perturbations. The desired ``change of linear momentum'' perturbations
can be chosen to be $\pounds_{Y_i} g$ where $Y_i$ is an asymptotic boost.
These are manifestly stable.
If the black hole is part of a family parametrized\footnote{Even if
the black hole is not
part of a family parameterized by $(M, J_A)$, it should be possible to
show existence of a basis of stable ``change of mass'' and
``change of angular momentum'' perturbations.} by $(M, J_A)$, then
the ``change of mass'' and
``change of angular momentum'' perturbations can be chosen to be
the perturbations towards other members of this family. These are
also manifestly stable. Thus, we conclude that the black hole
is stable to a dense set of axisymmetric perturbations in the subspace
of $\mathcal K$ comprised by solutions to the constraints.

Furthermore, the results of section 3 strongly suggest that
for perturbations $\gamma \in \mathcal T$, $\bE$
and $\E$ should decay to zero on the
slices ${\S} (t)$ as $t \rightarrow \infty$, and thus,
at late times, $\gamma$ should approach a perturbation
towards a stationary black hole.

\medskip

\noindent {\bf Case (b):} Let $\gamma \in {\mathcal T}$ be such that
$\E(\gamma, \gamma) < 0$. We obtain a contradiction with the
possibility that $\gamma$ approaches a perturbation, $\gamma_0$,
towards a stationary black hole on ${\S}(t)$ as $t \rightarrow
\infty$ as follows. If $\gamma$ approached a stationary
solution, then the Bondi news and the shear of the horizon should
approach zero at asymptotically late times, in which case as $t
\rightarrow \infty$ we should have $\bE(\gamma, {\S}(t))
\rightarrow \E(\gamma, {\S}(t)) \rightarrow \E(\gamma_0) =
0$. However, this is a contradiction because for all $t$ we have
$\bE(\gamma, {\S}(t)) \leq \bE(\gamma, \Sigma) = \E(\gamma, \Sigma) < 0$.

Although this contradiction does not prove instability, the fact that
$|\bE(\gamma, {\S}(t))|$ can only increase with time suggests
that the amplitude of the perturbation does not decrease with time. But then
the fluxes through
the horizon $\H^+$, and
null-infinity $\I^+$, should also not decrease, causing
$|\bE(\gamma, {\S}(t))|$ to increase further, etc.
It therefore seems highly plausible that
$\bE(\gamma, {\S}(t)) \rightarrow -\infty$ as $t \rightarrow \infty$
and that the
amplitude of the perturbation grows without bound, as has  been previously argued in
a different context in~\cite{friedman1, friedman, schutzfried,  friedman2}.

\bigskip

Thus, we conclude that in Case (a) the black hole is stable, whereas in
Case (b), the black hole is unstable. In other words,
we have argued that {\em the necessary
and sufficient condition for black hole stability is that the canonical energy
$\E$ be positive
semi-definite on the space ${\mathcal T} = \U \cap \V$ of smooth solutions
with $\delta M = \delta J_A =
\delta P_i = 0$ that
satisfy our horizon boundary conditions $\delta \vartheta|_B  = \delta
\epsilon|_B = 0$.}

\bigskip

We now briefly indicate some possible ways of
improving the above arguments. To prove
stability in Case (a), we would like to establish boundedness of $\gamma$ itself
rather than have boundedness/conservation of a norm on $\gamma$. However,
this is clearly not possible until we have imposed suitable gauge conditions on $\gamma$
that uniquely fix its evolution. Indeed, our present gauge conditions restrict $\gamma$ only near
infinity and near $\H^+$, so they allow pure gauge perturbations for which $\gamma$
becomes arbitrarily large. If suitable gauge conditions are imposed
on $\gamma$, then it is not
implausible that $\E$ would provide a conserved norm on $\gamma \in {\mathcal T}$
that would be equivalent
to a Sobolev norm. The canonical energy of the perturbation $(\pounds_t)^k \gamma$ would
similarly provide a family of conserved norms on $\gamma \in {\mathcal T}$ that plausibly would
be equivalent to higher Sobolev norms. If so, boundedness of $\gamma$ itself would
be proven.

It does not seem feasible to try to directly convert the above argument for instability
in Case (b) into a mathematically rigorous proof. However, it is possible that a proof of
instability in Case (b) [as well as stability in Case (a)] could be obtained along the
following lines. Again, to get started, it should be necessary to suitably fix a gauge so as to obtain
deterministic dynamics. Consider, first, the case of perturbations off of a static black hole.
On account of the time reflection symmetry of the background, $\E$ will be invariant under
time reflections, so the integral expression for $\E$ in terms of initial data will not contain any
``cross-terms'' between $\delta p$ and $\delta h$. Thus, $\E$ can be written as a sum
of a ``kinetic energy'' (quadratic in $\delta p$) and a ``potential energy'' (quadratic in $\delta h$).
It is likely that the kinetic energy is always positive-definite. For the case of spherically symmetric
perturbations (with matter fields) off of a static, spherically symmetric background, it was shown in~\cite{seifert}
that the kinetic energy could be used to define
an inner product such that the potential energy could be expressed as the expectation value
of a self-adjoint operator $A$ that appears in the dynamical evolution equations.
In this case, it can be seen directly from the dynamical evolution
equations that positivity of $A$ is equivalent to stability. However, since the kinetic energy is positive,
positivity of $A$ is equivalent to positivity of $\E$, thus rigorously establishing the equivalence of dynamical stability and positivity of $\E$. It is not inconceivable that similar results
could be proven for perturbations of an arbitrary static black hole, although, obviously, a number of technical
details would have to be sorted out. A similar strategy could also be applied to the
stationary-axisymmetric case---making use of the $t-\psi_A$ reflection isometry to decompose
initial data into its ``time symmetric" and ``time antisymmetric" parts---but it would be less
obvious in this case that the ``kinetic energy'' would have to be positive.

\section{Proof of the Gubser-Mitra conjecture}

Up to this point, we have restricted our considerations to the
stability of black holes. In this section, we extend our
considerations to black branes of the form \eqref{ds}.
By making the same type of dynamical instability argument as given
in subsection 4.2 above, we will show
that if a family of black holes is thermodynamically unstable, then
the corresponding family of black branes must be dynamically unstable.

To begin, suppose we have a family of stationary, axisymmetric black holes,
$g_{ab}(M,J_A)$, in $D \geq 4$ dimensions,
that satisfy the asymptotic and gauge conditions of subsection 2.1.
Let $\vec \xi= (M_0,J_{0A})$ denote the
parameter values of a particular
black hole in this family, and consider the
one-parameter subfamily
\ben
g_{ab}(\lambda) := g_{ab}(\vec \xi + \lambda \vec v) \ ,
\een
where $\vec v$ is an arbitrary vector in the parameter
space $(M, J_A)$. It is obvious that for this family, we have
\ben\label{2}
\frac{d^2 M}{d\lambda^2} (0)  = \frac{d^2 J_A}{d\lambda^2} (0) = 0 \ .
\een
Thus, by \eqref{variation2}, for the perturbation determined by
$\vec v$ we have
\ben
\E = - \frac{\kappa}{8 \pi} \frac{d^2 A}{d \lambda^2} (0)
\een
On the other hand, we have
\ben\label{1}
\frac{d^2 A}{d\lambda^2} (0) = {\rm Hess}_A |_{\vec \xi}(\vec v, \vec v)  \ ,
\een
where ${\rm Hess}_A|_{\vec \xi}$ denotes the Hessian of
$A$ [see eq.(\ref{Hess})] at parameter value $\vec \xi$. Thus, we see that one can find a perturbation to
a black hole in the family that makes $\E$ negative if and only if
the Hessian ${\rm Hess}_A$ has a positive eigenvalue.

However, a nontrivial perturbation to a black hole in the family clearly does not
have $\delta M = \delta J_A = 0$, so a negative value of $\E$---or,
equivalently, a positive eigenvalue of ${\rm Hess}_A$---does not provide
any information about stability. However, consider now the
$(D+p)$-dimensional spacetime $\tilde \M = \M \times {\mathbb T}^p$ with
metric $\tilde g_{ab}$ defined as above in eq.~\eqref{ds} with each
$z_i$ a $2\pi l$-periodic coordinate parameterizing the corresponding
``extra-dimension''. (Here and in the following, a tilda
dentotes a quantity associated with the $(D+p)$-dimensional
spacetime $\tilde \M$.)  This metric represents a
``uniform black brane'', where uniform refers to the fact that
each $(\partial/\partial z_i)^a$ is a Killing field of this spacetime,
and where brane refers to the fact that the horizon cross
section is now $\tilde B = B \times {\mathbb T}^p$. A black brane spacetime
is ``asymptotically Kaluza-Klein (KK)'' rather than asymptotically flat,
but the ADM conserved quantities $(\tilde{M}, \tilde{\bf P},
\tilde{\bf J}, \tilde{\bf C})$ can be defined as
in the asymptotically flat case (see eq.(\ref{HX})). In addition, we have
conserved quantities (``KK charges''), $\tilde{\bf T}$, associated with
the asymptotic symmetries $(\partial/\partial z_i)^a$, given by
\ben
\delta \tilde T_i := \int_{\infty} [\delta Q_{\partial/\partial z^i}(\tilde g) - i_{\partial/\partial z^i} \theta(\tilde g, \delta \tilde g)] \ .
\een

We now prove the following proposition:

\begin{prop} Let $g_{ab}(M,J_A)$ be a family of black holes that is
  thermodynamically unstable at $(M_0,J_{0A})$, i.e., there exists a
  perturbation $\vec v$ within the black hole family for which $\E <
  0$, which will be the case if ${\rm Hess}_A$ has a positive eigenvalue
    at $(M_0,J_{0A})$. Then, for any black brane corresponding to $g_{ab}(M_0,J_{0A})$
  via eq.(\ref{ds}) with sufficiently large $l$, one can find a perturbation for which
  $\tilde \E < 0$ and $\delta {\tilde M} = \delta {\tilde J}_A =
  \delta {\tilde P}_i =
  \tilde T_i = 0$, $\delta \tilde \vartheta |_{\tilde B} = \delta \tilde \epsilon |_{\tilde B} =0$.
\end{prop}

\medskip
\noindent
{\em Proof:} For notational simplicity, we prove the proposition for
$p=1$ extra dimension, the general case is completely analogous.
Let $(\delta p^{ab}, \delta h_{ab})$ denote the initial data of a
perturbation $\vec v$ in the black hole family for which $\E
< 0$. Without loss of generality, we may assume that our Cauchy
surface $\Sigma$ is chosen to be maximal in the background and
perturbed spacetime\footnote{Existence of a maximal slice in the background
spacetime is shown in \cite{chrwal}. For the perturbed spacetime, existence
of a slice that preserves maximality follows from the fact that the perturbed
lapse required to achieve this condition
satisfies a suitable elliptic equation.},
so that $p^{ab} h_{ab} = 0$,
and $\delta p^{ab}h_{ab} + p^{ab} \delta h_{ab} = 0$.
We wish to construct initial data
$(\delta \tilde p^{ab}, \delta \tilde h_{ab})$ in the
$(D+1)$-dimensional black brane spacetime with $\tilde \E < 0$ for
which $\delta {\tilde M} = \delta {\tilde J}_A = \delta {\tilde P}_i =
\delta \tilde T = \delta \tilde A = 0$ and for which
$\delta \tilde \vartheta = 0$
at $\tilde B$.

Our ansatz for $(\delta \tilde p^{ab}, \delta
\tilde h_{ab})$ is
\ben
\label{firsttry}
\begin{split}
\delta \tilde p^{ab} = &\Big(
  \delta p^{ab} + \frac{1}{D-1} p^{ab} \phi + h^\half \{ D^{(a} X^{b)} + D^{(a} \zeta \fz^{b)}
  - \frac{1}{D}(h^{ab} + \fz^a \fz^b) D^c X_c \\
  &+ik/l \ \fz^{(a} X^{b)} + \frac{D-1}{D} \ ik/l \ \zeta \ \fz^{(a} \fz^{b)} - \frac{1}{D} \ ik/l \ \zeta h^{ab} \}
   \Big)\ \e^{iz \cdot k/l}
\ , \\
\delta \tilde h_{ab} = &\Big( \delta h_{ab} - \frac{1}{D-1}
  h_{ab} \phi \Big) \ \e^{iz \cdot k/l}
\end{split}
\een
where $k \in \mathbb Z$ and $\phi,\zeta,X^a$
are tensor fields on $\Sigma$ (so they are independent of
$z$, and tangent to $\Sigma$). The terms in $\delta \tilde p^{ab}$ involving $\zeta,X^a$ may be
written alternatively as $\tilde D^{(a} \tilde X^{b)}-
\frac{1}{D} \tilde h^{ab} \tilde D^c \tilde X_c$, where
\ben
\tilde X^a = (\zeta \ \fz^a + X^a) \ \e^{ik \cdot z/l} \ .
\een
If $k \in \mathbb Z$ is not zero, then
it is easily seen that the surface integrals defining
$\delta \tilde M$, $\delta \tilde {J}_A, \delta \tilde P_i, \delta \tilde T$
and $\delta \tilde A$ vanish because the $z$-dependence is $\e^{iz \cdot k/l}$.

The tensor fields $\phi,\zeta,X^a$ in our ansatz~\eqref{firsttry} are now
chosen so that the linearized momentum and Hamiltonian constraints are satisfied. It can be seen that the satisfaction of the linearized momentum constraint is equivalent to
\ben\label{mom1}
\tilde D_a \Big(\tilde D^{(a} \tilde X^{b)}-\frac{1}{D} \tilde h^{ab} \tilde D^c \tilde X_c \Big) = \half ik/l \ h^{-\half} \delta h_{cd} p^{cd} \e^{ik \cdot z/l} \fz^b \ .
\een
This is an equation on $\tilde \Sigma$, but because all tensor fields have the same dependence $\e^{ik \cdot z/l}$ on $z$, we can rewrite it as the following system of equations on $\Sigma$:
\ben\label{mom}
\begin{split}
&D_a \Big( D^{(a} X^{b)}-\frac{1}{D} h^{ab} D^c X_c \Big) - \half \ k^2/l^2 \ X^b + \frac{D-2}{2D} \ ik/l \  D^b \zeta = 0 \ , \\
&D^a D_a \zeta - 2\frac{D+1}{D} \ k^2/l^2 \ \zeta + \frac{D-2}{D} \ ik/l \ D^a X_a =
ik/l \ h^{-\half}\delta h_{cd} p^{cd} \ .
\end{split}
\een
The satisfaction of the Hamiltonian constraint is equivalent to the
linearized Lichnerowicz equation for $\phi$,
\ben\label{York}
\begin{split}
&\left(-D^a D_a + h^{-1} p_{ab} p^{ab} +   k^2/l^2 \ \frac{(D-1)}{(D-2)}\right) \ \phi = \\
&=  \frac{(D-1)}{(D-2)} \ k^2/l^2 \ \delta h_a{}^a - 2 \frac{(D-1)}{(D-2)} \ p^{ab} D_a X_b h^{-\half} \ .
\end{split}
\een
Together, the eqs.~\eqref{York},~\eqref{mom} form a system of linear, inhomogeneous, elliptic\footnote{Indeed, the principal symbol of the equation for $X^a$ in~\eqref{mom}
is $\sigma(\xi)^a{}_b = h^{cd} \xi_d \xi_c \ \delta^a{}_b - (1/D) \ h^{ac} \xi_c \xi_b$, which is an invertible linear map in each tangent space $T\Sigma$ for any $T^* \Sigma \owns \xi_a \neq 0$.} PDE's for the unknown tensor fields $\phi,\zeta,X^a$ on $\Sigma$. We need to impose boundary conditions at $B$. To find these, we consider the linearized expansion $\delta \tilde \vartheta$ on $\tilde B$ in the black brane spacetime. We would like this to vanish. It is calculated using eq.~\eqref{bndy} as
\ben
\begin{split}
\delta \tilde \vartheta =&\Big(
-\frac{3}{2} \ \phi \ h^{-\half} p_{ab}(h^{ab} - \eta^a \eta^b)
 -\frac{D-2}{2} \ \eta^a D_a \phi + \half \ \phi \pounds_\eta h_{ab} \ (h^{ab} - \eta^a \eta^b)\\
&-\eta^a \eta^b D_a X_b
+\frac{1}{D} \ D^c X_c + \frac{1}{D} \ ik/l \ \zeta
\Big) \ \e^{ik \cdot z/l} \\
=& \Big(-\frac{D-2}{2} \ \eta^a D_a \phi + 2 \ H \phi -\eta^a \eta^b D_a X_b
+\frac{1}{D} \ D^c X_c + \frac{1}{D} \  ik/l \ \zeta \Big)\ \e^{ik \cdot z/l} \\
=& \Big(-\frac{D-2}{2} \ \pounds_\eta \phi + 2 \ H\phi
 -\frac{D-1}{D} \ \pounds_\eta (\eta^a X_a) + \\
 & \ \ \ \frac{1}{D} \ H (\eta^a X_a)
 + \frac{1}{D} \ {\mathcal D}_a [(h^{ab} - \eta^a \eta^b) X_b]
 + \frac{1}{D} \  ik/l \ \zeta \Big)\ \e^{ik \cdot z/l}
  \quad \text{at $\tilde B$,}
\end{split}
\een
where $\eta^a$ is the normal to $B$ within $\Sigma$, ${\mathcal D}_a$ is the
derivative operator intrinsic to $B$, and $H$ is the
trace of the extrinsic curvature of $B$ within $\Sigma$.
In the first line, use has been made of the fact that $p_a{}^a = 0$ and $\delta \vartheta =0$ on $B$ in the original spacetime. In the
second line, we used that $0=\vartheta =(h^{ab}-\eta^a \eta^b)(
h^{-\half}p_{ab} + \half \pounds_\eta h_{ab})$ on $B$
in the original spacetime. In the last step we have, without loss of generality,
extended $\eta^a$ of $B$ such that it is geodesic. Because $B$ is the bifurcation surface
 of a Killing horizon, $H=0$ on $B$. In view of this,
 $\delta \tilde \vartheta=0$ on ${\tilde B}$ will follow
if we impose on $\phi,\zeta,X^a$ the boundary conditions
\ben
\begin{split}
\pounds_\eta \phi |_B &= 0 \ , \\
(h^{ab}-\eta^a \eta^b) X_b |_B &= 0 \ , \\
\pounds_\eta (X^b \eta_b) |_B &=0 \ , \\
\zeta |_B &= 0 \ .
\end{split}
\een
In other words, we impose Neumann conditions on $\phi$, Dirichlet conditions
on $\zeta$, Neumann conditions on the orthogonal components $\eta^c X_c$ of $X^a$, and
Dirichlet conditions on the tangent components $(h^{ab} - \eta^a \eta^b) X_b$ of $X^a$ at $B$.
We need to analyze the
existence and properties of the solutions to~\eqref{York},~\eqref{mom}.
The existence of a solution $(\zeta,X^a) \in \rho W^1_\rho(\Sigma)$ (and in fact
$\in \rho W^k_\rho$ for any $k$) follows from
lemma~\ref{lemma4} in appendix~B, because~\eqref{mom} is equivalent to~\eqref{mom1},
and the lemma gives a smooth solution $\tilde X^a \in \rho W^1_\rho(\tilde \Sigma)$ to the latter equation. Next, we analyze the existence and properties of
the solutions to~\eqref{York}. We need to distinguish the cases $D=4,5$ and $D \ge 6$.

\medskip
\noindent
\underline{$D \ge 6$:} In this case, the right side of \eqref{York} is in $L^2$. Existence of a smooth solution $\phi$ in the Sobolev space $\rho W^1_\rho$ (and in fact $\in
\rho W^k_\rho$ any $k$) satisfying Neumann boundary conditions can be proven as usual by considering the weak formulation of the boundary value problem
\ben\label{weak}
\begin{split}
&\int_\Sigma \Big( D^a \phi D_a \psi + h^{-1} p^{ab} p_{ab} \phi \psi + \frac{(D-1)}{(D-2)} \ k^2/l^2 \ \phi \psi \Big) h^\half \\
= & \frac{(D-1)}{(D-2)}
\int_\Sigma \Big(  \ k^2/l^2 \ h^\half \delta h_a{}^a \psi
- 2  \ p^{ab} D_a X_b \psi
\Big)
\end{split}
\een
for all smooth $\psi$ with compact support in $\overline \Sigma$, and making use of the Poincare-type inequality
\ben\label{poincare1}
c \int_\Sigma \rho^{-2} \psi^2 \ h^\half \le \int_\Sigma D^a \psi D_a \psi \ h^\half \ ,
\een
for some $c>0$.
Since these arguments are standard and very similar to those in
the proof of lemma~\ref{lemma3} in appendix~B, we do not repeat them here.

It remains to be shown that the normalized canonical energy $\tilde \E
= 1/(2\pi l) \ \tilde W_{\tilde \Sigma}(\delta \tilde g, \pounds_t
\delta \tilde g)$ on the $(D+1)$-dimensional black brane spacetime
approaches the canonical energy $\E<0$ of the original black hole perturbation
for $l \rightarrow \infty$. For this, it is important to understand how the
norms of $\phi_l, \zeta_l, X_l{}^a$ behave for large $l$ (here we put an ``$l$'' on the
tensor fields to indicate their dependence on $l$).  We use that the source term in~\eqref{mom1} is of
order $O(\rho^{D-1})$, so $\rho$ times the source is in $W^1_\rho(\tilde \Sigma)$
in $D \ge 6$ and tends to zero in that norm\footnote{In $D=4,5$ this is still true if we impose a gauge, as below,
that $\delta h_{ab}-\frac{1}{D-1}h_{ab}\delta h$ is of order $O(\rho^{-(D-2)})$,
because $p_a{}^a=0$.}. From lemma~\ref{lemma4} of appendix~B
we infer that the $L^2(\tilde \Sigma)$-norms of $\rho^{-1} \tilde X_l{}^a, \rho^{-1} \tilde D_b \tilde X_l{}^a, \rho \tilde D_a \tilde D_b \tilde X_l{}^c$ tend to zero of order $O(1/l)$ as $l \to \infty$. This translates into the
statement that the $L^2(\Sigma)$ norms of $\rho^{-1} \zeta_l, \rho^{-1} X_l{}^a, D_b X_l{}^a, D_a \zeta_l, \rho/l \ D_a \zeta_l, \rho/l \ D_a X_l{}^b$, $1/l \ \zeta_l, 1/l \ X^a_l$ as well as $\rho D_a D_b \zeta_l, \rho D_a D_b X_l{}^c$ tend to zero of order $O(1/l)$ as $l \to \infty$. It follows
immediately from~\eqref{York} that the $L^2$-norm of
$1/l \ \phi_l$ remains bounded as $l \to \infty$. Then,
combining~\eqref{weak} with~\eqref{poincare1} for $\phi=\psi=\phi_l$,
we find that $\rho^{-1} \phi_l, D_a \phi_l \to 0$ in the sense of
$L^2$ as $l \to \infty$. By standard bootstrap and scaling arguments for
elliptic PDE's of the nature under consideration (see e.g.~\cite{cd}
Appendix~A for a discussion and references), it follows similarly $\rho D_a D_b \phi_l \to 0$ in $L^2$.

Using now
the conditions $\delta p^{ab} h_{ab} + p^{ab} \delta h_{ab}=0$,
$p^c{}_c = 0$ (from which follow the same for the corresponding
tilde tensor fields on $\tilde \Sigma$), the boundary conditions at $B$, and
eliminating terms that are total derivatives with respect to $z$,
we obtain\footnote{Note that $\E$ below is extended to complex-valued
perturbations in such a way that it is anti-linear in the first argument; of
course we could equally work with real valued perturbations by taking
the real and imaginary parts of eqs.~\eqref{firsttry}.} from \eqref{Eexpr}
\ben\label{edecomp}
\begin{split}
& 16\pi \ \tilde \E - 16 \pi \ \E=\\
&{\rm Re}
\int_\Sigma N \Big\{
\half R_{ab}(h) \delta_1 h_c{}^{c*} \delta_2 h^{ab} + \half R_{ab}(h) \delta_1 h^{ab*} \delta_2 h_c{}^c
-2 R_{ac}(h) \delta_1 h^{ab*} \delta_2 h^c{}_b - \\
& \hspace{1cm} \frac{1}{4} \delta_1 h^{ac*} D_a D_c \delta_2 h_b{}^b -
\frac{1}{4} D_a D_c \delta_1 h_b{}^{b*} \delta_2 h^{ac} - \frac{1}{4} \delta_1 h^{ac*} D^b D_b \delta_2 h_{ac} - \frac{1}{4} \delta_2 h^{ac*} D^b D_b \delta_1 h_{ac} + \\
& \hspace{1cm}  \half \delta_1 h^{ac*} D_b D_a \delta_2 h_c{}^b + \half D_b D_a \delta_1
h_c{}^{b*} \delta_2 h^{ac} - \frac{3}{4} D_a D^a (\delta_1 h^{bc*} \delta_2 h_{bc}) -\\
& \hspace{1cm} \frac{3}{4} D_a(\delta_1 h^{ab*} D_b \delta_2 h_c{}^c +
\delta_2 h^{ab} D_b \delta_1 h_c{}^{c*}) + \frac{1}{4} D^a D_a(\delta_1 h_b{}^{b*} \delta_2 h_c{}^c) +\\
& \hspace{1cm} D_a(\delta_1 h_c{}^{a*} D_b \delta_2 h^{cb} + \delta_2 h^a{}_c D_b \delta_1
h^{cb*})
+ \half D_a(\delta_1 h_c{}^{b*} D_b \delta_2 h^{ac} + \delta_2 h_c{}^b D_b \delta_1 h^{ac*})-\\
& \hspace{1cm} \frac{1}{4} D^a (\delta_1 h_d{}^{d*} D^b \delta_2 h_{ab} + \delta_2 h_d{}^d D^b
\delta_1 h_{ab}{}^*)
\Big\} \ h^\half + \\
& {\rm Re}
\int_\Sigma N \Big\{
\half p^{ab} p_{ab} \delta_1 h_a{}^{a*} \delta_2 h_b{}^b - \frac{3}{2}
p^c{}_b p^{ab} (\delta_1 h_d{}^{d*} \delta_2 h_{ac} + \delta_2 h_d{}^d \delta_1 h_{ac}{}^*)+\\
& \hspace{1cm} 2 \delta_1 p^{ab*} \delta_2 p_{ab} + 2 p^{ac} p^{bd} \delta_1 h_{ab}{}^* \delta_2 h_{cd} - \frac{3}{2} \delta_1 h_d^{d*} \delta_2 p^{ab} p_{ab} -\\
& \hspace{1cm} \frac{3}{2} \delta_2 h_d{}^d \delta_1 p^{ab*} p_{ab}
+ 4 p^c{}_b (\delta_1 h_{ac}{}^* \delta_2 p^{ab} + \delta_2 h_{ac} \delta_1 p^{ab*})
\Big\} \ h^{-\half} -\\
& {\rm Re}
\int_\Sigma N^a \Big\{
-\delta_1 p^{bc*} D_a \delta_2 h_{bc} - \delta_2 p^{bc} D_a \delta_1 h_{bc}{}^*
+ 2\delta_1 p^{cb*} D_b \delta_2 h_{ac} + \\
&\hspace{1cm} 2\delta_2 p^{cb} D_b \delta_1 h_{ac}{}^*
+ \delta_1 h_{ac}{}^* D_b \delta_2 p^{bc} + \delta_2 h_{ac} D_b \delta_1 p^{cb*}-\\
&\hspace{1cm} p^{cb}(\delta_1 h_{ac}{}^* D_b \delta_2 h_c{}^d +
\delta_2 h_{ad} D_b \delta_1 h_c{}^{d*})
+ \half p^{cb}(
\delta_1 h_{ad}{}^* D^d \delta_2 h_{cb} + \delta_2 h_{ad} D_b \delta_1 h_c{}^{d*}
)
\Big\}-\\
&{\rm Re} \int_B \mu^\half \ \frac{D-2}{D-1} \Big( 2\delta h^{ab} - \frac{1}{D-1} \phi h^{ab}\Big)^*  \phi (h_{ab}-\eta_a \eta_b)
+\\
&{\rm Re}
\int_\Sigma N \Big\{
(D_a \zeta)^* D^a \zeta + \frac{2}{D^2} (D_a X^a)^* D_b X^b + \frac{2(D-1)^2}{D^2} \ k^2/l^2 \
\zeta^* \zeta + k^2/l^2 \ X_a^* X^a +\\
& \hspace{1cm} \half \ k^2/l^2 \ \Big(\delta h_{ab} - \frac{1}{D-1} \phi h_{ab}\Big)^*
\Big(\delta h^{ab} -
\frac{1}{D-1} \phi h^{ab}\Big)
\Big\} \ h^\half + \\
&{\rm Re} \int_\Sigma N^a \Big\{
ik/l \ (\delta h_{ac}-\frac{1}{D-1} \phi h_{ac}) (D^c \zeta +  ik/l \ X^c)^*
 \Big\} \ h^\half \ .
\end{split}
\een
where, in this equation, $*$ denotes complex conjugation. Here we have
defined
\ben
\begin{split}
\delta_1 h_{ab} &= -\frac{1}{D-1} \phi h_{ab}\\
\delta_2 h_{ab} &= 2\delta h_{ab} - \frac{1}{D-1} \phi h_{ab}\\
\delta_1 p^{ab} & = \frac{1}{D-1} p^{ab} \phi + h^\half \{ D^{(a} X^{b)} - \frac{1}{D}(
ik/l \ \zeta + D^c X_c) \}\\
\delta_2 p^{ab} & = 2\delta p^{ab} + \frac{1}{D-1} p^{ab} \phi + h^\half \{ D^{(a} X^{b)} - \frac{1}{D}(
ik/l \ \zeta + D^c X_c) \} \ .
\end{split}
\een
From this expression, together with the decay of the norms of the
   tensor fields $\zeta_l, \phi_l, X_l{}^a$ which we have described, it follows\footnote{It
   is essential here to use the conditions
 $p_{ab} = O(\rho^{-(D-2)}), R_{ab}(h) = O(\rho^{-(D-1)})$ as well as $\delta h_{ab} = O(\rho^{-(D-3)}), \delta p^{ab} = O(\rho^{-(D-2)})$. One can see that $\phi_l,\zeta_l,X_l^a$
 and their derivatives only enter in the combinations such that their norm tends to zero in $L^2$, or such that the additional $1/l$-factors give convergence to 0.} that the terms
on the right side of \eqref{edecomp} go to zero
as $l \to \infty$. Thus, we have
${\tilde \E} \rightarrow \E < 0$ as $l \to \infty$,
as we desired to show.

\medskip
\noindent
\underline{$D=4,5$:} In this case, the right side of \eqref{York} need not be in $L^2$, and the previous
argument for obtaining a solution $\phi$ does not work directly.
But in this case, we can apply the following modification to the above argument.
We now write $\phi = \chi \delta h_a{}^a + v$, where $\chi$ is equal to $1$ near infinity, and equal to $0$ in an open neighborhood of $B$. Imposing \eqref{York} on $\phi$ gives an equation for $v$ of the same type, but with a source in $L^2$ (and with the boundary condition
$\pounds_\eta v = 0$ at $B$). Existence of a smooth solution $v\in W^1$ again follows, and therefore
existence of a solution $\phi$ follows, too.
Now let $l \to \infty$ and denote the corresponding sequence of solutions by $v_l$.
From the equation satisfied by $v_l$ it is easy to see that $D_a v_l$ is Cauchy in $L^2$,
and, by~\eqref{poincare1}, it follows that $\rho^{-1} v_l$ also is Cauchy. Hence $\rho^{-1} v_l, D_a v_l$ are
convergent sequences in $L^2$, with limit $\rho^{-1} v \in W^1$. But it is easy to see
that $v$ must be equal to $-\chi \delta h_a{}^a$ since they both satisfy the same boundary value problem,
so $\phi_l$ again satisfies $\rho^{-1} \phi_l, D_a \phi_l, \rho D_a D_b \phi_l \to 0$ in $L^2$. It then follows that all terms~\eqref{edecomp} again go to zero. The only terms for which the argument is
somewhat different than for the case of $D \ge 6$ are the last two involving
$\delta h_{ab} - \frac{1}{D-1} \phi_l h_{ab}$, because $\delta h_{ab}$ is not
in $L^2$ now. For that term,
we write $\delta h_{ab} - \frac{1}{D-1} \phi_l h_{ab}$ as $(\delta h_{ab} - \frac{1}{D-1} \delta h_c{}^c h_{ab})
+ \frac{1}{D-1}(\delta h_c{}^c - \phi_l) h_{ab}$. Although $\delta h_{ab}$ is not in $L^2$
in dimensions $D=4,5$, we may assume that we are in a gauge\footnote{For
example, for perturbations from a Schwarzschild to another Schwarzschild
black hole, putting the metric perturbation into ``isotropic coordinates''
near infinity would provide such a gauge, and similarly for perturbations
within the Kerr family.} such that
$\delta h_{ab} - \frac{1}{D-1} \delta h_c{}^c h_{ab}$ falls off faster by one power of $\rho^{-1}$,
implying that it is in $L^2$. Furthermore, we have $(-D^a D_a + h^{-1} p_{ab} p^{ab}) \phi_l=
\frac{(D-1)}{(D-2)} [k^2/l^2 \ (\phi_l-\delta h_a{}^a)-2p_{ab} D^b X^a_l]$. Because $\rho^{-1} \phi_l, D_a \phi_l, D_b X^a_l \to 0$ in $L^2$,
it then follows (multiply by $(\phi_l-\delta h_a{}^a)^*$ and integrate by parts)
that $l^{-1}(\phi_l- \delta h_a{}^a) \to 0$ in $L^2$. Thus, again $\tilde \E \to \E <0$.

\medskip

Since $\delta \tilde A = 0$, we can further change the initial data by a gauge transformation [i.e. adding $\sigma \tilde \L^*(\tilde Y)$ with
a suitable vector field $\tilde Y^a$ tangent to $\tilde B$, see eq.~\eqref{timeevolution0}] so that $\delta \tilde \epsilon |_{\tilde B} = 0$. For example, take $\tilde Y^a =
\frac{1}{D-1} \ (ik/l)^{-1} \ \phi \fz^a \ \e^{ik\cdot z/l}$ on $\tilde \Sigma$ near $\tilde B$, and vanishing in a neighborhood of infinity, and extend it off $\tilde \Sigma$ such that $\pounds_t \tilde Y^a = 0$. By proposition~3, this will not
change $\tilde \E$.
\qed

\medskip

Essentially all of the analysis of sections 2, 3, and 4 can be applied straightforwardly
to black branes. One important modification,
however, is that in the definition of ``perturbations toward stationary black branes,'' allowance must
be made for the possibility that the horizon Killing field $K^a(\lambda)$ may
correspond to a linear combination
of the Killing fields $(\partial/\partial z^i)^a$ in addition to $t^a$ and $\psi^a_A$. Thus, definition
2.1 must be modified by including  linear combinations of the Killing fields $(\partial/\partial z^i)^a$
in eq.(\ref{bkvf3}). Correspondingly, in Proposition 4,
the requirement that $\gamma_1$ also satisfy $\delta T_i = 0$
must be added to the hypothesis. In section 4, the definition of $\W$, eq.(\ref{W}), (and, hence, $\V$)
must be similarly
modified to allow $X^a$ to coincide near infinity
with a combination of the Killing fields $(\partial/\partial z^i)^a$
as well as rotational Killing fields and spatial translations.
Correspondingly, in statement (2) of Proposition 5, the condition $\delta T_i = 0$ must be added
to the conditions $\delta M = \delta J_A = \delta P_i = 0$. Aside from these changes, the
results of the previous sections apply.

We have just shown that for a black hole family such that the Hessian matrix ${\rm Hess}_A$ has a positive eigenvalue,
there exists a perturbation of the
corresponding black brane that satisfies
$\delta \tilde M = \delta \tilde {J}_A = \delta \tilde P_j = \delta \tilde T_i = 0$ together with
$\delta \tilde \vartheta |_{\tilde B} = \delta \tilde \epsilon |_{\tilde B} = 0$,
and, for sufficiently large $l$, has ${\tilde \E} < 0$. Since the flux analysis of section 3
for asymptotically flat spacetimes carries
over to the asymptotically Kaluza-Klein case, we may make the same arguments as given
for Case (b) of section 4.2 to conclude that the black brane must be unstable. Indeed
for $D \geq 6$ the perturbation we have constructed lies in the black brane analog of
$\V$, so we may make precisely the same arguments for instability. However, these arguments
can also be made for the case
$D=4,5$, since the black brane analog of Proposition 4 of section 2.3 yields a contradiction
with the perturbation approaching a perturbation towards a stationary black brane at late times.
Thus, we conclude that for any family of black holes that are thermodynamically unstable,
the corresponding family of black branes is
dynamically unstable to sufficiently long wavelength perturbations.

\section{Equivalence to the local Penrose inequality}

In this section, we show that for black holes, the satisfaction of the local Penrose inequality
is equivalent to the positivity of canonical energy for perturbations with
$\delta M = \delta J_A = \delta P_i = 0$ and hence, by our previous arguments, is equivalent
to dynamical stability.

Let $\bar g_{ab} (M,J_A)$ be a family of stationary, axisymmetric, and
asymptotically flat black hole metrics on $M$. Let $g_{ab}(\lambda)$
be a one-parameter family of axisymmetric metrics such that $g_{ab}(0)
= \bar g_{ab} (M_0,J_{0A})$. Without loss of generality, we may assume
(by applying an asymptotic Lorentz boost, if necessary) that the
linear momentum of $g_{ab}(\lambda)$ vanishes, $P_i(\lambda) =0$.  Let
$M(\lambda), J_A(\lambda)$ denote the mass and angular momenta of
$g_{ab}(\lambda)$. Let $\bar g_{ab} (\lambda) = \bar g_{ab} (M
(\lambda),J_A (\lambda))$ denote the one-parameter family of
stationary black holes with the same mass and angular momenta as
$g_{ab}(\lambda)$.  Let $\mathcal{A}(\lambda)$ denote the area of the
apparent horizon of $g_{ab}(\lambda)$ on an initial slice and let
$\bar{\mathcal{A}}(\lambda)$ denote the area of the event horizon of
$\bar{g}_{ab}(\lambda)$. Then the following proposition holds:

\begin{prop}
There exists a one-parameter family $g_{ab}(\lambda)$ for which
\ben
\mathcal{A}(\lambda) > \bar{\mathcal{A}}(\lambda)
\label{Ainequal}
\een
to second order in $\lambda$ if and only
if there exists a perturbation $\gamma'_{ab}$
of $\bar g_{ab} (M_0,J_{0A})$
with $\delta M = \delta J_A = \delta P_i = 0$ such that $\E(\gamma') < 0$.
\end{prop}

\noindent
{\em Proof:} On account of our gauge condition $\delta
\vartheta|_B = 0$, the apparent horizon of $g_{ab}(\lambda)$ coincides
with $B$ to first order in $\lambda$. To find the apparent horizon of
$g_{ab}(\lambda)$ to second order in $\lambda$, we would, in general, need
to displace $B$ by a second order gauge transformation similar to
\eqref{ch}. However, since the extrinsic curvature of $B$ vanishes for
the background metric $g_{ab}(0)$, the area will be unchanged
to second order under such
a displacement. Consequently, to second order in $\lambda$, we have
$\mathcal{A}(\lambda) = A(\lambda)$, where $A(\lambda)$ denotes the area
of $B$ in the metric $g_{ab}(\lambda)$.
Similarly, to second order in $\lambda$, we have
$\bar{\mathcal{A}}(\lambda) = \bar{A} (\lambda)$.

Obviously, we have $A (0) = \bar{A}(0)$,
as the two families coincide for $\lambda=0$.
Likewise, by the first law of black hole mechanics, we have
\ben
\frac{d A}{d\lambda} (0) = \frac{8 \pi}{\kappa} \left[\frac{d M}{d\lambda} (0) -
\sum \Omega_A \frac{d J_A}{d\lambda} (0) \right]
= \frac{d \bar {A}}{d\lambda} (0) \, .
\een
Thus, what matters is the second derivative of the areas.
Let $\gamma_{ab} = (dg_{ab}/d\lambda)|_{\lambda = 0}$
and let $\bar \gamma_{ab} = (d\bar{g}_{ab}/d\lambda)|_{\lambda = 0}$.
Since the families have the same mass and angular momenta,
our second variation formula
\eqref{variation2} gives
\bena
\frac{\kappa}{8 \pi} \left[\frac{d^2 A}{d\lambda^2} (0) -
\frac{d^2 \bar{A}}{d\lambda^2} (0)\right] &=&
\E(\bar \gamma, \bar \gamma) -  \E( \gamma, \gamma)   \non\\
&=& -\E(\gamma',\gamma') + 2 \E(\gamma',\bar \gamma)\ .
\eena
where we have defined $\gamma'_{ab} = \bar \gamma_{ab}-\gamma_{ab}$.
However, $\gamma'_{ab}$ is clearly a linearized perturbation with
$\delta M = \delta J_A = \delta P_i = 0$, and $\bar \gamma_{ab}$ is manifestly
a perturbation towards a stationary black hole. Therefore, the second term vanishes
by Proposition 4 of section 2.3. Thus, we have
\ben
\frac{\kappa}{8 \pi} \left[\frac{d^2 A}{d\lambda^2} (0) -
\frac{d^2 \bar{A}}{d\lambda^2} (0)\right]  = - \E (\gamma',\gamma')
\een
from which the proposition follows immediately. \qed

As explained in the introduction, the satisfaction of the inequality
\eqref{Ainequal} is incompatible with the stability of the family of
black holes $\bar g_{ab} (M,J_A)$. It is therefore reassuring that we
have found that satisfaction of this inequality is equivalent to the
condition that we have obtained in this paper for dynamical instability.

\bigskip

\noindent
{\bf Acknowledgments:}
The impetus for this work arose from a talk given by Harvey Reall at
the workshop ``Numerical Relativity Beyond Astrophysics'' (Edinburgh,
July, 2011) on the research in~\cite{reallfig} and discussions by one
of us (R.M.W.) with Reall following that talk.  We would also like to thank
P.~Chrusciel and E.~Delay for explanations
about gluing constructions. S.H. would like to thank M. Marletta for
discussions about boundary value problems. The research of R.M.W.
was supported in part by NSF grants PHY 08-54807 and PHY 12-02718
to the University of
Chicago. S.H. acknowledges financial support
through ERC grant QC~\&~C~259562 .

\bigskip
\begin{appendix}


\section{Proof of Lemma~\ref{lemma2}}

\noindent
{\bf Lemma 2.}
Let $\delta g_{ab}$ be a solution to the linearized Einstein equations
satisfying our asymptotic flatness conditions and our gauge conditions
\eqref{varB} and \eqref{deleps} at $B$. Suppose in addition that
$\delta A = 0$ (so that, by \eqref{deleps}, we have $\delta \epsilon|_B = 0$)
and that $\delta H_X = 0$ for some asymptotic symmetry $X^a$. Then
$W_\Sigma(g; \delta g, \pounds_\xi g) = 0$
for all smooth $\xi^a$ such that (i) $\xi^a|_B$ is tangent to the generators
of $\H^+$ and
(ii) $\xi^a$ approaches a multiple of $X^a$ as $\rho \rightarrow \infty$.
Conversely, if $\delta g_{ab}$ is smooth and asymptotically flat
and if $W_\Sigma(g; \delta g, \pounds_\xi g) = 0$ for all such $\xi^a$, then
$\delta g_{ab}$ is a solution to the linearized Einstein equation
with  $\delta \vartheta|_B = \delta \epsilon|_B = 0$ at $B$ and with
$\delta H_X = 0$.

\medskip

\noindent
{\em Proof:}
For a solution $\delta g_{ab}$,
our fundamental variation formula \eqref{fundid2} yields
\bena
W_\Sigma(g; \delta g, \pounds_\xi g) &=& \int_\infty [\delta Q_\xi(g) - i_\xi \theta(g;\delta g)]
-\int_B [\delta Q_\xi(g) - i_\xi \theta(g;\delta g)]  \non\\
&=& -\int_B [\delta Q_\xi(g) - i_\xi \theta(g;\delta g)] + \delta H_\xi \ .
\eena
Since $\xi^a \rightarrow cX^a$ for some constant $c$, we have $\delta H_\xi = c \delta H_X$.

We now evaluate the boundary term from $B$.
To do so, we write $n^a = (\partial/\partial u)^a, \ell^a = (\partial/\partial r)^a$, where $(u,r)$
are the Gaussian normal coordinates of \eqref{gaussian}. The condition
that $\xi^a|_B$ is tangent to the generators of $\H^+$ implies that we can
decompose it as
\ben
\xi^a = fn^a + uZ^a + rY^a \, ,
\een
where $Y^a,Z^a$ are smooth but
otherwise arbitrary.  In Gaussian null coordinates~\eqref{gaussian}, the metric perturbation
takes the form
\ben\label{lgaussian}
\delta g_{ab} = -2 \ \nabla_{(a} u \ ( r^2 \ \delta \alpha \nabla_{b)} u + r \ \delta \beta_{b)}) + \delta \mu_{ab} \
\een
where $n^a \delta \beta_a = \ell^a \delta \beta_a = n^a \delta \mu_{ab} = \ell^a \delta \mu_{ab} = 0$.

We now calculate $i_\xi \theta(g; \delta g) |_B$ [see
eq.~\eqref{thetadef}]. For the pull-back to $B$, we
have
\ben
(i_\xi \theta)_{a_1\dots a_{D-2}} = \frac{1}{16\pi} \ f \ v^c  n_c \ \epsilon_{a_1\dots a_{D-2}} \ ,
\een
where $\epsilon_{a_1\dots a_{D-2}}$ is the unperturbed intrinsic volume form on $B$.
From the definition, eq.~\eqref{vdef}, of $v^a$
we have on $B$
\bena
v^a n_a &=& g^{fh} n^e \nabla_f \delta g_{he} - g^{he} n^f \nabla_f \delta g_{he} \\
&=& -g^{fh} n^e \nabla_f(2r \nabla_{(h}u \delta\beta_{e)}) + g^{eh} n^f \nabla_f(2r \nabla_{(h}u \delta\beta_{e)}) + g^{fh} n^e \nabla_f \delta \mu_{ab} - g^{eh} n^f \nabla_f \delta \mu_{ab} \ .
\non
\eena
The first two terms on the right hand side can give a non-zero result on $B$ (i.e. $r=0$)
only if the derivative $\nabla_f$ acts on $r$, but in that case the tensor contractions
are seen to give a zero result. For the last two terms on the right side, we get, again on $B$,
\bena
v^a n_a &=& 2 n^{(f} \ell^{h)} n^e \nabla_f \delta \mu_{he} - 2 n^{(e} \ell^{h)} n^f \nabla_f \delta \mu_{he}
+ \mu^{fh} n^e \nabla_f \delta \mu_{he} - \mu^{eh} n^f \nabla_f \delta \mu_{he} \non\\
&=& - \mu^{eh} n^f \nabla_f \delta \mu_{he} -\mu^{fh}(\nabla_f n^e) \delta \mu_{he} \non\\
&=& -\mu^{ab} \pounds_n \delta \mu_{ab} + \half \delta \mu^{ab} \pounds_n \mu_{ab} \non\\
&=& -\mu^{ab} \pounds_n \delta \mu_{ab} = -2\delta \vartheta \ .
\eena
Here, we have used in the first line the expression for $g_{ab}$ in gaussian null
coordinates~\eqref{gaussian}, whereas to go to the second line we used that at least
one vector $n^a, \ell^a$ gets contracted into $\delta \mu_{ab}$ when all vectors
are pulled through the derivative, and this gives a zero result. To go to the third line
we used standard expressions for the Lie-derivative of a tensor, and that
$\mu^a{}_c \mu^b{}_d \nabla^c n^d = \half \pounds_n \mu_{ab}$ on $B$, which follows from
\eqref{gaussian}. To go to the fourth line we  used that $\pounds_n \mu_{ab} = 0$, as the background is
stationary. Thus, we have shown that
\ben
\int_B i_\xi \theta = -\frac{1}{8\pi} \int_B f \delta \vartheta \ \epsilon \, .
\een
Next, we calculate $\delta Q_\xi$.  When pulled back to $B$, we have
\ben
(\delta Q_\xi)_{a_1\dots a_{D-2}} = \frac{1}{16\pi} n^a \ell^b \nabla_{[a}(\delta g_{b]c} \xi^c) \  \epsilon_{a_1\dots a_{D-2}} + \frac{1}{32\pi} n^a \ell^b \nabla_{[a} \xi_{b]} \ \delta g_c{}^c  \  \epsilon_{a_1\dots a_{D-2}} \ .
\een
Since $\delta g_{ab} \xi^b = u\delta \mu_{ab} Z^b + r\delta \mu_{ab} Y^b$, we have on $B$,
\bena
2n^a \ell^b \nabla_{[a}(\delta g_{b]c} \xi^c) &=& n^a \ell^b (\nabla_{a} u \delta \mu_{bc} Z^c-
\nabla_{b} u \delta \mu_{ac} Z^c + \nabla_{a} r \delta \mu_{bc} Y^c-
\nabla_{b} r \delta \mu_{ac} Y^c) \non\\
&=& l^b \delta \mu_{bc} Z^c +  n^b \delta \mu_{bc} Y^c = 0 \ ,
\eena
as $\delta \mu_{ab} n^a = 0=\delta \mu_{ab} l^a$.
Thus, the first term on the right side of $\delta Q_\xi |_B$ is
zero. To calculate the second term, we use, on $B$,
\bena
&&2n^a \ell^b \nabla_{[a} \xi_{b]} \non\\
&=& n^a \ell^b [(\nabla_a f) n_b + (\nabla_a u) Z^b + (\nabla_a r) Y_b - (\nabla_b f) n_a - (\nabla_b u) Z_a -
(\nabla_b r) Y_a + 2f\nabla_{[a} n_{b]}]\non\\
&=&
n^a Y_a + l^a Z_a + n^a \nabla_a f + \frac{1}{2}f \ell^a \nabla_a (n^b n_b) - f\ell_b n^a \nabla_a n^b \non\\
&=&
l^a Z_a + n^a Y_a + n^a\nabla_a f
\eena
as $n^a \nabla_a r = n^a n_a = O(r)$, and as $\nabla_b(n^a n_a) =\nabla_b( r^2 \alpha )= O(r)$, and
$n^a \nabla_a n^b = O(r)$
since  $n^a$ is tangent to affinely parameterized null geodesics on $\H^+$. Thus, we obtain
\ben
\int_B \delta Q_\xi = \frac{1}{64\pi} \int_B
(n^a Y_a + l^a Z_a + n^a\nabla_a f) \delta \mu_b{}^b \ \epsilon \, .
\een

Thus, combining this with the previous result, we
have shown that, with $\xi^a =fn^a + uZ^a + rY^a$:
\ben\label{wanttovanish}
\int_B [\delta Q_\xi(g) - i_\xi \theta(g;\delta g)] = -\frac{1}{8\pi} \int_B \Big[ f\delta \vartheta \ \epsilon - \frac{1}{4}(n^aY_a + l^a Z_a + n^a\nabla_a f) \delta \epsilon\Big] \ ,
\een
and, hence,
\ben
W_\Sigma(g; \delta g, \pounds_\xi g) = \frac{1}{8\pi} \int_B \Big[ f\delta \vartheta \ \epsilon -
\frac{1}{4}(n^aY_a + l^a Z_a + n^a\nabla_a f) \delta \epsilon \Big] + c \delta H_X
\label{fZYc}
\een
It follows immediately from this formula that if $\delta g_{ab}$ is a solution
for which $\delta \epsilon |_B = 0 = \delta \vartheta |_B$
and $\delta H_X = 0$, then $W_\Sigma(g; \delta g, \pounds_\xi g)=0$. Conversely, if
$W_\Sigma(g; \delta g, \pounds_\xi g)=0$ for all $\xi^a$ of compact support, it follows
immediately from \eqref{fundid} with $E=0$ that $\delta g_{ab}$ satisfies the linearized
constraints. It then follows immediately from \eqref{fZYc} that if
$W_\Sigma(g; \delta g, \pounds_\xi g)=0$ for all
$f, Z^a,Y^a, c$, then $\delta \epsilon |_B =  \delta \vartheta |_B = \delta H_X = 0$.
 \qed

\section{Proof of lemmas~\ref{lemma3} and~\ref{lemma4}}

\noindent
{\bf Lemma 3.}
Let $Y=(N_0^{},N^a_0) \in C^\infty_0 \oplus C^\infty_0$. Then there exists a solution
$X$ in the space $\cap_k (\rho^2 W^{k+1}_\rho(\Sigma) \oplus \rho W^k_\rho(\Sigma)) \subset C^\infty(\overline \Sigma) \oplus C^\infty(\overline \Sigma)$
to the following boundary value problem:
\ben\label{pde}
\L\L^*(X) = Y \qquad \text{in $\Sigma$,}
\een
and $X = (N,N^a)$ satisfies
\ben\label{bcon}
N^a = N \eta^a \ , \quad \delta \vartheta =
\delta \epsilon = 0 \quad \text{on $B=\partial \Sigma$.}
\een
Furthermore, if $D \ge 5$, the solution is
unique, whereas for $D=4$, the solution is also unique unless $t^a$ is
tangent to the generators of the horizon (i.e., the black hole is nonrotating),
in which case the solution is unique up to $X^a \to X^a + ct^a$.

\medskip
\noindent
{\em Proof:}
The key ingredient in the proof is a Poincare-type inequality for certain
tensor fields $X = (N, N^a) \in W^2_\rho \oplus W^1_\rho$.
Tensor fields $ X  \in W^2_\rho \oplus W^1_\rho$ have, by the usual
theorems about Sobolev spaces, a
well defined restriction to $B$ in the space
$W^{3/2}(B) \oplus W^{1/2}(B)$, and the
restriction map is continuous. Consequently, the space
\ben\label{Xdef}
\X := \{X = (N, N^a) \in W^2_\rho
\oplus W^1_\rho \ \ \mid \ \ N^a = N \eta^a  \ \ \ \text{on $B$}\}
\een
is a closed subspace of $W^2_\rho \oplus W^1_\rho$. We first give the proof
for $D \geq 5$ and then give the modifications to the proof for $D=4$.

\medskip
\noindent
\underline{$D \ge 5$:}
For $X \in \X$, we
claim the following inequality in $D \ge 5$:
\ben\label{poincare}
C\|
X \|_{W_{\rho}^2 \oplus W^1_{\rho}} \ge \|\L^* M_\rho (X)\|_{L^2 \oplus L^2} \ge  c \|
X \|_{W_{\rho}^2 \oplus W^1_{\rho}} \ , \quad \quad
M_\rho = \left(
\begin{matrix}
\rho^2 & 0 \\
0 & \rho
\end{matrix}
\right) \ ,
\een
for non-zero constants $c,C$. The key inequality is the second one; the first inequality is merely a straightforward consequence of the
definition of $\L^*$, together with
the asymptotic conditions~\eqref{decay}
on the background spacetime
and the fact that $D^a \rho D_a \rho \le C, \rho^{2} D^a D^b \rho D_b D_b \rho \le C$, for other constants $C$. Combining the equality~\eqref{poincare} in the usual
way with the Lax-Milgram theorem~\cite{gilbarg}, we then infer the existence of
$\tilde X \in \X$ satisfying
\ben\label{LaxMilgr}
0=\langle \L^* M_\rho \tilde X | \L^* M_\rho \tilde \psi \rangle_{\mathcal K} + \langle \tilde Y | \tilde \psi \rangle_{\mathcal K}
\qquad \text{for all $\tilde \psi \in \X$,}
\een
for any source $\tilde Y \in C^\infty_0 \oplus C^\infty_0$ in dimensions $D \ge 5$.
Thus, taking $\tilde Y = M_\rho Y$, and $X = M_\rho \tilde X$, we get
a distributional solution $X$ to our equation~\eqref{pde} in the space $M_\rho \X$.

The solution is actually more regular, because, as is well-known~\cite{corvino1,cd}, the operator $\L\L^*$ is an elliptic matrix operator of mixed type in the sense of Agmon-Douglis-Nirenberg;
the leading derivative terms (``principal symbol'') are in fact
$\sigma_{\L\L^*} = \sigma_{\L^*}{}^T \sigma_{\L^*}$, with $X^a = (N,N^a)$ and
\ben
16 \pi \
\sigma_{\L^*}
\left(
\begin{matrix}
N\\
N^c
\end{matrix}
\right)
=
\left(
\begin{matrix}
\xi_a \xi_b - h_{ab} \xi_d \xi^d &  h^{-\half}(-p_{ab} \xi_c + 2\xi_d h_{c(a} p_{b)}{}^d )\\
2p^{ab} - \tfrac{2}{D-2} h^{ab} p & 2 h^\half \ \delta^{(b}{}_c \xi^{a)}
\end{matrix}
\right) \left(
\begin{matrix}
N\\
N^c
\end{matrix}
\right)
\ .
\een
As a consequence, the standard techniques for elliptic operators (as adapted to
systems of mixed type such as ours, see~p.~210 of~\cite{morrey}, and taking into account the weight factors of $\rho$) then give that the solutions $\tilde X$ are in fact in the Sobolev spaces
$\tilde X \in W^{k+4}_{\rho} \oplus W^{k+2}_{\rho}$ for all $k$, away from the boundary of $\Sigma$.
Actually, this regularity extends to the boundary $\partial \Sigma = B$ by thm.~6.1 of~\cite{schechter}, which is applicable\footnote{Note that the arguments in this paper are entirely local near $B$,
so it does not matter that our domain is not bounded.} in view of~\eqref{poincare}. By standard embedding theorems for Sobolev spaces, $\tilde X$, and also $X$, are in $C^\infty(\overline \Sigma) \oplus C^\infty(\overline \Sigma)$. It is then
also straightforward to see that $X$ solves the desired boundary value problem. First,
because $\tilde X \in \X$, we have $\tilde N^a = \tilde N \eta^a$, and because
$\rho$ is constant near $B$, the same holds for $X^a$. Thus, we satisfy
the first of the required boundary conditions. Next, because
$X$ is already known to be a weak solution of the equation~\eqref{pde}, and because it is smooth, it must satisfy the equation
in the classical sense. Furthermore, performing a partial integration in eq.~\eqref{LaxMilgr},
with $X=M_\rho \tilde X, Y = M_\rho^{-1} \tilde Y$, and smooth $\tilde \psi$ compactly supported in
$\overline \Sigma$, then shows that be boundary terms must vanish.
The boundary terms are obtained from \eqref{adjoint1}.
 Because $\psi$ has compact support in $\overline \Sigma$, only a boundary term from $B$ arises, which, with $(\delta h, \delta p) := \L^* M_\rho \tilde X$, is given by $0=\int_B [\delta Q_\psi(g) - i_\psi \theta(\delta g)]$, where $\psi=M_\rho \tilde \psi$.
In view of eq.~\eqref{wanttovanish} this implies that $\delta \vartheta |_B = 0 = \delta \epsilon |_B$.
These are precisely the last two remaining boundary conditions in~\eqref{bcon}.

Thus, what remains is to demonstrate the weighted Poincare-inequality~\eqref{poincare}. We first consider
smooth $X$ with compact support in a compact set $K \subset \overline \Sigma$, where
$\rho$ may be assumed to be $1$. In other words, such $X$ are zero in the asymptotic region, but not necessarily on $B$. For such fields, we can find a constant $c>0$ such that
\ben\label{prpoin}
c\|X\|_{W^2 \oplus W^1} \le \|\L^*(X)\|_{L^2 \oplus L^2} + \|X\|_{W^1 \oplus W^0} +
\Bigg| \int_B h^\half (D_b N_a N^b - D_b N^b N_a)\eta^a  \Bigg|^{\half}
\een
For a proof, see lemma~2.6 of~\cite{cd}. Now we assume that $N^a = N\eta^a$ on $B$. This means that, near $B$, we can write $N^a = N\eta^a + x Y^a$,
with $x= {\rm dist}_{B}$ the geodesic distance to $B$, and $Y^a$
some smooth tensor field.  Then the boundary term becomes
\bena\label{littlecalc}
&&\int_B h^\half \ (D_b N_a N^b - D_b N^b N_a)\eta^a \non\\
&=& \int_B h^\half \ (N D_b N \eta^b + N^2 \eta^a \eta^b D_b \eta^a
+N Y_a \eta^a - N D_b N \eta^b - N^2 D_a \eta^a - N Y_a \eta^a)\non\\
&=& -\int_B N^2 H \ h^\half = 0 \ ,
\eena
where $H$ is the trace of the extrinsic curvature of $B$ inside $\Sigma$.
To simplify the calculation, we have assumed that $\eta^a$ is extended away from $B$ such that
$\eta^a D_a \eta^b = 0$, and we used $D_a x = \eta_a$. Because $B$ is the bifurcation surface,
$H=0$, and the boundary term vanishes. Hence,~\eqref{prpoin} holds without the boundary term
if $X$ satisfies $N^a = N\eta^a$ on $B$, i.e. if $X \in \X \cap (C^{\infty}_0(K) \oplus C^\infty_0(K))$.

Consider next an open region $O \subset \Sigma$ such that $\overline O \subset K$.
It is shown combining~Prop.~7.2 and~Lemma~4.2 of~\cite{cd}, that for all
$X \in C^\infty_0(\Sigma \setminus O) \oplus C^\infty_0(\Sigma \setminus O)$, we have
\ben
c\|X\|_{W^2 \oplus W^1} \le \|\L^*M_\rho(X)\|_{L^2 \oplus L^2}
\een
for some constant $c>0$. We now combine this and eq.~\eqref{prpoin} (without boundary term) to obtain the desired
inequality~\eqref{poincare}. Suppose, to obtain a contradiction, that there is a sequence
$X_n \in \X$, such that $\|X_n\|_{W^2_\rho \oplus W^1_\rho}=1$, but $\| \L^* M_\rho(X_n)\|_{L^2 \oplus L^2} \to 0$. Let $\chi_1 + \chi_2=1$ be a partition of unity such
that ${\rm supp} \chi_1 \subset K, {\rm supp} \chi_2 \subset \Sigma \setminus O$. Then we estimate
\bena\label{kondra}
\|X_n\|_{W^2_\rho \oplus W^1_\rho} &\le& \|\chi_1 X_n\|_{W^2_\rho \oplus W^1_\rho} + \|\chi_2 X_n\|_{W^2_\rho \oplus W^1_\rho} \non\\
&\le& C \|\L^* (\chi_1 X_n) \|_{L^2 \oplus L^2} + C\| \chi_1 X_n \|_{W^1 \oplus W^0}
+ C\|\L^* M_\rho (\chi_2 X_n) \|_{L^2 \oplus L^2} \non\\
&\le&  C \|\chi_1 \L^*( X_n) \|_{L^2 \oplus L^2} + C\|[\L^*, \chi_1] X_n\|_{L^2 \oplus L^2} +
C\| \chi_1 X_n \|_{W^1 \oplus W^0} + \non\\
&&
C\|\chi_2 \L^* M_\rho (X_n) \|_{L^2 \oplus L^2} + C\|[\L^*, \chi_2] M_\rho X_n\|_{L^2 \oplus L^2}\non\\
&\le&  C \|\L^* M_\rho( X_n) \|_{L^2(\Sigma) \oplus L^2(\Sigma)} + C\| X_n \|_{W^1(O) \oplus W^0(O)}
\eena
with possibly new constants in each line. In the last step we used that the commutator $[\L^*, \psi]$
with a smooth compactly supported function $\psi$ decreases the order of each entry of the matrix
operator $\L^*$ by one unit (unless the order of the entry is already $=0$), so that
$[\L^*, \psi]: W^2 \oplus W^1 \to W^1 \oplus W^0$ is bounded. Now, by assumption,
$\| \L^* M_\rho(X_n)\|_{L^2 \oplus L^2} \to 0$ for the first term on the right side.
On the other hand, since $X_n$ is by assumption bounded in $W^2(O) \oplus W^1(O)$, it follows
from the Rellich-Kondrachov compactness theorem~(see e.g.~\cite{gilbarg}) that $X_n$ (or a subsequence thereof) is
Cauchy in $W^1(O) \oplus W^0(O)$. Hence, the above inequality shows that a subsequence of $X_n$
is Cauchy in $\X \subset W^2_\rho(\Sigma) \oplus W^1_\rho(\Sigma)$, hence convergent with limit $X \in \X$. Since the norm of $X_n$ in this space is $=1$, $X \neq 0$, and by the continuity of $\L^* M_\rho: \X \to L^2 \oplus L^2$, we learn that $\xi:=M_\rho X \in \rho^2 W^2_\rho \oplus \rho W^1_\rho$ satisfies $\L^*(\xi)=0$. So $\xi^a$ must be equal, almost everywhere, to a non-trivial Killing
vector field. However, a nontrivial Killing field in an asymptotically
flat spacetime must approach an infinitesimal Poincare transformation
near infinity. Consequently, in $D \ge 5$ spacetime dimensions, there are no
non-zero Killing for which $M_\rho^{-1} \xi \in L^2 \oplus L^2$, hence
a contradiction.

\medskip
\noindent
\underline{$D =4$:} In this dimension, the last statement is not true for the timelike
Killing field $\xi^a=t^a$ if the background is such that $t^a=(N,N^a)$
is tangent to the generators of the horizon: Indeed,
$M_\rho^{-1} t$ then is in $L^2 \oplus L^2$, and it satisfies the boundary condition $N^a = N\eta^a$.
Consequently, the Poincare inequality~\eqref{poincare} is not true in $D=4$ e.g. for a Schwarzschild background. However, in that case we can modify our argument as follows. We now supplement the definition of the space $\X$
eq.~\eqref{Xdef} by the condition that elements of $\X$ should be $L^2 \oplus L^2$ orthogonal to $M_\rho^{-1} t$. Then,
in this space, the Poincare inequality holds, and the rest of the argument goes through, giving
a unique solution to $\L^* \L(X) = Y$ in $M_\rho \X$ with the desired boundary conditions. The general solution is then simply $X^a + ct^a$.
\qed

\vspace{1cm}

\begin{lemma}\label{lemma4}
Let $j^a$ be a smooth vector field on $\Sigma$ such that
$j^a \in \rho^{-1} W^{k-1}_\rho$ for any $k>0$. Then there exists a unique, solution $X^a
\in \rho W_\rho^k$ to the boundary
value problem
\ben
D_a(D^{(a} X^{b)} - \frac{1}{D-1} h^{ab} D_c X^c) = j^b
\een
and
\ben
(h^{ab}-\eta^a \eta^b) X_a|_B = 0 \ , \quad
\eta^a D_a (\eta^b X_b) |_B = 0 \ .
\een
Furthermore $\|\rho^{-1} X^a\|_{W^k_\rho} \le C\|\rho j^a\|_{W^{k-1}_\rho}$.

The same statement is true for the boundary value problem for $\tilde X^a$
on the slice $\tilde \Sigma =
\Sigma \times {\mathbb T}^p$ in the black brane spacetime with spatial metric $\tilde h_{ab} = h_{ab} + \sum (dz_i)_a (dz_i)_b$, with $D_a$ replaced by $\tilde D_a$,
and with the coefficient in the trace term in the operator replaced by $1/(D+p-1)$.
\end{lemma}

\medskip
\noindent
{\em Proof:} Define $(PX)_{ab} = D_{(a} X_{b)} - \frac{1}{D-1} h_{ab} D^c X_c$.
The key ingredient of the proof of is the weighted Poincare inequality
\ben\label{poincare2}
c \int_\Sigma \rho^{-2} X_a X^a \ h^\half \le \int_\Sigma (PX)^{ab} (PX)_{ab} \ h^\half
\een
for some $c>0$. This inequality holds for any smooth $X^a$ satisfying the boundary conditions
stated in lemma~\ref{lemma4}. In order to prove this inequality, we consider the identity
\ben\label{iden}
\begin{split}
& (PX)^{ab} (PX)_{ab} = \\
& \half \ (D_a X_b) D^a X^b + D_a(X^{[b} D_b X^{a]}) - \half \ R_{ab}(h)
X^a X^b + \frac{D-3}{2(D-1)} \ (D_c X^c)^2 \ ,
\end{split}
\een
which holds for any vector field. Now let $O = \{\rho < C\}$ for some large $C$, and assume
that $X^a$ has compact support in $\overline \Sigma \setminus O$. Integrating the above identity then gives
\ben
\begin{split}
\half \int_\Sigma (D_a X_b) D^a X^b \ h^\half &\le \int_\Sigma (PX)^{ab} (PX)_{ab} \ h^\half
+ \half \int_\Sigma R_{ab}(h)
X^a X^b \ h^\half \\
& \le \int_\Sigma (PX)^{ab} (PX)_{ab} \ h^\half
+ \half \sup_{\overline \Sigma \setminus O} \rho^2 \ (R_{ab}(h) R^{ab}(h))^\half
\int_\Sigma \rho^{-2} \ X^a X_a \ h^\half \\
& \le \int_\Sigma (PX)^{ab} (PX)_{ab} \ h^\half
+ \epsilon \
\int_\Sigma \rho^{-2} \ X^a X_a \ h^\half
\end{split}
\een
for any given $\epsilon>0$, provided $C$ is chosen sufficiently large (here we
use that $R_{ab}(h) = O(\rho^{-(D-1)})$). Now from prop.~C.5 of~\cite{cd},
we also have for sufficiently large $C$
\ben
c \int_\Sigma \rho^{-2} \ X^a X_a \ h^\half \le \int_\Sigma (D_a X_b) D^a X^b \ h^\half \ ,
\een
and some $c>0$. It follows that inequality~\eqref{poincare2} holds if
$X^a$ is supported outside a sufficiently large $O$. Consider next a compact set $K$
such that $K \supset \overline O$, and let $X^a$ now be smooth and supported in $K$. Integrating now our identity~\eqref{iden} over $K$, we obtain
\ben\label{help2}
c\|X\|_{W^1} \le \|PX\|_{L^2} + \|X\|_{L^2} + \Bigg| \int_B h^\half (D_b X_a X^b - D_b X^b X_a)\eta^a  \Bigg|^{\half}
\een
for some $c>0$. Let now $\X = \{ X^a \in \rho W_\rho \mid 0 =
(h^{ab} - \eta^a \eta^b) X_b|_B \}$. Then the same calculation as in~\eqref{littlecalc}
shows that the boundary term vanishes for $X^a \in \X$. Thus, for $X^a \in \X$ we
have eq.~\eqref{poincare2} if the support of $X^a$ is outside $O$, whereas we
have~\eqref{help2} without boundary term if the support of $X^a$ is in $K$.
By considering now a partition of unity for the covering $O \cup K = \overline \Sigma$,
and arguing precisely as above around~\eqref{kondra}, we conclude that
the Poincare inequality~\eqref{poincare2} holds for any $X^a \in \X$, unless
there is an $X^a \in \X$ such that $(PX)_{ab}=0$.
By the definition of $P$, such a $X^a$ would be a conformal Killing
vector field for $h_{ab}$, and would hence have to approach a
conformal Killing vector field on $\mr^{D-1}$ in the asymptotic
region. In particular, the slowest possible fall-off of $X^a$ would be
that of a translation. However, by the definition of $\X$, we would
also have to have $\rho^{-1} X^a \in L^2$, which is impossible. This
establishes that the Poincare inequality holds for any $X^a \in \X$.

Combining the Poincare inequality in the usual way with the Lax-Milgram theorem on $\X$
(see lemma~\ref{lemma3} for a similar argument), we get a weak solution $X^a \in \X$ to our boundary value problem in the weak sense, i.e. $\langle P\psi | PX \rangle = -\langle \psi | j \rangle$
for all $\psi^a \in \X$. This also gives $\|\rho^{-1} X\|_{W^1_\rho} \le C \|\rho j\|_{L^2}$.
The usual bootstrap and scaling arguments then show $\|\rho^{-1} X\|_{W^k_\rho} \le C \|\rho j\|_{W^{k-1}_\rho}$ for any $k$, so the weak solution is a strong solution satisfying the Neumann condition $\eta^a D_a (\eta^b X_b)|_B=0$, whereas the remaining boundary conditions are already built into $\X$. This completes the proof of the lemma in the asymptotically flat case. The asymptotically KK case is exactly the same.
\qed

\section{Strengthened version of property~(3) of Proposition~\ref{prop2}}

In this Appendix, we strengthen property (3) of proposition~\ref{prop2} with the
following lemma:

\begin{lemma}\label{lemma5}
Let $(\delta h, \delta p)$ be a perturbation in the dense subspace
$\V \cap (W^{k+1}_\rho \oplus W^k_\rho)$ for sufficiently large $k$. Then for any
$\delta >0$,
we can find a new perturbation $(\delta h', \delta p')$ in the same subspace
having the property that it vanishes in a neighborhood of spatial infinity,
and is such that $\|(\delta h, \delta p)-(\delta h', \delta p')\|_{W^{k+1}_\rho \oplus W^k_\rho} < \delta$.
In fact, if $(\delta h, \delta p) \in \V \cap (C^\infty(\overline \Sigma) \oplus C^\infty(\overline \Sigma))$,
then $(\delta h',\delta p')$ can be chosen in $\V \cap (C^\infty_0(\overline \Sigma) \oplus C^\infty_0(\overline \Sigma))$, i.e. smooth and vanishing near spatial infinity.
\end{lemma}
\noindent
{\em Proof:} The proof of this lemma follows from the (much more general)
techniques and results in~\cite{corvino1,corvino2}, and also~\cite{cd}.
We consider a sufficiently large $R>0$ and the corresponding annular domain
$\Omega = \{ p \in \Sigma \mid R < \rho(p) < 2R \}$, with $\rho>0$ as usual
a function which is equal to the radial distance in the asymptotic region.
Furthermore, we introduce a function $x: \Omega \to \mr$ which in a small neighborhood
of $\partial \Omega$ is given by $x = {\rm dist}_{\partial \Omega}$, and
satisfies $x>0$ in the interior. A new norm for functions or tensor fields $u \in C^\infty_0(\Omega)$
is then introduced by
\ben
\| u \|_{W^k_{0,\rho x^2, \e^{-1/x}}} := \left\{ \sum_{n=0}^k \int_\Sigma h^\half (D_{(a_1} \cdots D_{a_n)} u)
D^{(a_1} \cdots D^{a_n)} u \ (\rho \ x^2)^{2n} \e^{-2/x} \right\}^\half \ \ ,
\een
and $W^{k}_{0,\rho x^2,\e^{-1/x}}$ is defined to be the closure of such functions or tensor fields in this norm.
Now let $\chi$ be a smooth function which is equal to 1 in the interior of $\Sigma$,
which interpolates between $1$ and $0$ strictly inside $\Omega$, and which is $0$ for $\rho>2R$.
We can choose this function so that $(D_{a_1} ... D_{a_n} \chi) D^{a_1} ... D^{a_n} \chi \le CR^{-2n}$.
Our first try for the compactly supported initial data is $(\chi \delta h, \chi \delta p)$. Then
for each $\delta>0,k\in \mathbb{N}$ we can find a sufficiently large $R$ so that $\|(\delta h, \delta p) -
(\chi \delta h, \chi \delta p) \|_{W^{k+1}_\rho \oplus W^k_\rho} < \delta$. Of course,
$(\chi \delta h, \chi \delta p)$ does not satisfy the linearized constraints; $\L(\chi \delta h, \chi \delta p) =: Y$
is a non-zero element $Y \in W^{k-1}_{0,\rho x^2, \e^{-1/x}}(\Omega) \oplus W^{k-1}_{0,\rho x^2, \e^{-1/x}}(\Omega)$ (with the usual decomposition
of $Y^a$ into lapse and shift understood). To satisfy the linearized constraints, we wish to correct $(\chi \delta h, \chi \delta p)$
by a suitable element of the form $-\e^{-2/x} M_{x}^2 \L^*(X)$, with $M_{x} = {\rm diag}(x^2,x)$ as above, i.e.
we wish to set
\ben
(\delta h',\delta p') := (\chi\delta h, \chi\delta p) -\e^{-2/x} M_{x}^2 \L^*(X)
\een
The factor $\e^{-2/x}$ is inserted in order that we can continue the second term sufficiently
smoothly by $0$ to all of $\Sigma$. Such a perturbation is then clearly compactly supported, and,
to satisfy the linearized constraints, $X$ would have to satisfy
\ben\label{Ysolve}
\e^{2/x} \L \e^{-2/x} M_x^2 \L^*(X) = \e^{2/x} Y \ .
\een
The operator $\e^{2/x} \L \e^{-2/x} M_x^2 \L^*$ on the left hand side is
shown in~\cite{cd} to be a bounded map
$$
{\mathfrak k}^\perp \cap (W^{k+4}_{0,\rho x^2, \e^{-1/x}}(\Omega) \oplus W^{k+2}_{0,\rho x^2, \e^{-1/x}}(\Omega)) \to
{\mathfrak k}^\perp \cap (W^{k}_{0,\rho x^2, \e^{-1/x}}(\Omega) \oplus W^{k}_{0,\rho x^2, \e^{-1/x}}(\Omega))
$$
where ${\mathfrak k} = {\rm span}(t^a, \psi_1^a, ..., \psi_N^a)$ is the span of the Killing fields,
and $\perp$ means orthogonal complement in $L^2(\Omega, \e^{-2/x} h^\half)$. Furthermore,
the operator has a bounded inverse. It follows from $\delta M, \delta J_A=0$ for the perturbation
$(\delta h, \delta p) \in \V$ [cf.~Proposition~\ref{prop2}] that
\bena
&&\langle \xi | \e^{2/x} Y \rangle_{L^2(\Omega,h^\half \e^{-2/x})}
= \langle \xi | \L(\chi \delta h, \chi \delta p) \rangle_{L^2(\Sigma)}
= \\
&&\langle \L^*(\xi) | Y \rangle_{L^2(\Sigma) \oplus L^2(\Sigma)} - \delta H_\xi(\delta h, \delta p)
= 0 + 0 \non
\eena
for any $\xi \in \mathfrak k$. Hence $\e^{2/x} Y$
is in ${\mathfrak k}^\perp$. Furthermore, for sufficiently large $R$, the $W^{k}_{0,\rho x^2, \e^{-1/x}}(\Omega) \oplus W^{k}_{0,\rho x^2, \e^{-1/x}}(\Omega)$-norm is arbitrarily small.
So we can invert~\eqref{Ysolve} with $X \in W^{k+3}_{0,\rho x^2, \e^{-1/x}} \oplus W^{k+1}_{0,\rho x^2, \e^{-1/x}}$
having small norm, and hence with $-\e^{-2/x} M_{x}^2 \L^*(X) \in
\e^{-2/x}M^2_x(W^{k+1}_{0,\rho x^2, \e^{-1/x}}(\Omega) \oplus W^{k}_{0,\rho x^2, \e^{-1/x}}(\Omega))$
having small norm in $W^{k+1}_\rho(\Sigma) \oplus W^k_\rho(\Sigma)$, where we extend the tensor fields by $0$ outside $\Omega$. Consequently, for sufficiently large $R$,
the $W^{k+1}_\rho(\Sigma) \oplus W^k_\rho(\Sigma)$-norm of
\ben
\left(
\begin{matrix}
\delta h'\\
\delta p'
\end{matrix}
\right)
-
\left(
\begin{matrix}
\delta h\\
\delta p
\end{matrix}
\right) =
\left(
\begin{matrix}
(\chi-1)\delta h\\
(\chi-1)\delta p
\end{matrix}
\right) -\e^{-2/x} M_{x}^2 \L^*(X)
\in W^{k+1}_\rho(\Sigma) \oplus W^k_\rho(\Sigma)
\een
is as small as we like. By construction $(\delta h', \delta p')$ satisfies the linearized
constraints, and has $\delta M=\delta {\bf P} = \delta {\bf J} = 0$
and $\delta \vartheta|_B = \delta \epsilon |_B = 0$. Thus, it is in the space $\V$ by proposition~\ref{prop2}.
As shown in prop.~5.7 of~\cite{cd}, if $(\delta h, \delta p) \in \V$ is
even smooth, then $-\e^{-2/x} M_{x}^2 \L^*(X)$ is a smooth pair of tensor fields on $\Omega$, which can be continued by $0$ outside of $\Omega$ and the resulting tensor fields $(\delta h',\delta p')$ are in $C^\infty_0(\overline \Sigma)$.
\qed

\end{appendix}

\end{document}